\documentclass[default]{sn-jnl}
\usepackage{lineno,hyperref}
\usepackage{ragged2e}
\usepackage{etoolbox}
\usepackage{geometry}
\usepackage{multirow}
\usepackage{ulem}
\usepackage{amsmath, amssymb}
\usepackage{rotating}
\usepackage{morefloats}
\usepackage{scalerel}
\usepackage{graphicx,parskip}
\usepackage{stackengine}
\usepackage{comment}
\usepackage{caption}
\usepackage{natbib}
\usepackage{url}
\usepackage{tikz}
\setcitestyle{authoryear, round}
\usepackage{fancybox,color}
\usepackage{booktabs}
\usepackage[utf8]{inputenc}
\usepackage[english]{babel}
\usepackage{tikz}
\usetikzlibrary{trees,arrows.meta,positioning}
\definecolor{bestcol}{HTML}{0B84F3}   
\definecolor{groupbg}{HTML}{F5F7FA}   

\tikzstyle{start} = [rectangle, rounded corners, minimum width=6cm, minimum height=1cm, text centered, draw=black, fill=red!30]
\tikzstyle{criteria} = [rectangle, rounded corners,  minimum width=7cm, minimum height=1cm, text centered, draw=black, fill=orange!30]
\tikzstyle{filter} = [rectangle, rounded corners, minimum width=8cm, minimum height=1cm, text centered, draw=black, fill=blue!20]
\tikzstyle{screen} = [rectangle, rounded corners, minimum width=8cm, minimum height=1cm, text centered, draw=black, fill=green!20]
\tikzstyle{arrow} = [thick,->,>=stealth]

\begin{document}

\title[Article Title]{A comprehensive review and analysis of different modeling approaches for financial index tracking problem}

\author[1]{\fnm{Vrinda} \sur{Dhingra}}\email{vdmath.iitr@gmail.com}

\author[1]{\fnm{Amita} \sur{Sharma}}\email{amita.sharma@nsut.ac.in}

\author*[2]{\fnm{Anubha} \sur{Goel}}\email{anubha.goel@tuni.fi}







\abstract{Index tracking, also known as passive investing, has gained significant traction in financial markets due to its cost-effective and efficient approach to replicating the performance of a specific market index. This review paper provides a comprehensive overview of the various modeling approaches and strategies developed for index tracking, highlighting the strengths and limitations of each approach. We categorize the index tracking models into three broad frameworks: optimization-based models, statistical-based models and machine learning based data-driven   approach. A comprehensive empirical study conducted on the S\&P 500 dataset demonstrates that the tracking error volatility model under the optimization-based framework delivers the most precise index tracking, the convex co-integration model, under the statistical-based framework achieves the strongest return-risk balance, and the deep neural network with fixed noise model within the data-driven framework provides a competitive performance with notably low turnover and high computational efficiency. By combining a critical review of the existing literature with comparative empirical analysis, this paper aims to provide insights into the evolving landscape of index tracking and its practical implications for investors and fund managers.}

\keywords{Index tracking, tracking error minimization, tracking portfolio, cardinality constraints, machine learning, deep learning}

\maketitle

\section{Introduction}
\label{sec1: intro}
Fund management (or investment management) is a core function within financial services that professionally manages assets across multiple asset classes to meet stated investment objectives on behalf of investors. Investors include individuals or institutions, such as insurance companies, pension funds, and corporations. Fund management encompasses activities such as asset selection, trading, monitoring, reporting to stakeholders, and internal auditing/governance. The primary objective is to balance capital growth and income over the medium to long term, subject to risk, cost, and regulatory constraints.

Over the past decades, the landscape of investment has evolved significantly, driven by advancements in financial technology, increasing market complexity, and shifting investor preferences. A key dimension of this evolution is the management of portfolios relative to benchmark indices such as the S\&P 500 or the Dow Jones Industrial Average (DJIA). Broadly, investment strategies can be classified into two categories: \textit{active} and \textit{passive} management and this review focuses on passive index tracking within this taxonomy. To delineate scope, the empirical illustration uses the S\&P 500; the modeling frameworks surveyed are general and apply across asset classes. Below is a description of each:

\begin{itemize}
    \item[(a)] \textit{Active management} involves a hands-on approach where fund managers actively make decisions about buying and selling securities with the aim to outperform a specific benchmark index. The goal is to generate higher returns than the benchmark by leveraging the manager's expertise, research, and market insights. Active strategies can range from traditional stock-picking to more structured approaches such as enhanced indexing, which blends benchmark replication with small, tactical deviations aimed at improving return performance.
    
    \item[(b)] \textit{Passive management} seeks to replicate the performance of a benchmark index, typically by holding the same securities in the same weights---or, when full replication is costly, a representative subset that closely approximates the index. The primary objective is to match benchmark returns with lower fees, reduced turnover, and limited risk of underperforming the benchmark. Common passive investment vehicles include index funds and exchange-traded funds (ETFs).
\end{itemize}

\begin{table}[htp!]
    \centering
    \caption{Comparison of Active and Passive Management Strategies}
    \begin{tabular}{lll}
        \noalign{\smallskip}\hline\noalign{\smallskip}
        \textbf{Aspect} & \textbf{Active Management} & \textbf{Passive Management} \\
      \noalign{\smallskip}\hline\noalign{\smallskip}
        Approach & Research and analysis driven & Index replication \\
        \noalign{\smallskip}\hline\noalign{\smallskip}
        Goal & Outperform the market & Match market performance \\
        \noalign{\smallskip}\hline\noalign{\smallskip}
        Costs & Higher fees due to active trading and research & Lower fees due to minimal trading and research \\
        \noalign{\smallskip}\hline\noalign{\smallskip}
        Risk & Potentially higher risk due to market timing & Generally lower risk with broader diversification \\
        \noalign{\smallskip}\hline\noalign{\smallskip}
        Risk Exposure & Company and market risk & Market risk\\
        \noalign{\smallskip}\hline\noalign{\smallskip}
        Returns & Potential for higher returns & Typically matches market returns \\
        \noalign{\smallskip}\hline\noalign{\smallskip}
        Suitable For & Investors seeking above-market returns & Long-term investors seeking market-matching returns \\
       \noalign{\smallskip}\hline\noalign{\smallskip}
    \end{tabular}
    \label{tab:activepassive}
\end{table}

\begin{figure}[htp!]
\centering
\includegraphics[height=6.2cm, width=0.62\textwidth]{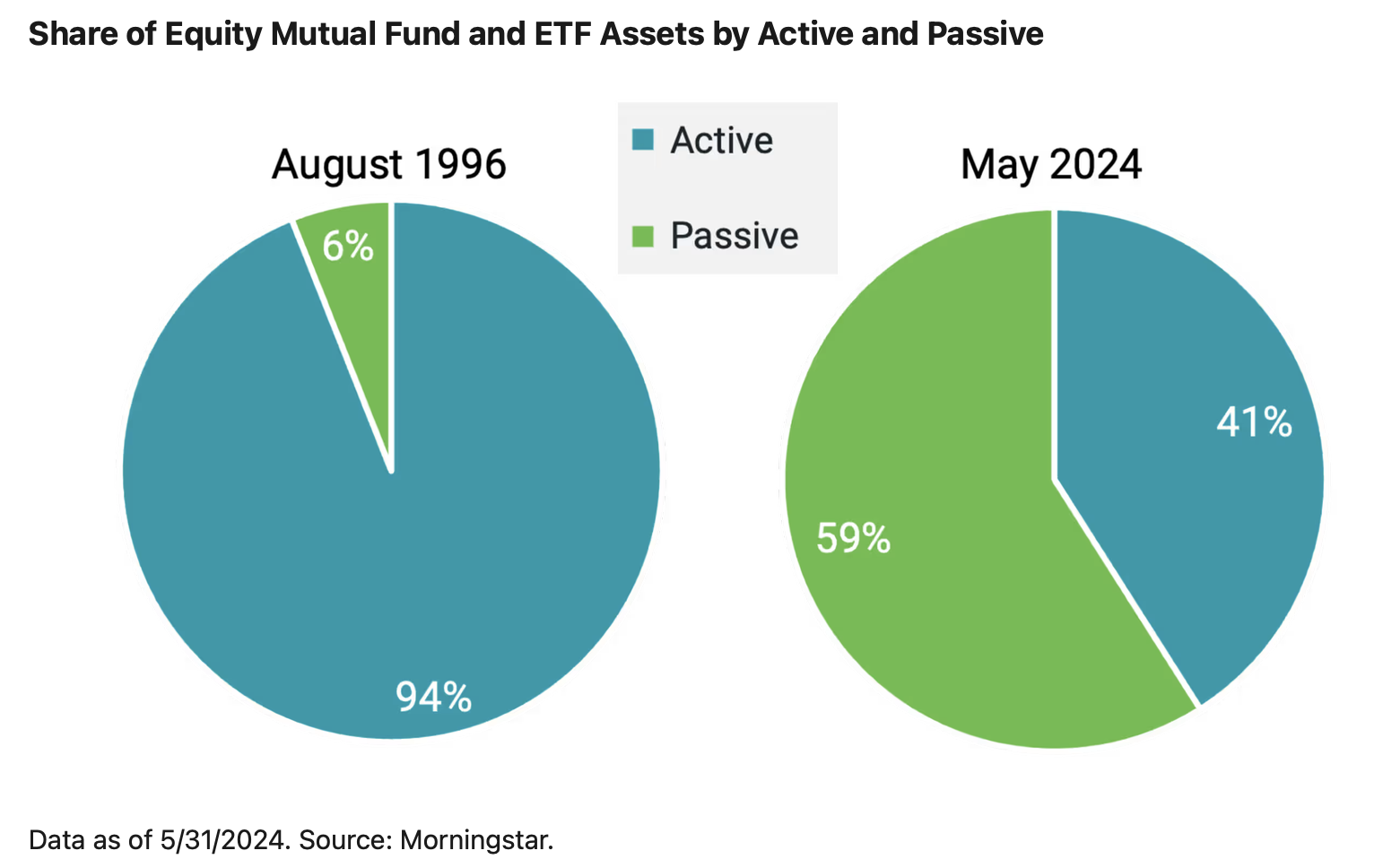}
\caption{Share of active and passive investment over three decades (Source: Morningstar).}
\label{fig:activepassive}
\end{figure}

Table \ref{tab:activepassive} summarizes the key differences between active and passive management across key dimensions. Over the years, passive investing has gained widespread popularity among investors and asset managers due to its simplicity, transparency, stability, and cost-effectiveness (see Figure \ref{fig:activepassive}). Empirical studies consistently show that a large proportion of actively managed funds fail to outperform their benchmarks over long horizons and net of fees, underscoring the appeal of passive alternatives \citep{activevspassive2003, activevspassive2018, activevspassive2020}. Within this context, index tracking---constructing a portfolio to replicate a benchmark’s risk–return profile while keeping tracking error and tracking difference small---has emerged as one of the most prominent passive investment approaches, offering a systematic and efficient means of replicating benchmark performance.

Index tracking portfolios such as index funds (IFs) and exchange-traded funds (ETFs) are the most widely adopted vehicles. An index fund is a pooled investment that mirrors a benchmark’s returns, typically via full or partial replication of its constituents, and prices once per day at net asset value (NAV). ETFs, by contrast, are similar in objective but differ in structure: they trade on stock exchanges throughout the day and use a creation–redemption process that keeps market prices close to NAV. In the market, one can invest in many different types of ETFs such as sector-based IF, broad market IF, market capitalization IF, Equal weighted IF, Smart Beta IF, International IF, and Debt IF among others.  

The origins of index tracking are commonly traced to the mid-1970s, when John Bogle of Vanguard Group launched the first index fund in 1976 \citep{bogle2011index}. Designed to replicate the performance of the S\&P 500, this fund provided a simple, low-cost way to obtain diversified exposure to the U.S. equity market. At the time, the concept marked a clear break from prevailing practices that emphasized active management and stock selection. Since then, index tracking has scaled globally, with providers offering products that track domestic and international benchmarks across equities, fixed income, and other asset classes. The product set broadened further with the advent of exchange-traded funds in the 1990s---e.g., the SPDR S\&P 500 Trust (SPY) in 1993---followed by regional and single-country exposures such as the U.S.-listed iShares MSCI Germany (EWG), alongside mutual-fund offerings like the Motilal Oswal Nifty 50 Index Fund (India).

\begin{figure}[htp!]
\centering
\includegraphics[height=6cm, width=0.5\textwidth]{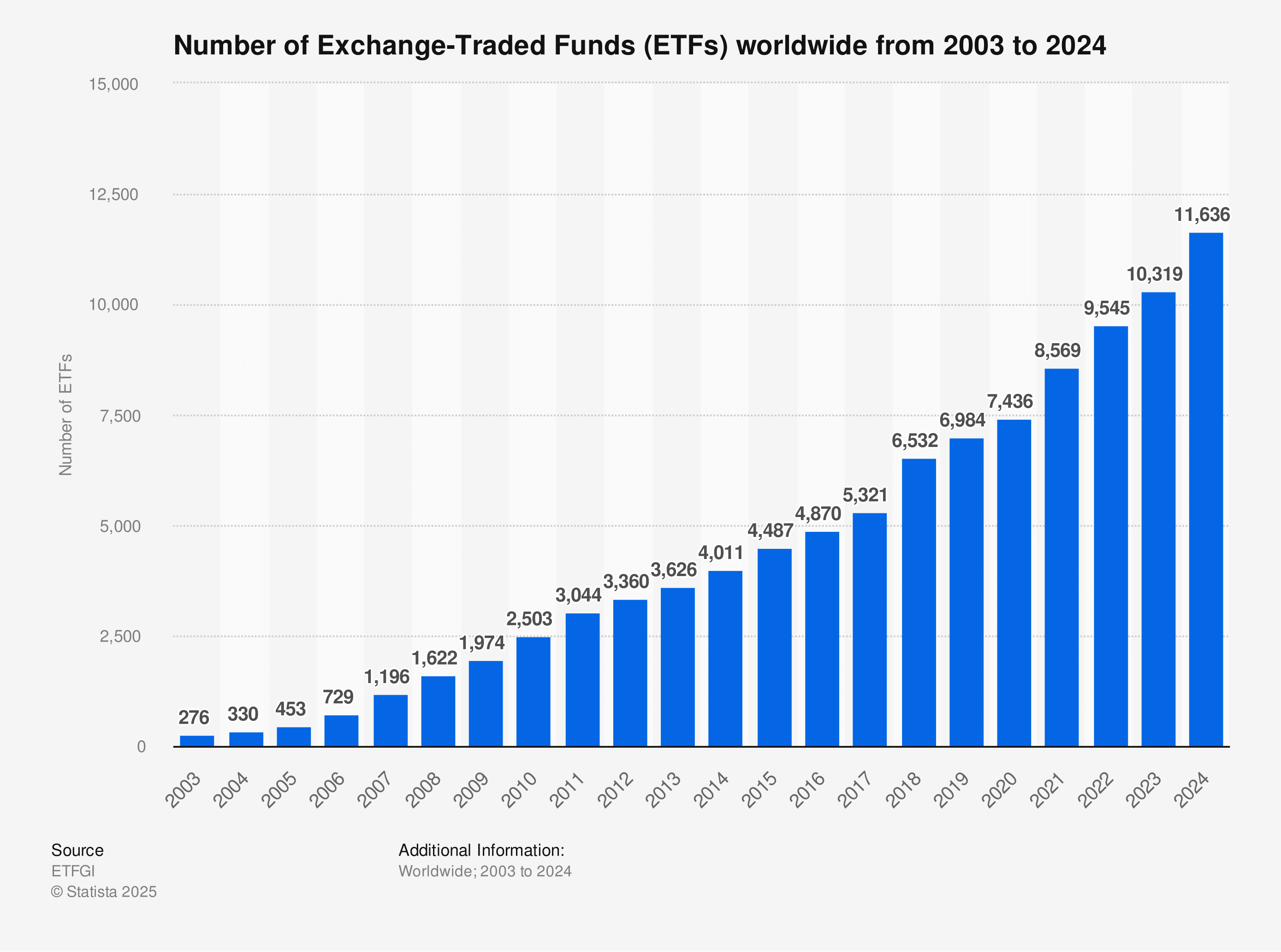}\hfill
\includegraphics[height=6cm, width=0.5\textwidth]{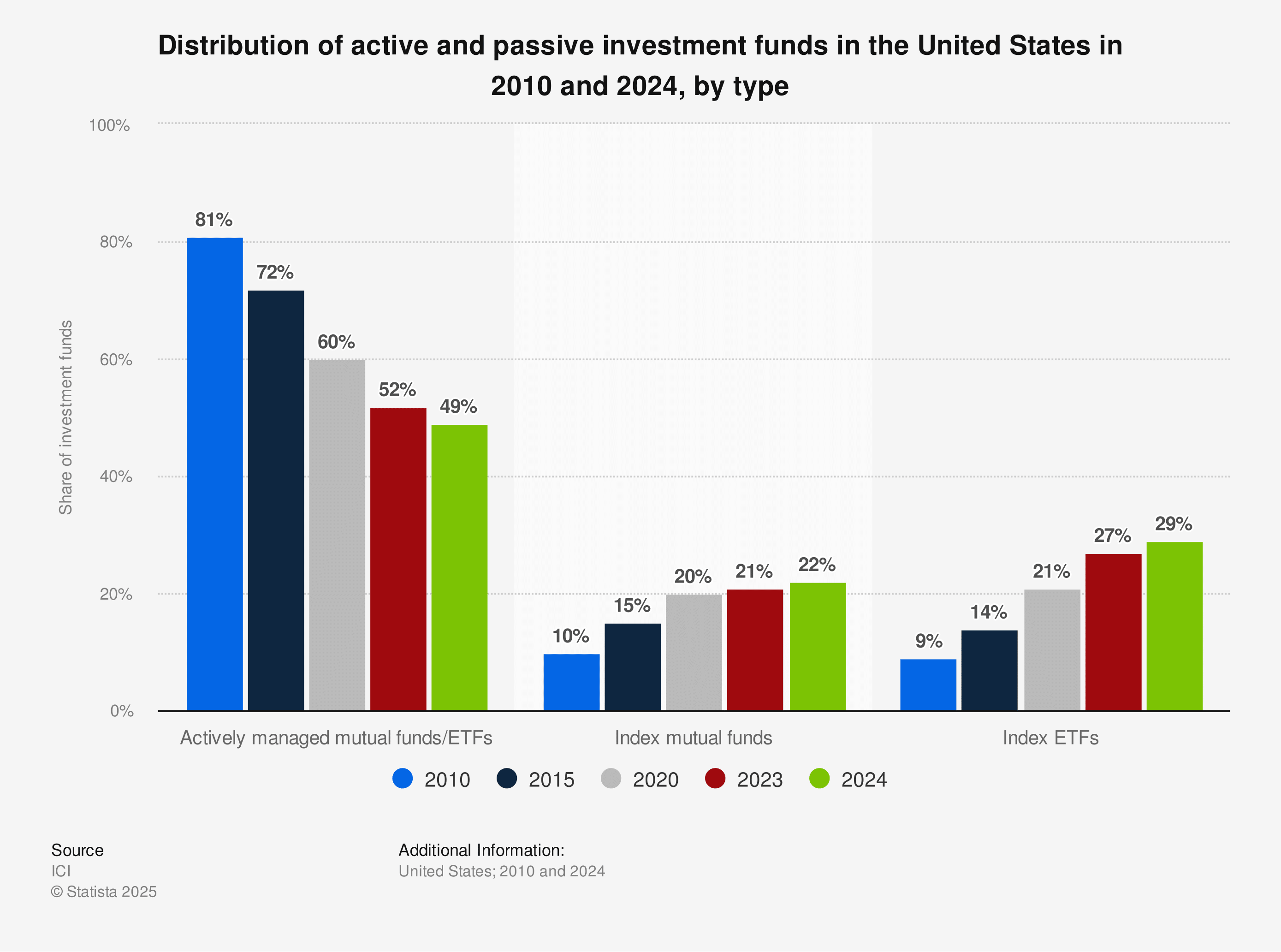}
\caption{Growth of passive investment over the years. \textit{Source: Statista 2025}}
\label{fig:globalETF}
\end{figure}

The scale of passive investing or index tracking specifically, has expanded markedly worldwide. As shown in Figure~\ref{fig:globalETF}, the number of ETFs increased from 276 in 2003 to 8{,}754 in 2022 worldwide, reflecting a structural shift toward low-cost, index-based vehicles. In the U.S., the share of actively managed funds declined from 81\% in 2010 to 52\% in 2023, while index mutual funds and ETFs collectively grew to nearly 60\% of the equity fund market.\footnote{Source: \url{https://www.americancentury.com/institutional-investors/insights/has-passive-investing-gotten-too-big/}} Parallel growth is evident in emerging economies. For instance, in India, ETF trading volume expanded from about 51 thousand crore INR in FY 2019--20 to 3.83 lakh crore INR in FY 2024--25, a more than sevenfold increase over five years.\footnote{Source: \url{https://www.zerodhafundhouse.com/blog/a-comprehensive-guide-to-exchange-traded-funds-etfs-in-india/}} These patterns highlight the global diffusion of passive investing, well established in developed markets and rapidly expanding in developing ones.





\textbf{Who should invest in index tracking portfolio?}
Index tracking portfolio aims to track market benchmark index, so their returns point to match those of the underlying index, with small differences due to fees and trading frictions. They are therefore, attractive to investors seeking predictable, broad market exposure with low costs and minimal portfolio management. These strategies offer an accessible and beneficial investment option for every investors, experienced or begineer. Unlike actively managed funds, which typically trade more and incur higher management fees as well as discretionary decisions, index funds are passively managed and provide diversification at scale. Because they are designed to match rather than beat the benchmark, these funds are efficient vehicle for long-term wealth accumulation and are particularly suited for cost-conscious, long-horizon investors or as the core allocation in a diversified equity portfolio. Investors should still expect full market ups and downs---these index trackers are not capital-protected.

\textbf{How are index tracking portfolios formed?}
The development of index tracking as a research area has its roots in the portfolio optimization literature, where the problem was first formalized as minimizing the tracking error between a constructed portfolio and its benchmark index. In practice, index replication can be achieved through two primary approaches: \textit{full replication} and \textit{partial replication}. Full replication involves holding all the securities in the benchmark index in the same proportions; it is feasible for narrow, liquid universes but becomes costly for broad indices with many constituents due to transaction, administrative, and liquidity frictions (see \cite{beasleyetal_2003_EJOR, CanakgozBeasley_2009EJOR} for a discussion).

Consequently, many index-tracking portfolios adopt \textit{partial replication}, selecting a representative subset of securities to approximate the index. This selection is typically governed by a \textit{cardinality constraint} that limits the number of assets in the portfolio, reducing trading costs and operational complexity while maintaining close alignment with the benchmark. The inclusion of such constraints renders the problem combinatorial and NP-hard, necessitating advanced computational methods and careful turnover control during rebalancing.

The earliest and most widely adopted formulation measures tracking error as the root mean squared error (RMSE) of excess returns, leading to a least-squares quadratic programming (QP) framework \citep{beasleyetal_2003_EJOR}. Over time, heuristic and metaheuristic techniques---genetic algorithms, simulated annealing, tabu search---have been proposed to handle these complex formulations efficiently \citep{beasleyetal_2003_EJOR, Guastaroba_2012EJOR_kernel, Derigs_metaheuristic_2003, Krink_2009_ANOR, Santanna_heuristicIT_2017}. For a comprehensive review of heuristic-based methods, see \cite{IMA_ITreview}.

The optimization framework has also been extended through alternative definitions of tracking error, including linear \citep{Rudolf_1999} or absolute deviations \citep{Guastaroba_2012EJOR_kernel}, as well as risk-based measures such as conditional value-at-risk \citep{Goel_MCVaR2018}. These developments aim to balance tractability, interpretability, and robustness---and to mitigate sensitivity to outliers and model misspecification. In parallel, statistical frameworks emerged that leverage relationships between index and constituent returns without necessarily solving a global optimization problem. Regression-based formulations \citep{CanakgozBeasley_2009EJOR, Li_quantregHDIT_2020}, regularization techniques \citep{Wu_nonneglasso_2014, Fastrich_qnorm_2014}, and cointegration \citep{Santanna_cointeg_2017, Santanna_cointeg_2020} or factor-based approaches \citep{CoriMarcellino_2006_FactorIT} expanded the toolkit, often yielding sparse, implementable portfolios suitable for high-dimensional settings.


More recently, artificial intelligence techniques have significantly broadened the methodological landscape of index tracking. Neural networks \citep{Ouyang_deepNN_2019, Zheng_stochasticNN_2020}, random forests \citep{randomforestIT_2022}, and deep autoencoders \citep{Zhang_autoencoder_2020} have been applied to capture complex, nonlinear dependencies between asset and index returns, allowing for more adaptive and potentially more efficient portfolio construction. \cite{dai2024deep} design a deep learning framework that learns dynamic trading policies under benchmark constraints and delivers superior out-of-sample tracking performance on S\&P 500 data. \cite{peng2023reinforcement} frame the problem using reinforcement learning, treating index tracking as an infinite-horizon sequential decision-making task with transaction costs; their deep reinforcement learning (RL) agent outperforms traditional models in long-run tracking accuracy and cost efficiency. \cite{zheng2020index} propose a stochastic neural network that incorporates cardinality constraints directly via reparameterization, enabling sparse and realistic index replicating portfolios with state-of-the-art tracking accuracy. In parallel, researchers have explored explainable artificial intelligence (XAI): \cite{zhang2024index} integrate SHAP (Shapley Additive Explanations) with deep autoencoders to produce interpretable stock selection mechanisms, showing improved performance and model transparency. Attention has also turned to robustness, \cite{bradrania2022state} introduce market-state aware models that adapt to structural shifts in market regimes, enhancing the resilience of index tracking strategies. These advances exemplify how cutting-edge AI methods---spanning deep learning, reinforcement learning, model interpretability, and adaptive generalization---can improve performance under real-world constraints, aligning index tracking with central challenges in modern AI research such as learning under structure, distributional shift, and constraint-aware inference.

\textbf{What do we offer in this article for passive management?}
In this review, we aim to provide a comprehensive analysis of the modeling approaches for index tracking (IFs/ETFs) developed over the past three decades. From traditional optimization-based methods to recent data-driven techniques, this paper not only chronicles the evolution of index tracking but also empirically evaluates these approaches using S\&P 500 data under common evaluation metrics (e.g., tracking error and turnover). A recent contribution by \cite{IMA_ITreview} provides a systematic review of solution methodologies for index tracking, covering exact and heuristic approaches through a bibliometric analysis of journals and focus areas. In contrast, the present review emphasizes the modeling frameworks themselves rather than the algorithmic solution techniques. Furthermore, our study complements the literature by providing an empirical comparison of these frameworks on real data, thereby offering practical insights into the relative effectiveness of different methodologies. An additional contribution of this article is the open access to the code, enabling transparency, reproducibility, and a standardized benchmark against which future researchers can compare their strategies. For the empirical comparison, we report daily results over almost 10 decades, starting from October 2012 to November 2022 and rebalance after every 3 months. All replication code and data-processing scripts are available at \url{https://github.com/vrindadhingra/index-tracking-review-}. This paper is thus intended as both a scholarly synthesis and a practical resource for academics and practitioners in passive investment, particularly index tracking.

The remainder of the article is organized as follows. Section \ref{Sec2} outlines the review methodology and article selection criteria. Section \ref{Sec3} discusses the evolution of index tracking approaches, categorizing milestones into optimization-based, statistical-based, and data-driven frameworks. Section \ref{Sec4} presents the empirical evaluation of representative models, and Section \ref{Sec5} concludes with key findings, implications, and directions for future research.

\section{Review Methodology}
\label{Sec2}
The methodology adopted in this review consists of three main stages:
(i) a systematic search for relevant research articles using well-defined keywords in a scientific database;
(ii) a structured filtering and screening process to obtain a final corpus of studies; and
(iii) a comprehensive empirical evaluation of the most prominent modeling frameworks for index tracking identified from the reviewed literature. The overall design of the review process is inspired by established survey methodologies in the artificial intelligence and financial optimization literature, particularly those adopted in \cite{AIML_review} and \cite{IMA_ITreview}.



\subsection{Selection and filtering of articles for review}









\begin{figure}[htp!]
\centering
\includegraphics[height=10cm, width=0.95\textwidth]{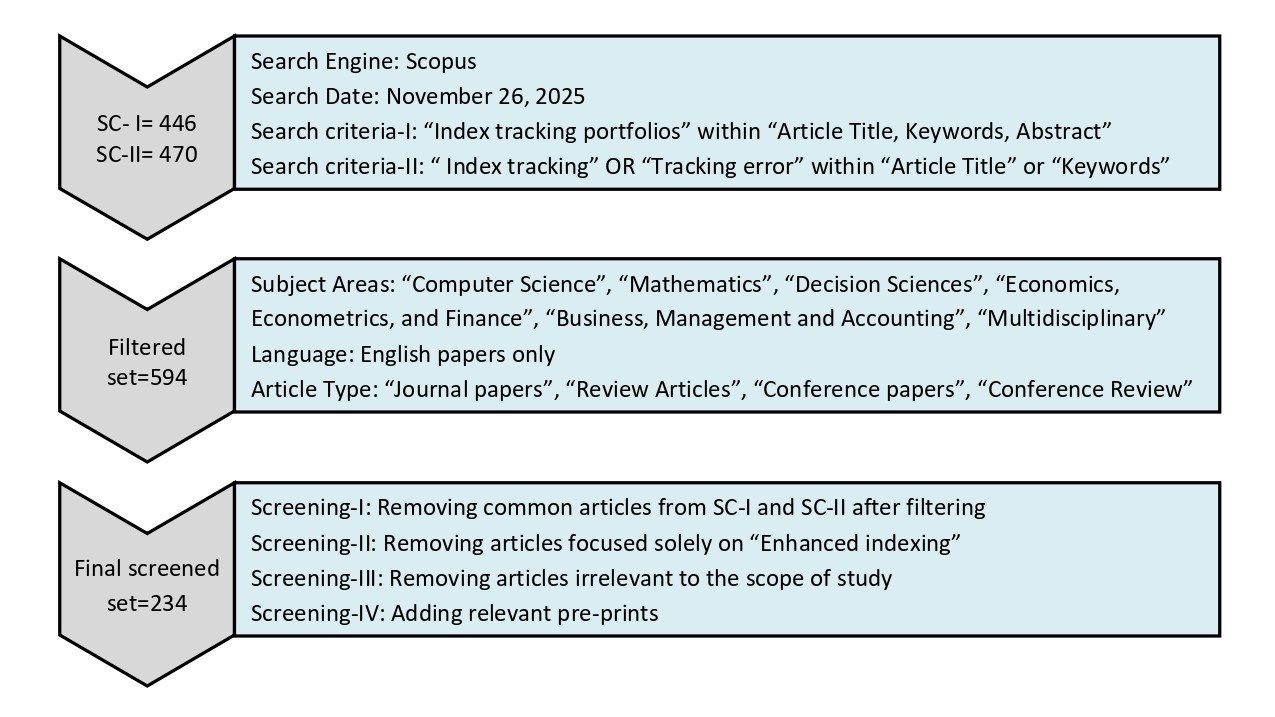}
\caption{Search and filtering of articles for review}
\label{fig:search}
\end{figure}

The initial step in constructing a comprehensive review involves identifying and selecting relevant research articles that have made significant contributions to the field of index tracking. As depicted in Figure \ref{fig:search}, the process for article selection is structured into three main stages: the \textbf{Search Step}, the \textbf{Filtering Step}, and the \textbf{Screening Step}. Each stage plays a critical role in narrowing the literature to a focused and high-quality set of studies that are directly aligned with the objectives of this review.


\begin{enumerate}
    \item[1.]{\textbf{Search Step}:} In the first stage, we conducted a broad literature search using the \textit{SCOPUS database}, chosen for its extensive coverage of peer-reviewed journals across finance, operations research, artificial intelligence, and applied mathematics. To ensure a comprehensive capture of index tracking–related research, two complementary search criteria were employed.

\begin{itemize}
    \item \textbf{Search Criterion-I (SC-I):} A keyword-based search was performed using the term \textit{``index tracking portfolios"} within the article title, abstract, and keywords. This primary search yielded \textbf{446 articles}, capturing studies that explicitly address index tracking portfolio construction.
    
    \item \textbf{Search Criterion-II (SC-II):} 
    To further expand the search and capture articles with alternative but related terminology, we conduct a secondary search using the keywords \textit{``index tracking"} or \textit{``tracking error"}. This search was more focused, specifically targeting the article title and keywords sections to ensure a higher relevance of articles discussing these core concepts. The secondary search yielded \textbf{470 articles}.
\end{itemize}

\noindent By applying these two criteria, we aim to ensure that all important and relevant articles on index tracking (IT) are included in the study.

\item[2.]{\textbf{Filtering step}}: Following the initial search, the collected articles were refined using SCOPUS's built-in filtering tools to eliminate studies outside the scope of this review. Specifically:

\begin{itemize}
    \item Only articles published in the English language were retained to ensure consistency in interpretation and analysis.
    \item As shown in Figure \ref{fig:search}, articles were further filtered based on subject area and article type, focusing on research that aligns with the scope of IT methodologies and financial analysis.
\end{itemize}

As a result of this filtering process, non-academic publications and studies not directly related to IT were excluded. After this stage, the literature set was reduced to 594 articles, comprising 375 articles from SC-I and 219 articles from SC-II.

\item[3.]{\textbf{Screening step}:} This is the final and most crucial stage in the selection process, ensuring that the articles included in this review are both unique and relevant. In this step:

\begin{itemize}
    \item Duplicate articles appearing across both search criteria or multiple filtered categories were removed.
    
    \item Each remaining article was examined to assess its relevance to pure index tracking. Studies focusing exclusively on enhanced indexation, unrelated portfolio optimization problems, or broader asset allocation frameworks were excluded.
    
    \item To incorporate the most recent developments, preprints indexed by SCOPUS were also reviewed. Of the 66 available preprints, \textbf{14} articles were identified as directly relevant to IT and not overlapping with already published works. These were included to ensure the review reflects the latest research trends.

\end{itemize}

\end{enumerate}

\begin{figure}[htp!]
\centering
\includegraphics[height=8cm, width=0.9\textwidth]{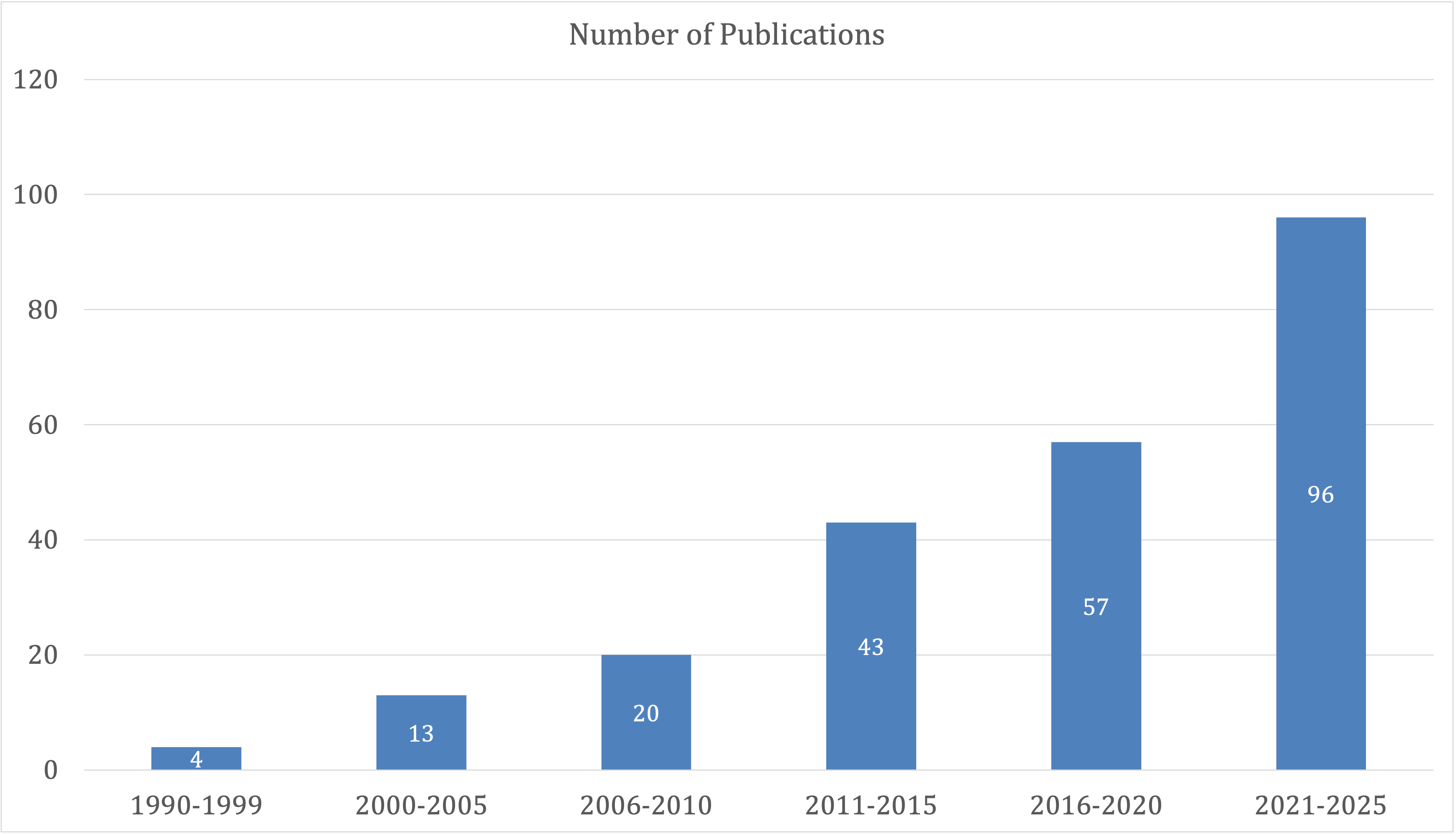}
\caption{Year-wise publication count}
\label{fig:pubcount_yearwise}
\end{figure}

The systematic application of the search, filtering, and screening steps resulted in a \textbf{final set of 233 articles}, forming a comprehensive and up-to-date body of literature on index tracking. Figure~\ref{fig:pubcount_yearwise} presents the year-wise distribution of the selected articles, highlighting the steady growth and increasing research interest in index tracking over time. In addition, Table~\ref{tab:top_articles_CC} reports the most highly cited articles within the selected corpus, offering insight into seminal contributions and influential studies that have shaped the field.

\begin{table}[htp!]
    \caption{Top articles on index tracking based on citation count}
    \begin{tabular}{llc}
        \noalign{\smallskip}\hline\noalign{\smallskip}
        \textbf{Author(s)} & \textbf{Title} & \textbf{TC} \\
        \noalign{\smallskip}\hline\noalign{\smallskip}
        \multirow{1}{*}{\cite{beasleyetal_2003_EJOR}} & An evolutionary heuristic for the index tracking problem & \multirow{1}{*}{266}\\
        \multirow{1}{*}{\cite{CanakgozBeasley_2009EJOR}} & Mixed-integer programming approaches for index tracking and enhanced   & \multirow{1}{*}{192} \\
        & indexation & \\
        \multirow{1}{*}{\cite{Rudolf_1999}} & A linear model for tracking error minimization & \multirow{1}{*}{141} \\
        \multirow{1}{*}{\cite{Guastaroba_2012EJOR_kernel}} & Kernel Search: An application to the index tracking problem & \multirow{1}{*}{139} \\
        \multirow{1}{*}{\cite{Dose_clustering_2005}} & Clustering of financial time series with application to index and enhanced  & \multirow{1}{*}{120} \\
        &  index tracking portfolio & \\
        \multirow{1}{*}{\cite{Frino_2001_SP500}} & Tracking S\&P 500 Index Funds & \multirow{1}{*}{109} \\
        \multirow{1}{*}{\cite{Gaivoronski_2005_benchtrack}} & Optimal portfolio selection and dynamic benchmark tracking & \multirow{1}{*}{102} \\
        \multirow{1}{*}{\cite{Ruiz_2009ANOR}} & Bond portfolio optimization problems and their applications to index & \multirow{1}{*}{97} \\
        & tracking: A partial optimization approach & \\
        \multirow{1}{*}{\cite{WuYangLiu_lasso_2014}} & Nonnegative-lasso and application in index tracking & \multirow{1}{*}{79} \\
        \multirow{1}{*}{\cite{KonnoWatanbe_1996}} & A hybrid optimization approach to index tracking & \multirow{1}{*}{77} \\
        \cite{Benidis_sparseHDIT_2018} & Sparse Portfolios for High-Dimensional Financial Index Tracking & 70\\
        \multirow{1}{*}{\cite{CoriMarcellino_2006_FactorIT}} & Factor based index tracking & \multirow{1}{*}{68} \\
        \multirow{1}{*}{\cite{ChenKwon_2012COR_robustIT}} & Robust portfolio selection for index tracking & \multirow{1}{*}{67} \\
        \multirow{1}{*}{\cite{Krink_2009_ANOR}} & Differential evolution and combinatorial search for constrained index-tracking & \multirow{1}{*}{62} \\
        \cite{StrubBaumann_rebalancingIT_2018} & Optimal construction and rebalancing of index-tracking portfolios & 59\\
        \multirow{1}{*}{\cite{Takeda_2013CMSc_sparseIT}} & Simultaneous pursuit of out-of-sample performance and sparsity in index  & \multirow{1}{*}{58} \\
        & tracking portfolios & \\
        \cite{FocardiFabozzi_clustering_2004} &	A methodology for index tracking based on time-series clustering	& 53\\
        \cite{Fastrich_qnorm_2014} & Cardinality versus q-norm constraints for index tracking & 48\\
       \multirow{2}{*}{\cite{Santanna_heuristicIT_2017}} & Index tracking with controlled number of assets using a hybrid heuristic & 39\\
& combining genetic algorithm and non-linear programming	 &\\
\noalign{\smallskip}\hline\noalign{\smallskip}
    \end{tabular}
    \label{tab:top_articles_CC}
\end{table}

\subsection{Sorting articles}

With the refined corpus of articles obtained from the screening process, the next step is to categorize and organize the literature according to the modeling frameworks employed for index tracking. Based on the methodological foundations adopted by prior studies, we broadly classify the selected articles into three major categories: optimization-based frameworks, statistical-based frameworks, and data-driven frameworks.

\begin{itemize}
    \item \textbf{Optimization-based frameworks:} This category comprises studies that formulate the IT problem as a mathematical optimization task. The primary objective in these approaches is to construct a tracking portfolio that closely replicates the benchmark index by minimizing a specified measure of tracking error or deviation, subject to constraints such as budget, cardinality, and risk exposure. 
    The literature in this stream explores a wide range of optimization techniques, including linear programming, quadratic programming, and combinatorial optimization. Commonly used tracking error measures include the mean squared error, sum of squared errors, standard deviation of tracking error, and mean absolute deviation. Depending on the choice of error metric and constraints, the resulting optimization problems may be linear or non-linear, often leading to computational challenges, particularly in large-scale settings.


    \item \textbf{Statistical-based frameworks:} This category includes studies that adopt statistical and econometric methodologies to design and analyze IT portfolios. These approaches focus on modeling the underlying relationship between asset returns and the benchmark index using techniques such as regression analysis, factor models, and cointegration methods. 
    Statistical-based frameworks emphasize interpretability and inference, enabling the identification of long-run equilibrium relationships, common risk factors, and explanatory variables that drive index movements. By exploiting the statistical structure of asset returns, these models aim to achieve effective tracking while maintaining economically meaningful portfolio compositions.


    \item \textbf{Data-driven frameworks:} This category encompasses research that leverages data-centric and machine learning methodologies for IT. Data-driven frameworks emphasize adaptability and predictive modeling, allowing for the dynamic incorporation of market information, historical data, and economic indicators. Studies in this stream employ a variety of machine learning techniques, including clustering algorithms, random forest models combined with regression, support vector regression, deep autoencoders, and neural networks. These approaches are particularly effective in capturing non-linear relationships and complex dependencies between asset returns and the benchmark index, and are often designed to enhance tracking performance under dynamic and volatile market conditions.

\end{itemize}

This classification is used to organize the subsequent review of index tracking methodologies.



\section{Evolution of Index Tracking framework}
\label{Sec3}

The intellectual foundations of index tracking are rooted in modern portfolio theory and equilibrium asset pricing. The mean–variance framework of \cite{markowitz1952} established the mathematical basis for constructing efficient portfolios, while the Capital Asset Pricing Model (CAPM) of \cite{sharpe1964} formalized the market portfolio as the optimal passive investment under equilibrium assumptions. These developments were further reinforced by the efficient market hypothesis of \cite{fama1970}, which argues that, in informationally efficient markets, persistent excess returns from active management are difficult to achieve. Empirical evidence supporting this view was provided by \cite{jensen1968}, who documented that actively managed mutual funds systematically underperformed market benchmarks on a risk-adjusted basis. 

Together, these theoretical and empirical insights laid the groundwork for passive investment strategies and directly motivated the emergence of index funds. This paradigm shift culminated in the launch of the Vanguard Index Fund in 1976, designed to replicate the performance of the S\&P~500 index. The success of this fund marked a turning point in asset management, establishing IT as a viable and cost-efficient alternative to active portfolio management.

Building on these conceptual foundations, \cite{Roll_1992} introduced the first formal optimization-based framework for IT by defining tracking error volatility (TEV) as the variance of the difference between portfolio and benchmark returns. By minimizing this quadratic deviation, Roll recast IT as a well-defined mean–variance optimization problem, thereby establishing a rigorous mathematical formulation for benchmark replication. This contribution marked a turning point, reframing IT as a precise mathematical optimization problem within the mean–variance paradigm.

Subsequent research extended this framework by proposing alternative measures of tracking error. In particular, \cite{Rudolf_1999} introduced linear tracking error metrics, including mean absolute deviation, mean absolute downside deviation, maximum deviation, and maximum downside deviation between the tracking portfolio and the benchmark. These formulations offered increased robustness to outliers and asymmetric return distributions. Nevertheless, quadratic tracking error measures—such as TEV—continued to dominate both academic studies and industry applications, owing to their smooth differentiability, analytical tractability, and compatibility with convex optimization techniques.

A major step toward practical IT was taken by \cite{beasleyetal_2003_EJOR}, who incorporated transaction costs and cardinality constraints into a root mean squared tracking error (RMSE) minimization framework. By explicitly limiting the number of assets held in the tracking portfolio, their formulation captured key real-world considerations faced by fund managers. However, the inclusion of cardinality and turnover constraints transformed the problem into a mixed-integer nonlinear program, rendering it NP-hard. To address this computational challenge, the authors proposed evolutionary population-based heuristics, demonstrating that near-optimal tracking portfolios could be obtained efficiently despite the combinatorial nature of the problem.

This contribution motivated a growing body of research on heuristic and metaheuristic solution methods for IT. Early developments include the simulated annealing approach of \cite{Derigs_metaheuristic_2003}, which employed a linear multi-factor model to estimate returns and covariances, and the factor-driven construction heuristic proposed by \cite{CoriMarcellino_2006_FactorIT}, where assets were selected sequentially based on their factor loadings relative to the benchmark. Subsequently, \cite{Zhuetal_2010_heuristic} introduced particle swarm optimization (PSO) to navigate the high-dimensional search space of IT portfolios more effectively.

Subsequent research extended these early heuristic approaches by developing hybrid algorithms that combine global exploration with local refinement, aiming to improve both convergence speed and solution quality. \cite{Scozzari_MSE_ANOR2013} proposed a hybrid evolutionary–local search framework for minimizing mean squared tracking error, demonstrating that embedding local improvement steps within evolutionary operators significantly enhances tracking accuracy. Their results highlighted the importance of balancing diversification and exploitation in large-scale IT problems.

Building on this idea, \cite{Santanna_heuristicIT_2017} introduced a family of specialized heuristics that integrate regression-based asset pre-selection with adaptive local search mechanisms. By reducing the dimensionality of the candidate asset universe prior to optimization, their approach achieved competitive tracking performance across multiple benchmark indices while substantially lowering computational time. More recent contributions continue this trend toward hybridization and algorithmic specialization. For example, \cite{heuristic2024} developed a hybrid simulated annealing framework tailored to cardinality-constrained IT, while \cite{metaheuristic_2025} proposed a harmony search–based metaheuristic employing problem-specific search operators and dual population initialization schemes to accelerate convergence in large asset universes.

Collectively, these studies establish heuristic and metaheuristic optimization as a central pillar of practical IT, particularly in settings where exact mixed-integer formulations become computationally infeasible. They also illustrate a gradual shift from generic metaheuristics toward domain-aware hybrid algorithms that explicitly exploit financial structure, sparsity, and benchmark dependence.


Alongside heuristic developments, another important research direction focused on improving the tractability of IT models through reformulation and convexification. In particular, several studies sought to replace quadratic tracking-error measures with linear or piecewise-linear alternatives, thereby transforming non-linear mixed-integer programs into linear or convex optimization problems that are easier to solve. A notable contribution in this direction is \cite{Guastaroba_2012EJOR_kernel}, which enabled more efficient solution methods while maintaining the integrity of the tracking objective.

While quadratic tracking error measures such as tracking error volatility remained dominant, alternative quadratic formulations were also explored to enhance numerical stability and computational efficiency. In particular, \cite{Xuetal_SES_OPMS2016} proposed the sum of errors squared (SES) as an alternative tracking objective, defined as the squared deviation between portfolio and benchmark returns aggregated over time. The resulting optimization model retains a quadratic structure but differs from TEV in its aggregation of deviations. To solve the associated problem efficiently, the authors developed a non-monotone projected gradient algorithm and demonstrated its effectiveness on large-scale IT instances. Despite its computational appeal, subsequent studies have observed that SES-based formulations may lead to highly concentrated portfolios when combined with cardinality constraints, highlighting a trade-off between numerical tractability and diversification.



In parallel with optimization-based formulations, a substantial body of literature has developed statistical frameworks for IT that exploit empirical relationships between index returns and constituent asset returns, often without explicitly solving a global tracking-error minimization problem. These approaches emphasize inference, interpretability, and structural dependence rather than direct optimization. One of the earliest contributions in this direction is the factor-based IT framework of \cite{CoriMarcellino_2006_FactorIT}, which models both the benchmark index and its constituents as being driven by a small number of latent common factors and idiosyncratic components. By identifying assets whose factor loadings closely resemble those of the index, the resulting portfolios are constructed to replicate index dynamics through shared systematic exposures rather than direct return matching.


Regression-based formulations constitute another prominent class of statistical IT models. These approaches seek to approximate benchmark returns as a linear combination of selected constituent asset returns, thereby aligning portfolio behavior with the index in expectation. A seminal contribution is due to \cite{CanakgozBeasley_2009EJOR}, who proposed a mixed-integer regression-based framework in which portfolio weights are chosen such that the regression of portfolio returns on index returns yields an intercept close to zero and a slope close to one.

Subsequent research extended regression-based IT by incorporating robustness and distributional considerations. In particular, quantile regression has been employed to address the limitations of least squares estimation under heavy-tailed and asymmetric return distributions. Early contributions by \cite{Mezali_quantileregITEI_2013} demonstrated that quantile-based tracking portfolios provide improved downside protection by focusing on conditional quantiles rather than mean behavior. This line of work was further developed in high-dimensional settings by \cite{Li_quantregHDIT_2020}, who introduced sparse quantile regression formulations that enable effective asset selection while maintaining robustness to extreme market movements. More recent studies, such as \cite{Aguilar_TPquantiles_2022}, explored multi-quantile and tail-probability tracking objectives, highlighting the relevance of quantile-based methods for risk-sensitive index replication.

Another influential strand of statistical IT research is built on regularization and sparse learning principles. These methods impose explicit penalties on portfolio weights to induce sparsity, thereby achieving asset selection without introducing binary decision variables. Early work by \cite{Wu_nonneglasso_2014} applied non-negative Lasso regression to IT, ensuring economically interpretable long-only portfolios while automatically controlling portfolio size. Extensions using Elastic Net penalties \citep{WuYang_elsticnet_2014} balance sparsity and stability by combining $\ell_1$ and $\ell_2$ regularization, mitigating the instability of pure Lasso solutions in the presence of highly correlated assets. More generally, $q$-norm regularization frameworks \citep{Fastrich_qnorm_2014} allow flexible control over sparsity levels and tracking accuracy. From an artificial intelligence perspective, these approaches can be viewed as early forms of supervised sparse learning in finance, where regularization plays a central role in preventing overfitting and enhancing generalization in high-dimensional asset spaces.

Cointegration-based approaches form another important class of statistical IT models, particularly suited for capturing long-run equilibrium relationships between asset prices and benchmark indices. Rather than matching short-term return fluctuations, these methods focus on identifying subsets of assets whose price processes share common stochastic trends with the index. Early contributions by \cite{Santanna_cointeg_2017} demonstrated that portfolios constructed from cointegrated assets exhibit strong long-term tracking properties, even when short-run deviations are present. This framework was further extended in \cite{Santanna_cointeg_2020}, which introduced convex cointegration formulations that improve numerical stability and scalability while preserving long-run alignment with the benchmark. Cointegration-based tracking portfolios are especially attractive in markets characterized by persistent common movements, as they provide robustness to transitory shocks and reduce sensitivity to short-term noise.

While early IT models primarily emphasized minimizing tracking error, subsequent research highlighted the importance of explicitly controlling downside risk and tail behavior, particularly during periods of market stress. To this end, several studies integrated coherent risk measures into the tracking framework. \cite{CVaRconstraint_Wang2012} introduced conditional value-at-risk (CVaR) constraints within a mean absolute deviation formulation, showing that downside risk could be effectively limited without materially degrading tracking accuracy. Building on this idea, \cite{Goel_MCVaR2018} proposed a two-tail mixed CVaR (TMCVaR) model that simultaneously penalizes extreme positive and negative deviations from the benchmark, thereby controlling both over- and under-performance. This formulation aggregates CVaR measures across multiple confidence levels, yielding improved robustness and more stable tracking behavior. More recent contributions, such as \cite{RiskIT_Sant'Anna2022} and \cite{AnisKwon_RABIT_2023}, further integrate dynamic or adaptive risk measures to jointly enhance tracking performance and risk resilience in volatile market environments.

In parallel with optimization- and statistical-based formulations, recent years have seen a growing body of research adopting data-driven and ML approaches for IT. This shift has been driven by the increasing availability of high-dimensional financial data and the need to model nonlinear, temporal, and structural dependencies among asset returns that are difficult to capture within classical parametric frameworks. Early data-driven contributions primarily relied on clustering techniques to identify representative subsets of assets whose collective behavior approximates that of the benchmark index. Notable examples include the hierarchical clustering approaches of \cite{FocardiFabozzi_clustering_2004} and \cite{Dose_clustering_2005}, which used Euclidean- and correlation-based distance measures to construct sparse tracking portfolios with reduced dimensionality.

Subsequent research extended these ideas by introducing more flexible similarity measures and richer representations of asset dynamics. \cite{Tang_softsubspace_clustering_2014} proposed soft subspace clustering, allowing assets to be grouped using adaptive feature weights rather than uniform distance metrics, thereby improving robustness in high-dimensional settings. Temporal structure was incorporated through k-medoids clustering with dynamic time warping (DTW) distances in \cite{Zhangetal_kmediod_clustering_2021}, enabling improved alignment of return trajectories across time. More recently, \cite{goel2024_TDA_clustering} introduced topological data analysis (TDA) to IT, using persistent homology to extract stable geometric features from asset return paths and form tracking portfolios based on structural similarity rather than pointwise distance.

A complementary strand of data-driven research focuses on learning nonlinear mappings between constituent asset returns and benchmark indices using supervised and unsupervised ML models. Autoencoder-based approaches \citep{Zhang_autoencoder_2020} learn low-dimensional latent representations that summarize common market structure, while deep latent representation models \citep{KimKim_deepLR_2020} and neural networks---including feedforward, stochastic, and recurrent architectures \citep{Zheng_stochasticNN_2020, KwakSongLee_NN_2021, Wangetal_LSTM_2024}---capture complex nonlinear and temporal dependencies. In parallel, tree-based and kernel methods have been explored for IT: random forests enable nonlinear interaction modeling and implicit feature selection \citep{Cao_RFstat_2022}, support vector regression (SVR) provides margin-based generalization in high-dimensional spaces \citep{Tengetal_SVR_2017}, and reinforcement learning frameworks sequentially adapt portfolio weights using reward signals linked to tracking performance \citep{peng2023_RL}.

Beyond prediction-oriented ML models, a growing literature emphasizes structural learning approaches that exploit similarity, geometry, and spectral properties of asset returns. \cite{zhangetal2025_reg_vs_clust} provide a systematic comparison of clustering techniques---including K-means, K-medoids, hierarchical clustering, and DTW-based methods---and show that while regression-based models often achieve superior tracking accuracy, correlation-based hierarchical clustering tends to generate portfolios with lower risk. Advances in TDA-based tracking have continued with sparse formulations that use persistent homology to guide regularization and asset selection without extensive cross-validation \citep{Goeletal_TDA2025}. Other recent contributions incorporate network, spectral, and factor-learning representations, such as combining Random Matrix Theory with network centrality measures to filter correlation matrices and identify influential assets \citep{francesca2025_RMT}, eigenvalue-matching approaches based on structured PCA for low-turnover tracking \citep{PCAIT_2025_cesarone}, and network-based models that embed adaptive community information to enhance risk-adjusted performance \citep{networkIT_2025}. Complementing these structural approaches, sparse learning methods---including non-convex $\ell_p$ regularization \citep{IT_normregularization_2025} and penalized Huber-loss regression \citep{HuberlossIT_2024}, continue to offer computationally efficient and robust solutions for high-dimensional IT.

The development of IT models reflects a gradual progression from early optimization-based formulations toward more flexible statistical and data-driven frameworks, shaped by both theoretical insights and practical implementation constraints. As discussed above, the existing literature can be organized into three broad and partially overlapping modeling paradigms: optimization-based approaches, statistical-based approaches, and data-driven or machine learning–based approaches. Each paradigm addresses the IT problem from a distinct methodological perspective and gives rise to different trade-offs in terms of tracking accuracy, interpretability, computational complexity, and scalability.

\begin{figure}[htp!]
\centering
\includegraphics[height=8cm, width=0.90\textwidth]{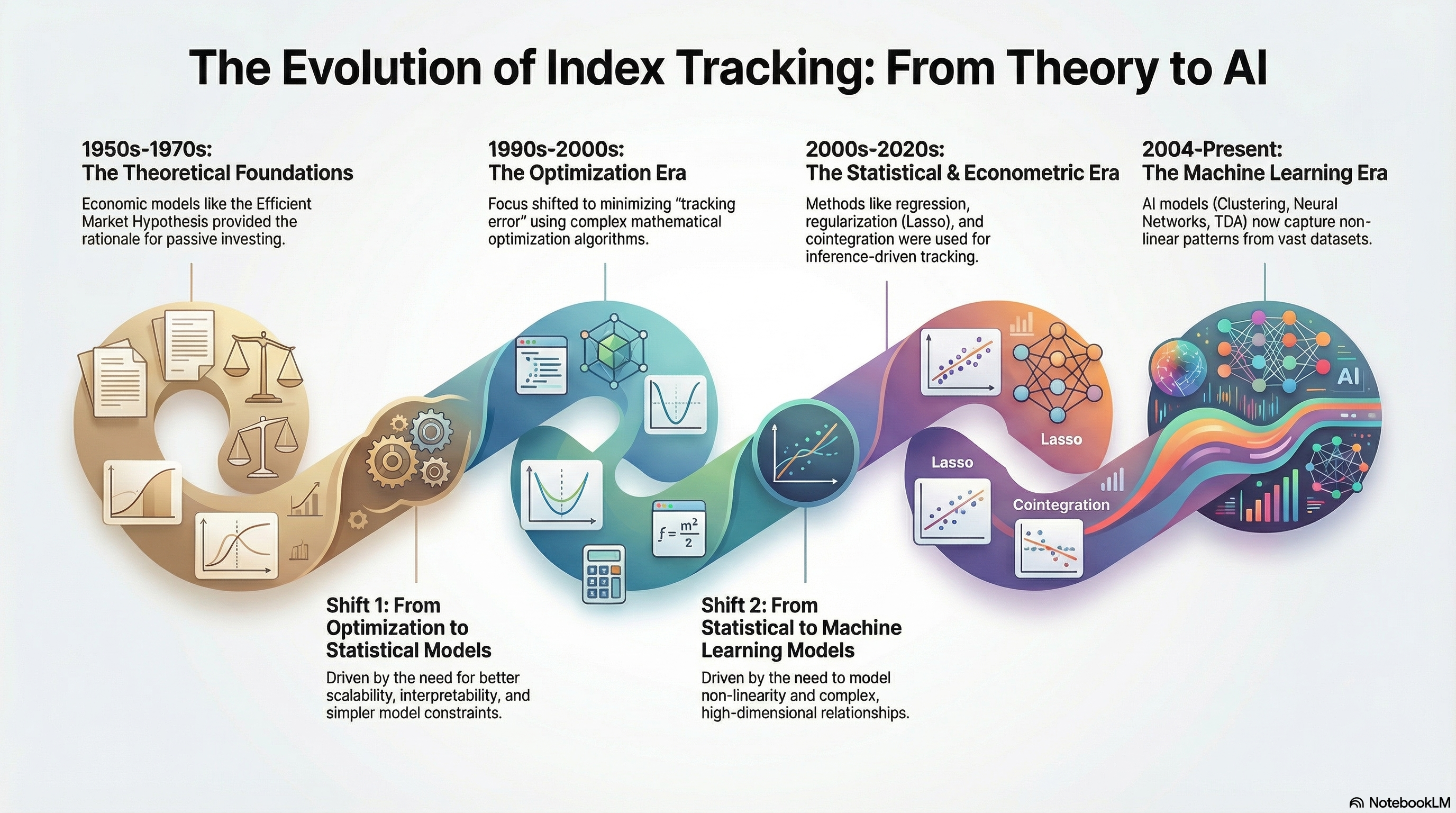}
\caption{A roadmap of the evolution of index tracking models: from theoretical development to AI (developed using Notebook LM)}
\label{fig:corr1_stat}
\end{figure}

In the remainder of this section, we review the principal modeling frameworks within each of these categories using a unified notation. Consider a universe of $n$ assets observed over $T$ time periods (or scenarios). Let $p_{i,t}$ denote the historical price of asset $i$ at time $t$, for $t=1,2,\ldots,T$. The return of asset $i$ at time $t$ is defined as
$
\displaystyle r_{i,t} = \frac{p_{i,t}-p_{i,t-1}}{p_{i,t-1}}.
$
Let $R = [r_{i,t}] \in \mathbb{R}^{T \times n}$ denote the matrix of asset returns. A tracking portfolio is represented by the weight vector $w=(w_1,w_2,\ldots,w_n)$, where $w_i$ denotes the proportion of capital invested in asset $i$. Denoting by $I_t$ the return of the benchmark index at time $t$, the return of the tracking portfolio at time $t$ is given by
$
\displaystyle R_{w,t} = \sum_{i=1}^n r_{i,t} w_i.
$

\label{sec: optimization framework}

\subsection{Optimization based framework} 
In this section, we discuss the important index tracking models developed over the years that use an optimization based framework for determining the index tracking portfolio. Table \ref{tab:OPT_top_articles} lists the best articles based on different tracking error metrics according to their citation score. We explain all these methods in details with their mathematical modeling. Typically, an optimization-based index tracking problem aims at minimizing the tracking error between the returns of the portfolio and the benchmark index. 

\begin{table}[htp!]
    \caption{Classification of top articles under optimization-based approaches for index tracking based on different tracking error metrics used}
    \begin{tabular}{lllc}
        \noalign{\smallskip}\hline\noalign{\smallskip}
       \textbf{Tracking Error} & \textbf{Author(s)} & \textbf{Title} & \textbf{TC} \\
        \noalign{\smallskip}\hline\noalign{\smallskip}
            \multirow{4}{*}{RMSE} & \cite{beasleyetal_2003_EJOR} & An evolutionary heuristic for the index tracking problem & \multirow{1}{*}{266} \\
           & \cite{Gaivoronski_2005_benchtrack} &	Optimal portfolio selection and dynamic benchmark tracking	& 102\\
	& \cite{Krink_2009_ANOR} &	Differential evolution and combinatorial search for constrained  & 62\\
 & & index-tracking & \\
        \noalign{\smallskip}\hline\noalign{\smallskip}
         \multirow{5}{*}{MSE} & \cite{Ruiz_2009ANOR}	& A hybrid optimization approach to index tracking & 77\\
	 & \cite{Santanna_heuristicIT_2017} & Index tracking with controlled number of assets using a hybrid  & 39\\
 & & heuristic combining genetic algorithm and non-linear programming & \\
	 & \cite{Scozzari_MSE_ANOR2013} & Exact and heuristic approaches for the index tracking problem & 46\\
  & & with UCITS constraints  & \\
   \noalign{\smallskip}\hline\noalign{\smallskip}  
          \multirow{4}{*}{SES} & \cite{Xuetal_SES_OPMS2016}& An efficient optimization approach for a cardinality-constrained  & 42\\
          & & index tracking problem & \\
	& \cite{Wang_SES_2018} & An index tracking model with stratified sampling and optimal 	& 6 \\
 & & allocation & \\
    \noalign{\smallskip}\hline\noalign{\smallskip}  
           \multirow{4}{*}{TEV/TESD} & \cite{Derigs_metaheuristic_2003} &	Meta-heuristic based decision support for portfolio optimization &	39\\
           & &  with a case study on tracking error minimization in passive &\\
           & & portfolio management & \\
& \cite{Mutunge_TEV_COR2018} & Minimizing the tracking error of cardinality constrained portfolios & 37\\
\noalign{\smallskip}\hline\noalign{\smallskip}  
           \multirow{2}{*}{MAD/AD} & \cite{Rudolf_1999} &	A linear model for tracking error minimization & 141\\
            & \cite{Guastaroba_2012EJOR_kernel} & Kernel Search: An application to the index tracking problem & 139 \\
          \noalign{\smallskip}\hline\noalign{\smallskip}  
           \multirow{1}{*}{MADD} & \cite{Rudolf_1999} &	A linear model for tracking error minimization & 141\\
            \noalign{\smallskip}\hline\noalign{\smallskip}  
           \multirow{1}{*}{Min-Max} & \cite{Rudolf_1999} &	A linear model for tracking error minimization & 141\\
            \noalign{\smallskip}\hline\noalign{\smallskip}  
           \multirow{1}{*}{DMin-Max} & \cite{Rudolf_1999} &	A linear model for tracking error minimization & 141\\
    \noalign{\smallskip}\hline\noalign{\smallskip}  
           \multirow{2}{*}{MCVaR} & \cite{Goel_MCVaR2018} & Index tracking and enhanced indexing using mixed conditional & 28\\
           & &  value-at-risk & \\
           \noalign{\smallskip}\hline\noalign{\smallskip} 
           
    \end{tabular}
    \label{tab:OPT_top_articles}
\end{table}
%

The standard optimization-based index tracking problem (IT-OPT), limiting the number of assets in the tracking portfolio is a mixed integer programming problem, given as:
\begin{center}
\begin{equation}
\label{eqn:minTE}
 \qquad \ \text{(IT-OPT) \  \qquad Min}  \quad \text{TE}(R_{wt}, I_t)  \qquad \qquad \qquad \qquad \qquad 
    \end{equation}
    subject to \ \ \qquad  \qquad 
    \begin{equation}
    \label{eqn:size}
       \qquad \qquad \qquad \qquad  \epsilon_{i} z_{i} \leq w_{i} \leq \delta_{i}z_{i}, \ \ i=1, \ldots, n,  
    \end{equation}
    \begin{equation}
    \label{eqn:budget}
       \quad  \sum\limits_{i=1}^{n} w_{i}=1,
    \end{equation}
    \begin{equation}
    \label{eqn:cardinality}
     \quad \sum\limits_{i=1}^{n} z_{i} \leq K,
    \end{equation}
    \begin{equation}
    \label{eqn:binaryZ}
     \qquad \qquad \qquad \ z_{i} \in \{0,1\}, \ \ i=1, \ldots, n.
     \end{equation}
\end{center}
\smallskip
Here, the objective in \eqref{eqn:minTE} aims to minimize the tracking error function TE$(R_{wt}, I_t)$ between the returns of the index $I_t$ and the that of the tracking portfolio $R_{wt}; \, t=1,2,\ldots,T$. The constraint in \eqref{eqn:cardinality} is the cardinality constraint that limits the number of assets in the tracking portfolio to be $K$, where $0 < K < n$. The variables $z_{i}$ are binary variables; if $z_{i}=1,$ then asset $i$ is included in the portfolio, and $z_{i}=0$ implies asset $i$ is not included in the portfolio. The constraint \eqref{eqn:size} is called the holding constraint or the size constraint that ensures that if asset $i$ is included in the portfolio, then the weight $w_{i}$ is bounded between user defined parameters, $\epsilon_{i}$ and $\delta_{i}.$

\noindent These constraints are present in almost all mathematical models associated with the index tracking problem. What changes is the TE function TE$(R_{wt}, I_t)$, which can take different structure forms, linear to non-convex. We now describe some commonly used TE functions, shortlisted from the literature, as described in Table \ref{tab:OPT_top_articles}.

\subsubsection*{Different tracking error functions and their corresponding optimization framework}
\begin{enumerate}

\item \textbf{Tracking Error Variance (TEV)}:
The variance-based tracking error is the earliest and most widely adopted measure of tracking performance in index tracking literature \citep{Roll_1992, LarsenResnick_1998_TEV, Rudolf_1999}. It quantifies the variability of the difference between the portfolio return and the benchmark return over time, reflecting the consistency of the replication. The tracking error variance (TEV) is defined as
\[
\text{TEV} = \sigma
^2(R_{wt} - I_{t}),
\]
and its square root, the tracking error standard deviation (TESD), is given by
\[
\text{TESD} = \sqrt{\sigma^2(R_{wt} - I_{t})}.
\]
In matrix form, if $x= w-b \in \mathbb{R}^n$ denotes the vector of active portfolio weights, where $b \in \mathbb{R}^n$ is the weight vector of the benchmark index and $\Sigma \in \mathbb{R}^{n \times n}$ the covariance matrix of asset returns, then TEV can be written as
\[
\text{TEV} = x^\top \Sigma x,
\]
subject to the standard portfolio constraints
\[
\sum_{i=1}^{n} w_i = \sum_{i=1}^{n} b_i = 1, \quad 0 \leq w_i \leq 1 
\]
which translates into
\[
\sum_{i=1}^{n} x_i = 0, \text{and}  -b_i \leq x_i \leq 1 - b_i \quad \text{for } i=1,2,\ldots,n.
\]
Thus, the final model that minimizes TEV (or equivalently, TESD) takes the following form:
 \vspace{-0.5cm}
    \begin{center}
    \begin{equation*}
    \label{eqn:minMAD}
    \qquad \ \text{(TEV) \  \qquad Min  \quad } x^\top \Sigma x  \qquad \qquad \qquad \qquad \qquad  \qquad \qquad \qquad
    \end{equation*}
    subject to \qquad \qquad \qquad \qquad \qquad \quad 
    \begin{equation*}
        \sum_{i=1}^{n} x_i = 0, \qquad \qquad \qquad \qquad
    \end{equation*}
    \begin{equation*}
         -b_i \leq x_i \leq 1 - b_i, \ i=1,2, \ldots, n
    \end{equation*}
      \eqref{eqn:cardinality}---\eqref{eqn:binaryZ} \qquad \qquad \qquad \qquad \qquad \qquad
    \end{center}

Minimizing TEV (or equivalently, TESD)  yields a portfolio that minimizes the volatility of its deviation from the benchmark index. Despite subsequent criticism for ignoring systematic bias, variance-based measures remain central in both academic research and industry practice due to their convexity, interpretability, and empirical robustness in realistic index tracking applications.


\vspace{0.5cm}

\item \textbf{Mean Squared Error (MSE):} 
While variance-based tracking error measures such as TEV gained early popularity due to their simplicity and analytical tractability, they were later criticized for being ``centered'' measures—that is, they quantify the dispersion of the portfolio's excess return around its mean, rather than around zero. As a result, a portfolio with a consistent bias (i.e., one that systematically over or under performs the index by a constant amount) could exhibit a low variance-based tracking error despite failing to accurately replicate the benchmark. To address this limitation, \cite{beasleyetal_2003_EJOR} introduced a generalized tracking error function that measures deviations directly from the benchmark returns without centering around the mean. This family of measures is expressed as
\[
\text{TE}_{\alpha} = \dfrac{1}{T} \left[ \sum_{t=1}^{T}\left|I_{t}-R_{wt}\right|^{\alpha}\right]^{(1/\alpha)} ,
\]
where $\alpha$ determines the degree of penalization for large deviations. A commonly adopted special case is the Root Mean Squared Error (RMSE), corresponding to $\alpha = 2$, defined as
\[
\text{RMSE} = \sqrt{ \frac{1}{T} \sum_{t=1}^{T}\left(I_{t}-R_{wt}\right)^{2}}.
\]
The inclusion of the square root renders the RMSE function non-linear and non-convex, leading to a mixed-integer non-linear programming (MINLP) formulation. Consequently, heuristic or metaheuristic algorithms are typically employed to obtain near-optimal solutions. For instance, \cite{beasleyetal_2003_EJOR} proposed a population-based heuristic for RMSE minimization, and subsequent studies have adopted similar approaches \citep{Gaivoronski_2005_benchtrack, Krink_2009_ANOR, Zhuetal_2010_heuristic, Andriosopoulos_2013_evolalgo}. 
Another variant of RMSE is the mean squared error (MSE) tracking error function, which is given as
\[
\text{MSE} = \frac{1}{T} \sum_{t=1}^{T}\left(I_{t}-R_{wt}\right)^{2} = \frac{1}{T} \sum_{t=1}^{T}\left(I_{t}- \sum_{j=1}^{n}r_{jt}w_{j}\right)^{2}.
\]
This objective function is convex and quadratic, resulting in the following convex quadratic programming problem:
\vspace{-0.5cm}
    \begin{center}
    \begin{equation}
    \label{eqn:minMAD}
    \qquad \ \text{(MSE) \  \qquad Min  \quad } \frac{1}{T} \sum_{t=1}^{T}\left(I_{t}- \sum_{j=1}^{n}r_{jt}w_{j}\right)^{2} \notag \qquad \qquad \qquad \qquad  
    \end{equation}
    subject to \eqref{eqn:size}-\eqref{eqn:binaryZ}. \quad  \qquad  \qquad \qquad \qquad  \qquad 
    \end{center}

We consider the MSE minimization problem for the empirical study in this review. 


\vspace{0.5cm}
    
\item \textbf{Sum of Errors Squared (SES):} The SES measure quantifies the deviation between the portfolio and benchmark returns by first aggregating all period-wise errors and then squaring the total. Specifically, the error in each period, defined as the difference between the benchmark index return and the portfolio return is summed across the entire investment horizon, and the square of this cumulative deviation forms the SES tracking error function. Mathematically, the SES function is expressed as
\[
\text{SES} = \frac{1}{T} \left[\sum_{t=1}^{T}\left(I_{t}-R_{wt}\right)\right]^{2}.
\]
and the corresponding tracking model becomes:
  \vspace{-0.5cm}
    \begin{center}
    \begin{equation}
    \label{eqn:minMAD}
    \qquad \ \text{(SES) \  \qquad Min  \quad } \frac{1}{T} \left[\sum_{t=1}^{T}\left(I_{t}-R_{wt}\right)\right]^{2}\notag \qquad \qquad \qquad \qquad 
    \end{equation}
    subject to \eqref{eqn:size}-\eqref{eqn:binaryZ}. \quad  \qquad  \qquad \qquad 
    \end{center}

Unlike the MSE objective, which penalizes deviations period by period, the SES measure emphasizes overall directional consistency between the portfolio and the benchmark. The function remains convex and quadratic in form; however, the inclusion of a cardinality constraint renders the optimization problem (SES), a NP-hard. \cite{Xuetal_SES_OPMS2016} addressed this challenge by employing a non-monotone projected gradient (NPG) algorithm, demonstrating its computational efficiency in identifying near-optimal tracking portfolios under the SES minimization.


   \vspace{0.5cm}

These quadratic loss functions (TEV, MSE, and SES) often overemphasize large deviations and may not accurately align with investor preferences. \cite{Rudolf_1999} argued that, due to the linear nature of fund managers' performance fees, linear deviations provide a more accurate representation of an investor’s risk attitude than squared deviations. Motivated by this reasoning, \cite{Rudolf_1999} proposed four linear tracking errors for index tracking, described in detail in the following. 

\medskip

\item \textbf{Mean absolute deviation (MAD)}: The first linear tracking error proposed by \cite{Rudolf_1999} minimizes the mean absolute deviations (MAD) between the tracking portfolio's returns and the benchmark returns, defined as
    $$\textbf{MAD}= \dfrac{1}{T} \sum\limits_{t=1}^{T} |R_{wt}-I_{t}|.$$
Because the absolute value function is non-differentiable, it is linearized by introducing two non-negative auxiliary variables, \(y_t^+\) and \(y_t^-\), which respectively represents the positive and negative deviations of the tracking portfolio return from the index returns.
It follows that if 
    $$R_{wt}>I_{t}, \quad \text{then} \quad y_t^{+}=R_{wt}-I_{t} \quad \text{(with } y_t^{-}= 0\text{)}, $$
    representing over performance of the tracking portfolio than the index, and if 
    $$R_{wt}<I_{t}, \quad \text{then} \quad y_t^{-} =I_{t}-R_{wt} \quad \text{(with } y_t^{+} = 0\text{)},$$
    representing under performance of the tracking portfolio than the index. Since only one of $y_t^+$ or $y_t^-$ is positive at any time $t$, these conditions can be combined into the single constraint as
    $$R_{wt} - y_t^{+} + y_t^{-} = I_{t}, \quad t=1,2,\ldots,T.$$
Thus, the index tracking problem using MAD as the objective is formulated as the following linear programming problem:
    \vspace{-0.5cm}
    \begin{center}
    \begin{equation}
    \label{eqn:minMAD}
    \qquad \ \text{(MAD) \  \qquad Min  \quad } \dfrac{1}{T} \sum\limits_{t=1}^{T} (y_{t}^{+}+y_{t}^{-})  \qquad \qquad \qquad \qquad \qquad  \qquad \qquad \qquad
    \end{equation}
    subject to \eqref{eqn:size}-\eqref{eqn:binaryZ} \quad  \qquad  \qquad \qquad \qquad  \qquad 
    \begin{equation}
        R_{wt}-y_{t}^{+}+y_{t}^{-}=I_{t}; \ t=1, 2, \ldots, T \quad
    \end{equation}
    \end{center}
This formulation being linear is convex and is amenable to solution via standard linear programming techniques, ensuring that the tracking portfolio closely replicates the index returns.

\vspace{0.5cm}

\item \textbf{Mean absolute downside deviation (MADD)}: While MAD accounts for both positive and negative deviations between portfolio and index returns, in many practical scenarios only the downside (or underperformance) of the index is of concern. In such cases, the tracking error is measured by the mean absolute downside deviation (MADD), as introduced in \cite{Rudolf_1999} that focuses on minimizing only the negative deviations in the MAD model, taking the following form:
    \vspace{-0.5cm}
    \begin{center}
    \begin{equation}
    \label{eqn:minMADD}
    \qquad \ \text{(MADD) \  \qquad Min  \quad } \sum\limits_{t=1}^{T} y_{t}^{-}  \qquad \qquad \qquad \qquad \qquad \qquad \qquad \qquad \qquad \qquad
    \end{equation}
    subject to \eqref{eqn:size}-\eqref{eqn:binaryZ} \quad  \qquad  \qquad \qquad \qquad  \qquad 
    \begin{equation}
        R_{wt}+y_{t}^{-} \geq I_{t}; \ t=1, 2, \ldots, T \qquad
    \end{equation}
    \end{center}
    This linear programming formulation emphasizes reducing the downside deviation, thus providing a robust metric for scenarios where it is critical to avoid under-performance relative to the benchmark index.
    
    \vspace{0.5cm}
    
\item \textbf{MinMax}: The third linear tracking error introduced by \cite{Rudolf_1999} is the MinMax. This tracking error aims at minimizing the maximum (largest absolute) deviation between the portfolio and the index returns, representing a robust ``worst-case'' optimization approach. Mathematically, this tracking error metric is defined as
$$\text{MinMax} = \min \max_{t} |R_{wt}-I_{t}|.$$
This formulation can be reformulated as a linear programming (LP) problem by introducing an auxiliary variable $\xi$ that bounds the absolute deviations, 
$$
\max_{t} |R_{wt}-I_{t}|=\xi \Longleftrightarrow |R_{wt}-I_{t}| \leq \xi \quad \Longleftrightarrow \quad R_{wt}-\xi \leq I_{t}, \quad R_{wt}+\xi \geq I_{t}, \quad \forall\, t = 1,2,\ldots,T.
$$
Thus, the MinMax optimization problem is expressed as the following linear program:
     \vspace{-0.5cm}
    \begin{center}
    \begin{equation*}
    \label{eqn:minMADD}
    \qquad \ \text{(MinMax) \  \qquad Min  \quad } \xi  \qquad \qquad \qquad \qquad \qquad \qquad \qquad \qquad \qquad \qquad \qquad 
    \end{equation*}
   \qquad  subject to \eqref{eqn:size}-\eqref{eqn:binaryZ} \quad  \qquad  \qquad \qquad \qquad  \qquad 
    \begin{equation*}
        R_{wt}-\xi \leq I_{t}; \ t=1, 2, \ldots, T, \qquad
    \end{equation*}
    \begin{equation*}
        R_{wt}+\xi \geq I_{t}; \ t=1, 2, \ldots, T. \qquad
    \end{equation*}    
    \end{center}
This LP formulation minimizes the worst possible tracking deviation over the entire period, yielding a robust tracking portfolio that limits extreme deviations from the index, making it particularly useful for conservative investors with strong aversion to large single-period errors.
    
    \vspace{0.5cm}

\item \textbf{Downside MinMax (DMinMax)}: The fourth linear tracking error by \cite{Rudolf_1999} is the Downside MinMax. This approach minimizes the maximum downside deviation, specifically focusing on scenarios where the tracking portfolio under-performs the benchmark index. This optimization emphasizes controlling the worst-case under-performance, providing a conservative strategy designed to minimize the largest observed negative deviations. Mathematically, the DMinMax tracking error is formulated as:
$$\text{DMinMax} = \min \max_{t} (I_{t}-R_{wt})^{+}.$$
This can be modeled as a linear programming problem by introducing an auxiliary variable \(\xi\) representing the maximum downside deviation, and imposing constraints only on downside scenarios, that is,$R_{wt}\leq I_{t}$:
 \vspace{-0.5cm}
    \begin{center}
    \begin{equation*}
    \label{eqn:minMADD}
    \qquad \ \text{(DMinMax) \  \qquad Min  \quad } \xi  \qquad \qquad \qquad \qquad \qquad \qquad \qquad \qquad \qquad \qquad \qquad 
    \end{equation*}
    \qquad subject to \eqref{eqn:size}-\eqref{eqn:binaryZ} \quad  \qquad  \qquad \qquad \qquad  \qquad 
    \begin{equation*}
        R_{wt}+\xi \geq I_{t}; \ t=1, 2, \ldots, T. \qquad
    \end{equation*}    
    \end{center}

This linear formulation directly limits the worst-case underperformance of the portfolio relative to the benchmark index. The DMinMax model is particularly suitable for risk-averse investors seeking to minimize substantial downside deviations from the index returns.


\vspace{0.5cm}

\item \textbf{Mixed CVaR based index tracking model}: Conditional Value-at-Risk (CVaR) is a coherent downside risk measure that captures the expected loss beyond a specified confidence level $\alpha$ \citep{Rockafellar_CVAR_2002}. Unlike variance-based measures, which treat all deviations symmetrically, CVaR focuses on the tail of the loss distribution, thus directly accounting for extreme downside risk. In portfolio optimization, minimizing CVaR ensures robustness against rare but severe market losses, making it a natural choice for risk-aware index tracking. However, as noted by \cite{Goel_MCVaR2018}, reliance on a single confidence level $\alpha$ may overlook other parts of the return distribution. In this regard, the mixed CVaR (MCVaR) framework aggregates multiple CVaR values across several confidence levels, thereby incorporating richer distributional information while retaining convexity and coherence.


\vspace{0.5em}
\noindent 
\cite{Goel_MCVaR2018} further extend this framework by proposing the \textbf{two-tail mixed CVaR (TMCVaR)} model for index tracking. The approach penalizes both downside and upside deviations of the tracking portfolio relative to the benchmark, corresponding respectively to under and over performance, thereby promoting balanced replication of the benchmark’s return dynamics. The two components are combined using a weight parameter $\delta \in (0,1)$, which controls the relative emphasis on each tail:
\[
\delta\, \text{MCVaR}(-X_{wt}) + (1 - \delta)\, \text{MCVaR}(X_{wt}),
\]
where $X_{wt}= I_{t} - R_{wt}$ denotes the excess return of the index $I_{t}$ over the tracking portfolio $R_{wt}$.

\vspace{0.5em}
\noindent 
Each MCVaR component is defined as a weighted sum of CVaR measures evaluated at $m$ different confidence levels, given as:
$$\text{MCVaR}(X_{wt}) = \sum_{k=1}^{m} \lambda^U_k \text{CVaR}_{\alpha^U_k}(X_{wt}),
$$
$$
\text{MCVaR}(-X_{wt}) = \sum_{k=1}^{m} \lambda^D_k \text{CVaR}_{\alpha^D_k}(-X_{wt}),
$$
where $\lambda^U = (\lambda^U_k, k = 1, \dots, m)$ and $\lambda^D = (\lambda^D_k, k = 1, \dots, m)$ are non-negative weights satisfying\\ $\sum\limits_{k=1}^{m} \lambda^U_k = \sum\limits_{k=1}^{m} \lambda^D_k = 1$. Using the linearization of CVaR, \cite{Goel_MCVaR2018} obtain the following optimization problem that minimizes the weighted two-tail MCVaR measure as a function of the portfolio weights:

\begin{center}
\begin{equation*}
\label{eqn:TMCVaR}
\textbf{(TMCVaR)} \quad \min_{\beta, w} \quad \delta \left(\sum_{k=1}^{m} \lambda^U_k \left[\beta^U_k + \frac{1}{(1 - \alpha^U_k)T} \sum_{j=1}^{T} u^U_{jk}\right]\right) + (1 - \delta) \left(\sum_{k=1}^{m} \lambda^D_k \left[\beta^D_k + \frac{1}{(1 - \alpha^D_k)T} \sum_{j=1}^{T} u^D_{jk}\right]\right)
\end{equation*}
subject to \qquad \qquad \qquad \qquad \qquad \qquad \qquad \qquad  \qquad \qquad \qquad  \qquad \qquad \qquad
\begin{equation*}
\label{eqn:TMCVaR_constraints1}
u^U_{jk} + \left(\sum_{i=1}^{n} r_{ij}w_i - I_j\right) + \beta^U_k \geq 0 \quad \text{for } k = 1, \dots, m, \, j = 1, \dots, T,
\end{equation*}
\begin{equation*}
\label{eqn:TMCVaR_constraints2}
u^D_{jk} - \left(\sum_{i=1}^{n} r_{ij}w_i - I_j\right) + \beta^D_k \geq 0 \quad \text{for } k = 1, \dots, m, \, j = 1, \dots, T,
\end{equation*}
\begin{equation*}
\label{eqn:TMCVaR_constraints3}
\sum_{i=1}^{n} w_i = 1, \qquad \qquad \qquad \qquad \qquad \qquad \qquad \qquad \qquad \qquad \qquad
\end{equation*}
\begin{equation*}
\label{eqn:TMCVaR_constraints4}
w_i \geq 0 \quad \text{for } i = 1, \dots, n, \qquad \qquad \qquad \qquad \qquad \qquad \qquad \qquad
\end{equation*}
\begin{equation*}
\label{eqn:TMCVaR_constraints5}
u^U_{jk}, u^D_{jk} \geq 0 \quad \text{for } k = 1, \dots, m, \, j = 1, \dots, T. \qquad \qquad \qquad \qquad 
\end{equation*}
\eqref{eqn:cardinality}---\eqref{eqn:binaryZ} \qquad \qquad \qquad \qquad \qquad \qquad \qquad \qquad \qquad \qquad \qquad \qquad \qquad 
\end{center}
where the variables $\beta^U_k$ and $\beta^D_k$ correspond to the value-at-risk (VaR) thresholds associated with the upper (upside) and lower (downside) tails at confidence levels $\alpha^U_k$ and $\alpha^D_k$, respectively. $u^U_{jk}$ and $u^D_{jk}$ are non-negative auxiliary variables representing the deviations of portfolio returns beyond the corresponding VaR thresholds in the upper and lower tails for each scenario $j$ and confidence level $k$. Thus, $u^U_{jk}$ and $u^D_{jk}$ linearize the piecewise CVaR components, while $\beta^U_k$ and $\beta^D_k$ capture the conditional quantiles of the loss distribution for the upside and downside risks, respectively. 

The above model is a linear program, thus retains computational tractability while capturing both tails of the deviation distribution. By minimizing the two-tail MCVaR, the framework effectively balances over- and under-performance relative to the benchmark, ensuring tighter replication with controlled tail risk. Empirically, \cite{Goel_MCVaR2018} report that the proposed TMCVaR-based index tracking model outperforms variance- and MAD-based formulations in terms of higher correlation with the benchmark and lower tracking error across multiple global indices.

\noindent


\end{enumerate}

\begin{table}[htp!]
    \caption{Thematic clusters of different statistical based approaches for index tracking}
    \begin{tabular}{lllc}
        \noalign{\smallskip}\hline\noalign{\smallskip}
       \textbf{Theme} & \textbf{Author(s)} & \textbf{Title} & \textbf{TC} \\
        \noalign{\smallskip}\hline\noalign{\smallskip}
         \multirow{1}{*}{Least squares regression} & \cite{CanakgozBeasley_2009EJOR} & Mixed-integer programming approaches for index tracking and & 192 \\
            & & enhanced indexation & \\
    \noalign{\smallskip}\hline\noalign{\smallskip} 
\multirow{6}{*}{Cointegration} & \cite{Santanna_cointeg_2017} & Index tracking and enhanced indexing using cointegration & \multirow{2}{*}{26}\\
& & and correlation with endogenous portfolio selection & \\
& \cite{Acosta_cointeg_2015} &	On the index tracking and the statistical arbitrage choosing  & \multirow{2}{*}{12}\\
& & the stocks by means of cointegration: the role of stock picking & \\
& \cite{Santanna_cointeg_2020} & Solving the index tracking problem based on a convex   & 5\\
& & reformulation for cointegration & \\
\noalign{\smallskip}\hline\noalign{\smallskip}
\multirow{4}{*}{Regularization} & \cite{Wu_nonneglasso_2014} & Nonnegative-lasso and application in index tracking	& 79\\
& \cite{Fastrich_qnorm_2014} & Cardinality versus q-norm constraints for index tracking & 48 \\
& \cite{WuYang_elsticnet_2014} & Nonnegative Elastic Net and application in index tracking & 44\\
& \cite{Santanna_lassoLS_2020} & Lasso-based index tracking and statistical arbitrage long-short  & 23\\
& & strategies & \\
\noalign{\smallskip}\hline\noalign{\smallskip}
\multirow{2}{*}{Factor based} & \cite{CoriMarcellino_2006_FactorIT} & Factor based index tracking & 68\\
& \cite{KwonWu_factorIT_2017} & Factor-based robust index tracking & 15\\
\noalign{\smallskip}\hline\noalign{\smallskip}
\multirow{4}{*}{Quantile Regression} & 	\cite{Mezali_quantileregITEI_2013} &	Quantile regression for index tracking and enhanced indexation & 33\\
& \cite{Li_quantregHDIT_2020} &	Efficient sparse portfolios based on composite quantile 	& 5\\
& & regression for high-dimensional index tracking & \\
&	\cite{Aguilar_TPquantiles_2022}	& Creating better tracking portfolios with quantiles &	2\\

\noalign{\smallskip}\hline\noalign{\smallskip}
    \end{tabular}
    \label{tab:STAT_top_articles}
\end{table}

\subsection{Statistical based index tracking optimization models}

Statistical modeling approaches leverage regression-based techniques and statistical properties inherent in financial data to construct index-tracking portfolios. Unlike traditional optimization models, these approaches focus on establishing statistical relationships between asset returns and the benchmark index, offering robust performance in capturing complex dependencies. Prominent methods in this category include least squares regression, quantile regression, co-integration analysis, and regularized regression models such as LASSO and Elastic Net; with top articles listed in Table \ref{tab:STAT_top_articles}. In the following, we briefly describe each of these statistical methodologies and discuss their formulations and applications in the index-tracking literature.

\begin{enumerate}

\item \textbf{Regression based models:}  Regression-based methods provide a natural and intuitive framework for modeling the relationship between individual asset returns and the benchmark index. By expressing each asset’s return as a linear function of the index return, regression captures how closely the asset co-moves with the market—quantified through the slope and intercept coefficients. In the context of index tracking, this approach is particularly valuable because it enables the construction of portfolios whose aggregated return dynamics mimic those of the benchmark. Such regression-based formulations thus bridge statistical estimation and portfolio optimization, offering both interpretability and tractability. Two popular regression based approaches include the least squares regression (LSR) and quantile regression (QR), discussed below. 


\subsubsection*{(a) Least squares regression approach}
The LSR approach, proposed by \cite{CanakgozBeasley_2009EJOR}, formulates the index tracking problem as a regression-based optimization model. Initially, each asset's returns are individually regressed against the benchmark index returns, resulting in a linear regression equation for each asset as:
$$r_{it} = \alpha_i + \beta_i I_{t} + \epsilon_{it}, \quad i = 1,2,\ldots,n; \quad t = 1,2,\ldots,T,$$
where \(\alpha_i\) and \(\beta_i\) are respectively the intercept and slope coefficients for asset \(i\), and \(\epsilon_{it}\) is the regression residual. From these asset-specific regressions, the tracking portfolio's overall intercept $(\hat{\alpha})$ and slope $(\hat{\beta})$ are defined as weighted sums of individual intercepts and slopes:
$$\hat{\alpha} = \sum_{i=1}^{n} w_{i}\alpha_{i}, \quad \hat{\beta} = \sum_{i=1}^{n} w_{i}\beta_{i}.$$

The goal of the LSR model for index tracking is to find an optimal subset of \(K\) assets whose weighted returns replicate the benchmark index as closely as possible, ideally achieving a portfolio intercept of zero $(\hat{\alpha} = 0)$ and slope of one $(\hat{\beta} = 1)$. However, for real-life data this may not be achievable. To achieve this, the authors suggest a number of ways, one of which they adopt is a two-stage approach, where the primary objective is to attain the desired intercept of zero and the secondary objective is to achieve the desired slope of one, that is : first minimize $|\hat{\alpha}-0|$ and then minimize $|\hat{\beta}-1|.$ Though the modulus objectives are non-linear, they can be linearized by introducing auxiliary variables $d$ and $e$, such that
\begin{align*}
 d \geq \hat{\alpha}, \qquad  d \geq -\hat{\alpha}, \qquad e  \geq \hat{\beta}-1, \qquad  \ e  \geq -\hat{\beta}+1, \qquad d \geq 0,  \qquad e \geq 0
\end{align*}
The problem thus reduces into two linear programs, solved in two stages as follows:\\
\textbf{Stage 1 (Minimizing Intercept):} In the first stage, the intercept \(\hat{\alpha}\) is minimized subject to cardinality and holding constraints:
     \vspace{-0.5cm}
    \begin{center}
    \begin{equation*}
    \label{eqn:minMADD}
    \qquad \ \text{(LSR1) \  \qquad Min  \quad } d  \qquad \qquad \qquad \qquad \qquad \qquad \qquad \qquad \qquad \qquad \qquad 
    \end{equation*}
    subject to \eqref{eqn:size}-\eqref{eqn:binaryZ}, \quad  \qquad  \qquad \qquad \qquad  \qquad 
    \begin{equation*}
        d \geq \sum\limits_{i=1}^{n} w_{i}\alpha_{i}, \quad  d \geq -\sum\limits_{i=1}^{n} w_{i}\alpha_{i}. \quad 
    \end{equation*} 
    \end{center}
    The optimal intercept value from this stage is denoted as \(\alpha^{\text{opt}}\). \\
   \noindent\textbf{Stage 2 (Minimizing Slope Deviation):} In the second stage, the model minimizes the deviation of the portfolio slope \(\hat{\beta}\) from one, given that the intercept is fixed at its optimal value from Stage 1: 
     \vspace{-0.5cm}
    \begin{center}
    \begin{equation*}
    \label{eqn:minMADD}
    \qquad \ \text{(LSR2) \  \qquad Min  \quad } e  \qquad \qquad \qquad \qquad \qquad \qquad \qquad \qquad \qquad \qquad \qquad 
    \end{equation*}
    subject to \eqref{eqn:size}-\eqref{eqn:binaryZ}, \quad $\hat{\alpha}= \alpha^{\text{opt}},$ \qquad  \qquad \quad 
    \begin{equation*}
        e \geq \sum\limits_{i=1}^{n} w_{i}\beta_{i}-1, \quad e \geq -\sum\limits_{i=1}^{n} w_{i}\beta_{i}+1.
    \end{equation*}
    \end{center}
The final portfolio obtained from Stage 2 provides the least squares regression-based tracking portfolio, which closely replicates the benchmark index by minimizing deviations in both intercept and slope.

\vspace{0.5cm}

\subsubsection*{(b) Quantile regression approach}
Quantile regression (QR) extends the standard least squares regression by modeling the quantiles of a response variable as a function of the independent variables. This approach allows one to capture distributional characteristics of asset returns beyond the conditional mean, making it particularly useful in the presence of skewness or heteroskedasticity. For a given quantile level $\tau \in (0,1)$, the $\tau$-th conditional quantile of the return on asset $i$ at time $t$ is modeled as
$$Q_\tau(r_{it} \mid I_t) = \alpha_{i\tau} + \beta_{i\tau} I_t,
\quad i = 1,\dots,n; \quad t = 1,\dots,T,$$
where \( \alpha_{i\tau} \) represents the intercept and \( \beta_{i\tau} \) represents the slope of the \( \tau \)-th quantile regression line. For each asset $i$, the coefficients \( \alpha_{i\tau} \) and \( \beta_{i\tau} \) are obtained by 
minimizing the quantile loss function:
\[
\min_{\alpha_i, \beta_i} \left[ \tau \sum_{t: u_{t} \ge 0} |u_{t}| + (1 - \tau) \sum_{t: u_{t} < 0} |u_{t}| \right],
\]
where \( u_{t} = r_{it} - \alpha_{i\tau} - \beta_{i\tau} I_t \) are the quantile regression residuals. This asymmetric loss penalizes underestimation and overestimation differently, depending on $\tau$.
The problem can be expressed equivalently as a linear program by introducing nonnegative variables \( u_{t}^+ \) and \( u_{t}^- \), representing the positive and negative parts of the residuals such that \(u_t = u_t^+ - u_t^-\):
\[
\min_{\alpha_{i\tau}, \beta_{i\tau}, u_{t}^+, u_{t}^-} \left[ \tau \sum_{t=1}^{T} u_{t}^+ + (1 - \tau) \sum_{t=1}^{T} u_{t}^- \right]
\]
\begin{align*}
\text{s.t.} \quad u_{t}^+ & \geq r_{it} - \alpha_{i\tau} - \beta_{i\tau} I_t, \\
u_{t}^- &\geq - (r_{it} - \alpha_{i\tau} - \beta_{i\tau} I_t), \\
 u_{t}^+ & , \ u_{t}^- \geq 0.
\end{align*}
Solving the linear problem yields the asset-specific optimal coefficients \( \alpha_{i\tau} \) and \( \beta_{i\tau} \).  At the portfolio level, the aggregated quantile regression line at level $\tau$ is approximated as a weighted combination of the asset-specific estimates: 


\[
\hat{\alpha}(\tau) = \sum_{i=1}^{n} w_i \alpha_{i\tau}, \quad 
\hat{\beta}(\tau) = \sum_{i=1}^{n} w_i \beta_{i\tau},
\]
These aggregated values provide a linear approximation to the \( \tau \)-th quantile of the portfolio return, allowing us to construct portfolios for index tracking (\( \tau = 0.5 \)) or enhanced indexation (\( \tau < 0.5 \)) \citep{Mezali_quantileregITEI_2013}.




The objective is to construct a tracking portfolio whose intercept is as close to zero as possible, and whose slope approximates to one at the chosen quantile level $\tau$. Following a two-stage optimization analogous to the least squares method \citep{CanakgozBeasley_2009EJOR}, the formulation takes a two stage formulation to first minimize $|\hat{\alpha}(\tau)-0|$ and then minimize $|\hat{\beta(\tau)}-1|$, both linear programs. 


\smallskip
\noindent\textbf{Stage 1 (Minimizing intercept at quantile \(\tau\)):}
The first stage minimizes the absolute intercept deviation \(|\hat{\alpha}(\tau)-0|\) to find a portfolio with a quantile regressed intercept as close to zero as possible. Similar to LSR, the problem is linearized by introducing auxiliary variable $d \geq 0$ and solving:
 \vspace{-0.5cm}
    \begin{center}
    \begin{equation*}
    \label{eqn:minMADD}
    \qquad \ \text{(QR1) \  \qquad Min  \quad } d  \qquad \qquad \qquad \qquad \qquad \qquad \qquad \qquad \qquad \qquad \qquad 
    \end{equation*}
    subject to \eqref{eqn:size}-\eqref{eqn:binaryZ}, \quad  \qquad  \qquad \qquad \qquad  \qquad 
    \begin{equation*}
        \sum_{i=1}^{n} w_i \alpha_{i\tau} - d \leq 0,\quad \sum_{i=1}^{n} w_i \alpha_{i\tau} + d \geq 0. \quad 
    \end{equation*} 
    \end{center}
Let $d^{*}$ denote the optimal intercept deviation from Stage 1.

\smallskip
\noindent\textbf{Stage 2 (Minimizing slope deviation at quantile \(\tau\)):} The second stage minimizes \(|\hat{\beta}(\tau)-1|\) to find a portfolio with a quantile regressed slope as close to one as possible, while retaining the optimal intercept deviation of $d^{*}$ from stage 1. Again, linearizing the problem by introducing variable $e \geq 0$ and solving:
 \vspace{-0.5cm}
    \begin{center}
    \begin{equation*}
    \label{eqn:minMADD}
    \qquad \ \text{(QR2) \  \qquad Min  \quad } e  \qquad \qquad \qquad \qquad \qquad \qquad \qquad \qquad \qquad \qquad \qquad 
    \end{equation*}
    subject to \eqref{eqn:size}-\eqref{eqn:binaryZ}, \quad $\sum_{i=1}^{n} w_i \alpha_{i \tau} = d^{*},$ \qquad  
    \begin{equation*}
    \sum_{i=1}^{n} w_i \beta_{i\tau} - e \leq 1,\quad \sum_{i=1}^{n} w_i \beta_{i\tau} + e \geq 1.
    \end{equation*}
    \end{center}
This two-stage quantile regression framework provides a robust tracking portfolio designed to replicate benchmark index returns at specific quantiles, allowing investors to better manage distributional risks. It can further accommodate transaction costs and other realistic constraints, demonstrating flexibility for practical portfolio management applications.

\vspace{0.5cm}

\item \textbf{Regression with regularization models:} Regression models with regularization techniques have gained considerable attention in index tracking literature due to their ability to efficiently handle high-dimensional datasets and prevent overfitting. Regularization introduces penalty terms into the objective function, promoting sparse solutions that result in simpler, more interpretable portfolios. Popular regularization methods in portfolio selection include LASSO (Least Absolute Shrinkage and Selection Operator) and elastic net, which enforce sparsity by penalizing portfolio weights through $\ell_1$ and combined $\ell_1$-$\ell_2$ norms, respectively. Such approaches are particularly beneficial for constructing sparse portfolios from large market indices by selecting a limited subset of representative assets while maintaining tracking accuracy. In what follows, $\|\cdot\|_1$ and $\|\cdot\|_2$ denote the $\ell_1$ and $\ell_2$ norms, respectively, where
$$\|x\|_1 = \sum_{i=1}^{n} |x_i| \quad \text{and} \quad \|x\|_2 = \sqrt{\sum_i x_i^2}.$$
\subsubsection*{(a) Non-negative LASSO}
Non-negative LASSO, introduced by \cite{Wu_nonneglasso_2014}, presents an effective two-stage approach for sparse portfolio construction in the index tracking context. The central idea is to first select a subset of $K$ representative assets (where $K$ is the desired cardinality) from a large pool, and subsequently determine the optimal weights that track the benchmark index efficiently.

In the first stage, asset selection is performed using the following non-negative LASSO regression model:
\begin{equation}
    \text{(NNL)} \qquad \qquad  \min_{w \geq 0}\frac{1}{T}\|I - R_{w}\|_{2}^{2} + \lambda\|w\|_{1},
\end{equation}
The parameter \(\lambda \geq 0\) controls the trade-off between portfolio sparsity (asset selection) and the accuracy of tracking the index returns. The model minimizes the mean squared tracking error between the portfolio and index returns while enforcing sparsity through the $\ell_{1}$ penalty. Since the objective combines a convex quadratic loss term with a linear regularization term under non-negativity constraints, this stage constitutes a convex quadratic programming problem. Since the choice of \(\lambda\) significantly affects asset selection, it is determined via a \textit{bisection search algorithm}. This iterative approach begins with an interval \([\lambda_{\text{min}}, \lambda_{\text{max}}]\), systematically narrowing this interval by repeatedly solving the non-negative LASSO problem at midpoint values \(\lambda_{\text{mid}}\), until a portfolio of the desired cardinality is obtained.

Once the set of candidate assets is selected from the first stage, the second stage involves refining portfolio weights by solving the following non-negative least squares regression problem:
\begin{equation}
\label{eqn:nnls}
 \text{(NNLS)} \qquad  \qquad    \min_{w_{\small{\mathcal{S}}} \geq 0}\frac{1}{T}\|I - R^{\mathcal{S}}\|_{2}^{2},
\end{equation}
where \(R_{\mathcal{S}}\) denotes the portfolio return corresponding to the assets selected in stage one, and \(w_{\mathcal{S}}\) are their proportional weights. This formulation minimizes the mean squared tracking error between the portfolio returns of the selected subset and the index returns. The problem remains a convex quadratic program, solvable efficiently using standard numerical solvers. In our implementation, we employ the \texttt{nnls()} function from the \texttt{nnls} package in \textsf{R}, which computes the non-negative least squares solution through an active-set algorithm. The coefficients are subsequently normalized for the portfolio weights to sum to one. This second-stage optimization improves tracking accuracy by determining the optimal allocation among the pre-selected assets.  

The two-stage non-negative LASSO methodology thus achieves a sparse, interpretable, and practically efficient index tracking portfolio, making it especially suitable for high-dimensional indexes such as the S\&P 500 and Russell 2000.


\subsubsection*{(b) Non-negative elastic net}

While the non-negative Lasso is effective for selecting a sparse set of assets, it may face challenges in scenarios where predictors are highly correlated. To address this limitation, \cite{WuYang_elsticnet_2014} introduced the non-negative Elastic Net (NN-EN), which combines the \(\ell_1\) and \(\ell_2\) penalties of the Lasso and ridge regressions, respectively. The general formulation of the non-negative Elastic Net is given as:

\begin{equation}
\label{eqn:NNEN}
\text{(NN-EN)} \qquad \min_{w \geq 0} \left\|I - R_{w} \right\|_2^2 + \lambda_1 \|w\|_1 + \lambda_2 \|w\|_2^2,
\end{equation}

where $ \lambda_1, \lambda_2 \geq 0$ are the regularization parameters controlling sparsity and shrinkage, respectively. The \(\ell_1\) norm (\(\|w\|_1\)) encourages sparsity, and the \(\ell_2\) norm (\(\|w\|_2^2\)) helps stabilize the solution by grouping correlated predictors, thereby enhancing the robustness of asset selection.

Similar to the non-negative Lasso, the NN-EN estimator typically involves a two-stage optimization strategy for index tracking:

\begin{itemize}
    \item \textbf{Stage 1 (Asset selection):} The (NN-EN) problem in Eqn. \eqref{eqn:NNEN} is solved to identify a sparse subset of candidate assets. The tuning parameters \( \lambda_1 \) and \( \lambda_2 \) are optimized through a grid-and-bisection search: for each $\lambda_2$ in a predefined grid, a bisection search is performed over $\lambda_1$ to ensure that the number of selected assets does not exceed a cardinality constraint $K$. This adaptive search mechanism ensures that the final model yields at most $K$ non-zero weights, balancing sparsity and accuracy.

    \item \textbf{Stage 2 (Optimal weight estimation):} Once the subset assets are selected, the optimal weights are computed by solving a constrained non-negative least squares (NNLS) problem  (i.e. Eq. \eqref{eqn:nnls}). This step (solving (NNLS) problem) determines the optimal portfolio weights that best replicate the index returns, using only the assets selected in Stage 1.
\end{itemize}

This hybrid NN-EN + NNLS procedure leverages the strength of NN-EN to select assets with stable, correlated return series while retaining interpretability through non-negativity and sparsity. Empirical results in \cite{WuYang_elsticnet_2014} demonstrate that the NN-EN approach achieves superior tracking performance relative to the non-negative Lasso, particularly in markets where asset returns exhibit strong cross-correlation.





\vspace{0.5cm}

\item \textbf{Co-integration based models:} Co-integration is a statistical concept, originally introduced by \cite{Granger_1981}, and further developed by \cite{EngleGranger_1987} that builds on the concept that a linear combination of two or more non-stationary time series might be stationary, indicating a shared long-term trend. In financial markets, where stock prices often exhibit non-stationary behavior, co-integration helps in determining a common trend among different stocks, facilitating in predicting future movements. In simpler words, if two stocks are co-integrated, it means they move together over time. Within the context of index tracking, co-integration can be applied to identify stocks whose prices move closely with respect to the market index. For time series \(X_t\) and \(Y_t\), if a linear combination \(\beta_1 X_t + \beta_2 Y_t\) is stationary, then \(X_t\) and \(Y_t\) are said to be co-integrated. We investigate the following two methods that utilize co-integration to generate tracking portfolios, proposed by \cite{Santanna_cointeg_2017} and \cite{Santanna_cointeg_2020}. 

\subsubsection*{(a) Simulations based co-integration approach} 
The theoretical foundation of cointegration, introduced by \cite{EngleGranger_1987} showed that even though individual time series may be non-stationary, certain linear combinations of them can be stationary. This implies that a portfolio composed of such assets maintains a long-term equilibrium with the benchmark index. Following this principle, \cite{Santanna_cointeg_2017} proposed a simulation-based co-integration framework for index tracking. The approach assumes that both the index and the constituent asset price series are integrated of order one, i.e., $I(1)$, and that their logarithms can be linearly related through a co-integrating vector. Formally, the long-run relationship is modeled as:
\[
\log(J_t) = \beta_0 + \sum_{i=1}^{n} \beta_i \log(p_{it}) + \varepsilon_t,
\]
where $J_t$ denotes the index price level at time $t$, $p_{i,t}$ is the price of asset $i$, $\beta_i$ are the co-integration coefficients, and $\varepsilon_t$ represents the deviation from the long-run equilibrium. If the residual series $\varepsilon_t$ is stationary, then the index and the selected assets are said to be co-integrated.


The methodology developed by \cite{Santanna_cointeg_2017} consists of repeatedly selecting random subsets of assets and testing whether the corresponding linear combination with the index satisfies the co-integration condition. Specifically, the approach involves the following steps, performed over each training window:

\begin{enumerate}
    \item[(1)] \textbf{Random asset selection:} Randomly select a subset $S$ of $K$ assets from the full universe of $n$ assets (where $K$ is the desired cardinality). 

    \item[(2)] \textbf{Co-integration regression:} Estimate the long-run relationship for the selected subset $S$ using ordinary least squares:
    \[
    \log(J_t) = \beta_0 + \log(P^{S}_{t}) \beta + \varepsilon_t,
    \]
    where $\log(P^{\mathcal{S}}_t)$ are the log prices of selected assets in $S$ and $\varepsilon_t$ captures deviations from the long-run equilibrium between the index and the selected assets. 
    
  \item[(3)] \textbf{Stationarity test:} The stationarity of the residuals $\varepsilon_t$ is tested using the Augmented Dickey–Fuller (ADF) test. Subsets for which the null of a unit root is rejected at the $5\%$ level ($p < 0.05$) are retained as candidate co-integrated portfolios.
    
  \item[(4)] \textbf{Simulations based portfolio selection:} The above procedure is repeated for a sufficiently large number of iterations (random subsets). Among all subsets that satisfy the ADF stationarity condition, the optimal portfolio is chosen as the one with the smallest sum of squared residuals (SSR), i.e., $\min \sum_t \varepsilon_t^2$. The regression coefficients \(\beta_j\) from the optimal regression are then normalized to determine the portfolio weights as 
    \[
       w_j = \frac{\beta_j}{\sum_{i=1}^{K} \beta_i}, \quad j = 1, \ldots, K.
    \] 
   The normalized weights \(w_j\) are assigned to the corresponding selected subset of assets, while all non-selected assets receive a weight of zero.

\end{enumerate}

The simulation-based search allows for an efficient exploration of possible asset combinations, accommodating large universes where exhaustive co-integration testing would be computationally infeasible. The resulting co-integrated portfolios have been shown to provide stable tracking performance and improved out-of-sample behavior, particularly in non-stationary markets where traditional variance-based approaches fail to capture long-run dependencies.



\subsubsection*{(b) Convex co-integration model}

To overcome the limitation of the traditional co-integration based index tracking model that require an ex-ante stock selection, \cite{Santanna_cointeg_2020} developed a convex mixed-integer non-linear programming (MINLP) formulation that endogenously determines the composition of the tracking portfolio. The model extends the framework of \cite{Santanna_cointeg_2017} by embedding a cardinality constraint directly within the optimization problem, thereby allowing simultaneous estimation of the co-integrating vector and the subset of assets included in the portfolio.

Starting from the co-integration regression framework introduced earlier,
\[
\log(J_t) = \beta_0 + \sum_{i=1}^{n} \beta_i \log(p_{it}) + \varepsilon_t,
\]
the residual term $\varepsilon_t$ represents deviations from the long-run equilibrium. The optimization objective is to minimize the sum of squared residuals over all time periods, ensuring that the selected assets jointly produce a stationary combination with the benchmark index. This leads to the following convex MINLP formulation 
\begin{center}
   \begin{equation*}
    \label{eqn:cvx-coint}
      \text{(Cvx-CoInt) \ } \quad \min \; \sum_{t=1}^{T} \left[\log(J_t) - \beta_0 - \sum_{i=1}^{n} \beta_i \log(p_{it}) \right]^2  \qquad \qquad \qquad \qquad \qquad 
    \end{equation*}
    subject to \qquad \qquad \qquad \qquad \qquad \qquad \qquad \qquad \quad  
    \begin{equation*}
     \qquad \quad \quad \qquad  \sum_{i=1}^{n} \mu_i = K, \quad \mu_i \in \{0,1\}, \quad \beta_i \ge 0, \quad \beta_i \le \mu_i, \quad i = 1, \ldots, n,
    \end{equation*} 
    \end{center}
where $\mu_i$ is a binary decision variable indicating whether asset $i$ is included in the tracking portfolio, and $K$ specifies the desired number of assets. The non-negativity constraints on $\beta_i$ enforces a no short-selling restriction, ensuring practical applicability.

The convexity of the objective function can be established by expressing it in a matrix form as
\[
y = A x + b,
\]
where \(y = \log(J)\), \(A = \log(P)\) is the matrix of log-prices of \(n\) assets, and \(x = \beta\) is the coefficient vector. Defining the least-squares objective function as
\[
f(x) = \|y - A x\|_2^2 = (y - A x)^\top (y - A x),
\]
the Gradient and Hessian with respect to \(x\) are given by
\[
\nabla f(x) = -2A^\top (y - A x), \qquad
\nabla^2 f(x) = 2A^\top A.
\]

The second-order conditions confirm that the problem is convex for any $A$ since $A^\top A$ is positive semi-definite \citep{BoydVandenberghe_2004}. The resulting convex MINLP reformulation eliminates the random selection step used in the earlier simulation-based approach, producing tracking portfolios that are both stable and computationally efficient. Empirical evidence presented by \cite{Santanna_cointeg_2020} shows that this convex optimization-based formulation achieves lower turnover and smaller tracking errors, both in-sample and out-of-sample, compared to traditional co-integration portfolios.
\vspace{0.5cm}



\vspace{0.5cm}

\item \textbf{Factor-based model:} Factor based models, introduced by \cite{CoriMarcellino_2006_FactorIT} represent a powerful statistical framework for constructing low-dimensional index tracking portfolios. These models exploit the idea that a small number of latent common factors drive the co-movements of asset prices, enabling the construction of parsimonious tracking portfolios that effectively replicate the benchmark index’s behavior. By identifying the assets most closely aligned with these factors, one can build portfolios that effectively replicate the benchmark index.  Unlike traditional optimization based models that operate directly on returns, this approach first models the underlying structure of log prices to identify representative assets before performing the final tracking optimization using simple returns.

Formally, let $X$ be the $(T \times n)$ matrix representing the log prices of $n$ candidate assets observed over $T$ time periods. Principal Component Analysis (PCA) is applied to extract the primary common factors affecting asset prices:
\[
X = F L^{T} + E,
\]
where $F$ is the $(T \times k)$ matrix of common factors (principal components), $L$ is the $(n \times k)$ matrix of factor loadings, and $E$ denotes the $T \times n$ matrix of idiosyncratic errors. The number of extracted factors $k$ is typically chosen in such a way that they explain a substantial portion of the variation in asset prices. The factor-based index tracking procedure involves the following detailed steps:

\begin{enumerate}
    \item \textbf{Factor extraction:} PCA is first applied to the log-price matrix $X$, yielding the factor matrix $F$. These extracted factors capture systematic market behavior and serve as the basis for asset selection.


    \item \textbf{Asset selection via linear regression:} For asset selection, each asset's log prices $X_i$ are regressed against the extracted factor series $F$ using ordinary least squares (OLS) regression
    \[
        X_i = F \beta_i + \varepsilon_i,\quad i=1,\dots,N,
    \]
    where $\beta_i$ denotes the asset-specific loadings onto common factors and $\varepsilon_i$ represents idiosyncratic component, typically assumed to follow a zero-mean noise process. The coefficient of determination $R^{2}_{i}$ from each regression quantifies the degree to which the factors explain the $i$th asset's price movements, $i=1,\ldots,N$. Assets are ranked by their $R_i^2$ values, and the top $K$ ( where $K$ is the desired cardinality) assets are selected and indexed by the subset 
    \[
        \mathcal{S} = \{ i_1, \dots, i_K \}.
    \]
   This ensures that only assets most closely aligned with the underlying factor structure are retained. 

\item \textbf{Determining final optimal weights:} Once the representative assets are identified, their simple returns are used to determine the final portfolio weights. The final portfolio weights $w$ are obtained by solving the MSE minimization problem: 
    $$\min_{w} \; \| R^{\mathcal{S}}_{t} w - I_{t} \|_2^2, \quad \text{subject to} \quad  w_i \ge 0, \quad \sum_{i \in \mathcal{S}} w_i = 1,$$ \\
    where $R^{\mathcal{S}}$ is the matrix of returns of the selected assets. 
    
\end{enumerate}

The factor-based method offers robust and interpretable tracking portfolios, particularly effective in high-dimensional contexts, as demonstrated empirically by \cite{CoriMarcellino_2006_FactorIT}. This approach effectively combines dimensionality reduction, data-driven asset selection, and constrained optimization, enhancing both in-sample explanatory power and out-of-sample tracking performance. We next explain data-driven methods for index tracking. 

\vspace{0.5cm}

\begin{table}[htp!]
    \caption{Thematic clusters of different statistical and machine learning based approaches for index tracking}
    \begin{tabular}{lllc}
        \noalign{\smallskip}\hline\noalign{\smallskip}
       \textbf{Theme} & \textbf{Author(s)} & \textbf{Title} & \textbf{TC} \\
        \noalign{\smallskip}\hline\noalign{\smallskip}
            \multirow{4}{*}{Clustering} & \cite{Dose_clustering_2005} &
Clustering of financial time series with application to & \multirow{2}{*}{120} \\
& & index and enhanced index tracking portfolio & \\
& \cite{FocardiFabozzi_clustering_2004} &	A methodology for index tracking based on time-series & 53 \\
& & clustering & \\
& \cite{WuKwanCosta_clustering_2017} & A constrained cluster-based approach for tracking &  22\\
& &  the S\&P 500 index  & \\
\noalign{\smallskip}\hline\noalign{\smallskip}
\multirow{2}{*}{Deep learning} & \cite{KimKim_deepLR_2020} & Index tracking through deep latent representation learning & 33\\ 
& \cite{Ouyang_deepNN_2019} & Index tracking based on deep neural network	& 29\\
\noalign{\smallskip}\hline\noalign{\smallskip} 
\multirow{4}{*}{Neural networks} & \cite{Zheng_stochasticNN_2020} & Index tracking with cardinality constraints: A stochastic neural  & 21\\
& & networks approach & \\
& \cite{KwakSongLee_NN_2021} & Neural network with fixed noise for index-tracking portfolio 	& 24\\
& & optimization & \\
\noalign{\smallskip}\hline\noalign{\smallskip} 
\multirow{2}{*}{Random forest} & \cite{randomforestIT_2022} & Combining random forest and multicollinearity & 2\\
& & modeling for index tracking &	\\
\noalign{\smallskip}\hline\noalign{\smallskip}
\multirow{4}{*}{Topological learning} & 	
\cite{goel_sparseTDAIT_2024} & Sparse Portfolio Selection via topological data & \\
 & & analysis based clustering	& \\
 & \cite{Goeletal_TDA2025} & Risk reduced sparse index tracking portfolio: A topological &  \\
 & &  data analysis approach  & \\
\noalign{\smallskip}\hline\noalign{\smallskip}
    \end{tabular}
    \label{tab:ML_top_articles}
\end{table}
\end{enumerate}


\subsection{Data-driven models for index tracking}
In this section, we explore various data-driven methodologies that have emerged prominently in the index tracking literature. Specifically, we focus on selected state-of-the-art approaches spanning deep learning techniques, factor-based models, clustering techniques, random forests, and support vector regression. Table \ref{tab:ML_top_articles} outlines the top data-driven models for index tracking, based on their citations. We now detail the few techniques we have short-listed for the analysis in our review. 


\begin{enumerate}
    \item \textbf{Clustering based approach}\\
    Clustering analysis, also known as data segmentation is a technique employed to group assets into homogeneous clusters based on similarity, ensuring that assets within each cluster behave more similarly than those in different clusters. Early works by \cite{FocardiFabozzi_clustering_2004} and \cite{Dose_clustering_2005} introduced clustering techniques into index tracking and enhanced indexing frameworks. In particular, \cite{Dose_clustering_2005} applied hierarchical clustering to the data set S\&P 500, using two key similarity measures, as follows:
    \begin{itemize}
    \item \textbf{Correlation distance on returns:} This measure evaluates the similarity of asset return series using the metric
    \[
    d(X, Y) = \sqrt{2\,(1 - c_{XY})},
    \]
    where \(c_{XY}\) is the correlation coefficient between the returns of assets $X=\left( x_{t} \right)$ and $Y=\left( y_{t} \right), \ t=1, 2, \ldots, T$. 

    \item \textbf{Percentage difference of prices:} This alternative measure focuses on asset price levels by comparing percentage differences. The distance is defined as
    \[
    d(X, Y) = \min\{d_1(X,Y), d_2(X,Y)\},
    \]
    with
    \[
    d_1(X, Y) = \min_{a \in \mathbb{R}} \frac{1}{T} \sum_{t=1}^T \left( \frac{x_t - a y_t}{x_t} \right)^2,
    \]
    and
    \[
    d_2(X, Y) = \min_{a \in \mathbb{R}} \frac{1}{T} \sum_{t=1}^T \left( \frac{x_t - a y_t}{a y_t} \right)^2.
    \]

\end{itemize}

Hierarchical clustering starts with each asset in its own cluster and then repeatedly merges the most similar pairs of clusters until there is just one cluster containing all the assets. The results of this process are represented in a diagram called a \textit{dendogram}. Based on the clustering output, asset selection is carried out using one of two approaches: 

\begin{itemize}
    \item \textbf{Clust1:} The assets are partitioned into $K$ clusters, where $K$ is the desired cardinality and then one asset is randomly chosen from each cluster, giving equal representation to each. 
    \item \textbf{Clust2:} The assets are grouped into fewer than $K$ clusters, and selections are made from them in proportion to the cluster size.
\end{itemize}
Both clustering-based asset selection methods use the correlation distance metric defined as
\[
d_{ij} = \sqrt{2\left(1 - \rho_{ij}\right)},
\]
where $\rho_{ij}$ denotes the Pearson correlation between the return series of assets $i$ and $j$. This distance captures similarities in return co-movements and is scale-invariant. Hierarchical clustering is performed using complete linkage, which forms compact and well-separated clusters. 
This clustering-based selection is then followed by an optimization step that minimizes the mean-squared tracking error between the returns of the selected assets and that of the index. 
Indexing the selected set of assets by $S$, the final portfolio weights $w$ are obtained by solving the following MSE minimization problem as:
    $$\min_{w} \; \| R^{\mathcal{S}}_{t} w - I_{t} \|_2^2, \quad \text{subject to} \quad \sum_{i \in \mathcal{S}} w_i = 1\;\; w_i \ge 0,; i \in \mathcal{S}\, \quad $$ \\
where $R^{\mathcal{S}}$ is the matrix of returns of the selected assets. This approach thus combines data-driven segmentation with mathematical programming to construct a robust tracking portfolio. 
\vspace{0.5cm}

    
\item \textbf{Sparse support vector regression} \\
Support vector regression (SVR), an extension of support vector machine (SVM) for regression tasks, is built upon the principle of structural risk minimization and is well known for its ability to achieve high generalization performance even with limited data. Owing to its flexibility in modeling non-linear relationships, SVR has been widely used in various fields such as text categorization, image processing, and biomedical data analysis. In the context of index tracking,  \cite{Takeda2010_SVR_IT} applied the SVR model to index tracking while \cite{deleone_svr} formulated a corresponding 0–1 mixed-integer version to insure cardinality restrictions. However, both of these formulations suffered from scalability issues. To address this, \cite{Tengetal_SVR_2017} proposed two sparse SVR models that incorporate explicit cardinality constraints to construct parsimonious and computationally efficient tracking portfolios.

\smallskip
\begin{enumerate}
\item \textbf{Sparse $\epsilon$-SVR model}:\\
The sparse $\epsilon$-SVR model aims to minimize the tracking error while enforcing sparsity through an $\ell_0$-norm constraint. The model is formulated as
\begin{align}
(\text{SVR-IT}) \quad 
& \min_{w,\, \xi} \; \frac{1}{2}\|w\|_2^2 + C_1 \sum_{t=1}^{T} \left( c(\xi_t^{u})+c(\xi_t^{l}) \right)  \notag\\[6pt]
& \text{subject to} \notag\\
& R_{wt}w-I_{t} \leq \epsilon + \xi_t^{u}, \notag \\
& I_{t}-R_{wt}w \leq \epsilon + \xi_t^{l}, \notag \\
& \xi_t^{l}, \xi_t^{u} \geq 0, \quad t=1,\ldots, T, \notag \\
& e^{T} w = 1,\notag \\
& \|w\|_{0} \leq K, \notag\\
& 0 \leq w_j \leq u,\quad j=1,\dots,n.\notag
\end{align}
where \(\xi_t^{u}\) and \(\xi_t^{l}\) are slack variables capturing deviations from the allowable error tolerance. $\epsilon \geq 0$ denotes the maximum deviation tolerance where the parameter \(C_1 > 0\) is the weight of the deviations larger than a given $\epsilon$, \(c(\xi)\) is the loss function of deviation $\xi \in \mathbb{R}_{+},$ and \(u\) represents an upper bound on asset weights (typically set to prevent excessive concentration). The term \(\|w\|_0\) denotes the number of non-zero elements in \(w\), enforcing sparsity by limiting the number of assets in the portfolio. 

For convenience, the loss function \(c(\xi)\) adopts the Gaussian form \(c(\xi) = \frac{1}{2}\xi^2\), yielding a smooth quadratic optimization landscape, resulting in the following optimization model
\begin{align}
(\epsilon\text{-SVR}) \quad & \min_{w,\xi} \quad \frac{1}{2}\|w\|^2 + C_1 \sum_{t=1}^{T}\left[(R_{wt} w - I_t - \epsilon)_{+}^{2}+(I_t - R_{wt} w - \epsilon)_{+}^{2}\right] \notag\\[6pt]
& \text{subject to} \notag\\
& e^{T} w = 1,\notag \\
& \|w\|_{0} \leq K, \notag\\
& 0 \leq w_j \leq u,\quad j=1,\dots,n.\notag
\end{align}
In the $\epsilon$-SVR model, the maximum deviation tolerance $\epsilon$ is a prior parameter. If there does not exist a good estimation of $\epsilon$, then this parameter can be treated as a variable, leading to a new model, $\nu$-SVR, explained below. 

\smallskip
\item \textbf{Sparse $\nu$-SVR model}:\\
The second variant introduces flexibility by treating the tolerance level \(\epsilon\) as a decision variable, balancing its value against tracking accuracy through an additional parameter \(C_2\). This yields the following sparse version of the \(\nu\)-SVR model:

\begin{align}
(\nu\text{-SVR}) \quad & \min_{w,\epsilon,\xi} \quad \frac{1}{2}\|w\|^2 + C_1 \sum_{t=1}^{T}\left[(R_{wt} w - I_t - \epsilon)_{+}^{2}+(I_t - R_{wt} w - \epsilon)_{+}^{2}\right] + C_2 \epsilon \notag\\[6pt]
& \text{subject to} \notag\\
& e^{T} w = 1,\notag \\
& \|w\|_{0} \leq K, \notag\\
& \epsilon \geq 0,\notag\\
& 0 \leq w_j \leq u,\quad j=1,\dots,n.\notag
\end{align}

\end{enumerate}

\noindent In this second formulation, the parameter \(C_2\) explicitly controls the trade-off between tracking accuracy and model complexity by regulating the allowable tolerance \(\epsilon\).

\noindent Both sparse-SVR models are inherently non-convex due to the presence of the $\ell_0$-norm. To efficiently solve them, \cite{Tengetal_SVR_2017} adopted the \textit{penalty proximal alternating linear minimization} (PALM) algorithm combined with a sparse projection operator. This iterative procedure alternates between solving convex subproblems in \(w\) and thresholding to maintain sparsity, achieving convergence to a stationary point. This sparse SVR methodology provides effective and computationally practical solutions to index tracking problems, particularly beneficial in scenarios demanding precise control over portfolio cardinality and tracking accuracy.


\medskip

\item \textbf{Random forest} \\
Combining machine learning and statistical modeling techniques provides powerful tools for constructing sparse and efficient portfolios, particularly suited for high-dimensional index tracking problems. While SVR offers one such approach to generate sparse tracking portfolios, another promising machine learning framework is based on random forest (RF). Owing to their ensemble nature, RF models can effectively capture nonlinear dependencies, handle multicollinearity, and provide built-in measures of variable importance, making them highly suitable for identifying a small subset of influential assets in large financial universes. 

In this context, \cite{Cao_RFstat_2022} proposed two hybrid two-stage methodologies that integrate the predictive power of RF with the stability of Ridge regression. The first approach combines RF-based clustering with ridge regression, whereas the second employs RF regression followed by ridge regression. In both strategies, the RF model serves as the machine-learning component that performs asset selection or grouping, while ridge regression constitutes the statistical modeling stage used to estimate optimal portfolio weights. These hybrid frameworks illustrate how the fusion of ensemble learning and regularized regression can yield sparse, robust, and computationally efficient index-tracking portfolios.


\paragraph{Notation}
Let $\{h(R, \theta_k), \, k = 1, 2, \ldots, K\}$ denote the ensemble of $K$ decision trees in the RF, where $R \in \mathbb{R}^{T \times n}$ represents the matrix of asset returns across $T$ time periods, and $\theta_k$ denotes the parameters of the $k$-th tree. Each tree $h(R, \theta_k)$ is trained on a bootstrap sample drawn with replacement from the original data. At each node of a decision tree, a random subset of $m$ features ($m \ll n$) is selected, and the feature that maximizes an impurity-reduction criterion, such as information gain, Gini index, or mean squared error \citep{RFgini_2024} is used for splitting.

The aggregated RF prediction for an observation $R_t$ is obtained by averaging the outputs of all $K$ trees:
\[
H(R_t) = \frac{1}{K}\sum_{k=1}^{K} h(R_t, \theta_k).
\]
The RF model provides a measure of variable (feature) importance by assessing the contribution of each feature to the overall prediction accuracy. This feature-importance ranking serves as the basis for asset selection in the subsequent hybrid RF-based index tracking models.

\subsubsection*{Random forest clustering combined with ridge regression (RF cluster+ridge)}

The RF cluster + ridge model proposed by \cite{Cao_RFstat_2022} integrates ensemble learning with penalized regression to identify influential assets and determine stable portfolio weights. The RF component captures nonlinear dependencies and performs feature selection via clustering and variable-importance measures, while ridge regression serves as the statistical optimization step for estimating smooth and robust portfolio weights.

\paragraph{Step 1. Random forest classification for asset selection}
A RF classifier is trained to predict the \textit{direction} of the index return (i.e., whether it is positive or negative) using the historical returns comprising of all $n$ assets. Each input instance consists of a vector of asset returns on a given day, and the corresponding output label is defined as 1 if the index return is positive, and 0 otherwise.

The classifier is an ensemble of $L$ decision trees $\{h(R_{t}, \theta_l), \, l= 1, 2, \ldots, L\}$, where $\theta_l$ denotes the parameters of the $l$-th tree. Each tree is trained on a bootstrap sample drawn with replacement from the original data. At every node of a tree, a random subset of $m \ll n$ features is selected, and the feature that maximizes an impurity-reduction criterion (e.g., the Gini index or classification error) is chosen for splitting.

For a given observation $R_t$, the final prediction of the ensemble is made by majority voting over all trees:
\[
H(R_t) = \arg\max_{y \in \{0,1\}} \sum_{l=1}^{L} I\big(h(R_t, \theta_l) = y\big),
\]
where $I(\cdot)$ is the indicator function. This ensemble construction reduces variance and avoids overfitting while providing the foundation for feature-importance estimation.

\paragraph{Step 2. Feature-importance evaluation}
Once the RF classifier is trained, the importance of each asset is quantified using the Mean Decrease in Accuracy (MDA). This metric measures the drop in predictive accuracy when the values of a particular feature (i.e., asset return series) are randomly permuted, thus breaking their relationship with the target variable. Formally, for asset $i$, the MDA score is computed as
\[
\text{MDA}_i = \text{Acc}_{\text{base}} - \text{Acc}_{\text{perm}}^{(i)},
\]
where $\text{Acc}_{\text{base}}$ is the baseline accuracy of the RF on the validation data, and $\text{Acc}_{\text{perm}}^{(i)}$ is the accuracy obtained after randomly permuting the $i$-th feature across the samples. A higher $\text{MDA}_i$ implies greater relevance of asset $i$ in predicting index return direction. 

\paragraph{Step 3. Asset-subset selection}
Assets are ranked in descending order of their $\text{MDA}_i$ scores, and the top $K$ assets (where $K$ is the desired cardinality) are selected to form the reduced candidate set $\mathcal{S}\subseteq \{1,2,\ldots,n\}$, where $|\mathcal{S}| = K$. 
This subset represents the assets most strongly associated with index performance.

\paragraph{Step 4. Ridge regression for portfolio optimization}
Using the selected asset subset $\mathcal{S}$, ridge regression is employed to estimate the optimal tracking portfolio weights by minimizing the mean-squared tracking error with an $\ell_2$ penalty to control multicollinearity.  
The optimization problem is formulated as
\[
\hat{w}_{\mathcal{S}} =
\arg\min_{w_{\mathcal{S}}\ge 0}
\left\{
\frac{1}{T}\sum_{t=1}^{T}\left(I_t - \sum_{i\in\mathcal{S}} r_{it}w_i\right)^2
+ \lambda\sum_{i\in\mathcal{S}} w_i^2
\right\},
\]
where $\lambda>0$ is a regularization parameter controlling shrinkage.  

The regularization parameter $\lambda$ is selected via $k$-fold cross-validation by minimizing the out-of-sample mean-squared error.  
The resulting weight estimates $\hat{w}_{\mathcal{S}}$ are then normalized to satisfy the portfolio constraints
\[
\sum_{i\in\mathcal{S}} \hat{w}_i = 1, \qquad \hat{w}_i \ge 0,
\]
ensuring a fully invested and long-only portfolio.

This integrated RF cluster + ridge framework exploits the feature-selection ability of RF to identify the most informative assets while using ridge regression to produce stable and interpretable portfolio weights. The combination effectively captures nonlinear dependencies in asset returns and yields sparse yet robust tracking portfolios, making it particularly suitable for high-dimensional index-tracking problems.

\subsubsection*{Random forest regression combined with ridge regression (RF regression+ridge)}
Similar to the clustering-based method, the RF regression combined with ridge regression approach proposed by \cite{Cao_RFstat_2022} employs a two-stage strategy that integrates the nonlinear predictive capability of RF with the robustness of ridge regression. Unlike the classification-based strategy, this model directly performs regression to predict the magnitude of index returns, thereby capturing both linear and nonlinear dependencies between asset returns and benchmark index performance.

\paragraph{Step 1. Random forest regression for asset selection}
A RF regression model is trained to predict the index return series $\{I_t\}_{t=1}^{T}$ using historical asset returns $\{r_{it}\}$ as predictors.  
Let $\{h(R,\theta_l),\, l=1,2,\ldots,L\}$ denote the ensemble of $L$ regression trees, each fitted on a bootstrap sample of $R\in\mathbb{R}^{T\times n}$ (the matrix of asset returns).  
For a given observation $R_t$, the predicted index return is obtained by averaging the outputs of all $L$ trees
\[
\hat{I}_t = \frac{1}{L}\sum_{l=1}^{L} h(R_t, \theta_l).
\]
This ensemble approach enables the model to learn complex, nonlinear relationships and interactions among the asset returns, while maintaining generalization through bootstrap aggregation and random feature selection. As in the classification-based approach, the trained model also provides feature-importance scores, which are used to identify a sparse subset of assets for the next stage of the model.

    \paragraph{Step 2. Feature-importance evaluation:}
The contribution of each asset to predicting the index is quantified via the percentage increase in mean squared error (PIM), which evaluates how prediction accuracy deteriorates when the $i$-th feature is randomly permuted.  
Formally, for asset \(i\), the PIM$_{i}$ score is given by
\[
\text{PIM}_i = 100 \times 
\frac{1}{L} \sum_{l=1}^{L}
\frac{\text{MSE}_{l}^{(i,\mathrm{perm})} - \text{MSE}_{l}}{\text{MSE}_{l}},
\]
where $\text{MSE}_{l}$ denotes the mean squared error of the $l$-th regression tree, and $\text{MSE}_{l}^{(i,\mathrm{perm})}$ is the error after permuting the $i$-th feature. Higher $\text{PIM}_i$ values indicate stronger predictive influence of asset $i$ on index returns; $i=1,\ldots,n$.

\paragraph{Step 3. Asset-subset selection}
Assets are ranked according to their $\text{PIM}_i$ values, and the top $K$ assets are selected to form the subset $\mathcal{S}\subseteq\{1,2,\ldots,n\}$ with $|\mathcal{S}|=K$.  
This data-driven selection identifies the most informative assets for replicating the benchmark index.

\paragraph{Step 4. Ridge regression for portfolio optimization}
Portfolio weights are estimated using the same ridge-regularized tracking error minimization procedure described in the RF cluster + ridge framework.

%


By coupling the nonlinear predictive power of RF regression with the regularization strength of ridge regression, this hybrid model produces sparse yet highly stable index-tracking portfolios.  
It efficiently identifies influential assets, mitigates multicollinearity, and improves out-of-sample tracking performance, making it particularly suitable for large-scale, high-dimensional index-tracking applications.
 
\vspace{0.5cm}

\item \textbf{Deep autoencoders} \\
Autoencoders are a type of neural network designed to learn efficient representations of data. This is achieved by compressing high-dimensional input data into a low-dimensional latent space and then reconstructing the original data from this compressed form. The training process aims to minimize information loss, which can be measured using various loss functions. A common choice is the $\ell_2$ -norm difference between the input and output vectors, although other loss functions can also be used (\cite{Ouyang_deepNN_2019}; \cite{KimKim_deepLR_2020}; \cite{Zhang_autoencoder_2020}).

In the context of index tracking, autoencoders are trained using return data from stocks that make up a market index. For a given period \( t \), the return data is represented as a vector \( \mathbf{r}_t =[r_{1t}, r_{2t}, \ldots, r_{nt}] \) where $r_{it}$ is a return from $i$th asset at $t$. 

The overall loss function for training the autoencoder is defined as the sum of the squared differences ($\ell_2$-norm) between the original return vectors and the reconstructed return vectors across all time periods:
\[
L = \sum_{t=1}^T \|\mathbf{r}_t - \hat{\mathbf{r}}_t\|_2^2
\]
where \( t \) is the index for the training samples, \( \mathbf{r}_t \) is the original vector of returns for time period \( t \), and  \( \hat{\mathbf{r}}_t \) is the reconstructed vector of returns for time period \( t \).

To evaluate the contribution of each stock to the reconstruction, the individual reconstruction loss for each stock is calculated. This helps identify which stocks carry the most communal information related to the benchmark index. The individual loss for stock \( i \) is given by:

\[
L_i = \sum_{t=1}^T \|r_{it} - \hat{r}_{it}\|_2^2
\]
where \( \hat{r}_{it} \) is the reconstructed return of asset \( i \) at time \( t \). The smaller the \( L_i \) value, the less information is lost for the \( i \)-th asset, indicating it is more similar to the overall index. This information is used to rank the assets based on their communal information with the benchmark index. By selecting the top \( K \) assets with the highest communal information, one can construct a tracking portfolio of \( K \) assets that closely mimics the index. In \cite{Zhang_autoencoder_2020}, the authors consider six autoencoder architectures for asset selection, discussed as follows:

\begin{enumerate}
    \item \textbf{Single hidden layer autoencoder (SH-AE)}: This baseline model consists of an encoder–decoder structure with a single hidden (bottleneck) layer comprising of fewer neurons than the input layer. The encoder compresses the standardized asset return data into a low-dimensional latent representation, while the decoder attempts to reconstruct the original inputs from this compressed code. The reconstruction error, measured as the MSE between the input and reconstructed outputs for each asset, serves as a ranking criterion—assets with the lowest reconstruction errors are considered most representative of the underlying market structure.

    \medskip
        
    \item \textbf{Sparse autoencoder (SP-AE)}: This model retains the same structure as the single hidden layer autoencoder but introduces an $\ell_1$ penalty on the hidden-layer activations to enforce sparsity. This encourages only a few neurons to be active for any given input, thereby promoting feature selectivity and reducing redundancy. By emphasizing the most salient features of asset returns, the sparse autoencoder improves the robustness and interpretability of asset ranking. 

    \medskip
        
    \item \textbf{Contractive autoencoder (CON-AE):} 
    The contractive autoencoder incorporates a Jacobian-based penalty into the loss function to make the learned representations invariant to small perturbations in the input. Specifically, it penalizes the sensitivity of the encoder to changes in the input by adding a contraction term proportional to the Frobenius norm of the encoder’s Jacobian matrix:
    \[
    \left\| \frac{\partial h(\mathbf{r}_t)}{\partial \mathbf{r}_t} \right\|_F^2,
    \]
    where \( h(\cdot) \) denotes the encoder mapping. In practice, this Jacobian norm is approximated by the squared Frobenius norm of the encoder weight matrix, serving as an efficient surrogate for local sensitivity. This regularization promotes stability and smoothness in the latent space, making the model particularly effective in identifying assets whose return dynamics exhibit consistent long-term structure.

    \medskip
    
    \item \textbf{Stacked autoencoder (STCK-AE):} 
    The stacked autoencoder extends the basic autoencoder architecture by employing multiple nonlinear hidden layers in both the encoder and decoder components. This hierarchical deep structure—typically comprising successive layers of sizes 64, 32, and 16 neurons in the encoder enables the model to extract complex, higher-order correlations among asset returns. By progressively compressing the input data through multiple nonlinear transformations, the model captures both global and local structures in the return dynamics, providing richer latent representations that enhance the identification of assets contributing most to the market’s underlying structure.

    \medskip

    \item \textbf{Denoising autoencoder (DEN-AE):} 
    The denoising autoencoder enhances robustness by deliberately corrupting the input data with random noise and training the network to reconstruct the original, uncorrupted version. Let the corrupted input be represented as 
    \[
    \tilde{\mathbf{r}}_t = \mathbf{r}_t + \eta_t,
    \]
    where $\eta_t \sim \mathcal{N}(0, \sigma^2)$ denotes the Gaussian noise added to the original return vector $\mathbf{r}_t$. The network is trained to minimize the reconstruction error between $\tilde{\mathbf{r}}_t$ and $\mathbf{r}_t$, thereby learning noise-invariant and stable representations of asset returns. This process improves the model’s generalization capability and resilience to stochastic fluctuations commonly observed in financial time series.

    \medskip
    
    \item \textbf{Variational autoencoder (VAR-AE):}
    The variational autoencoder introduces a probabilistic latent representation by learning an encoder that maps inputs to a distribution over latent variables rather than a single code. Given a standardized return vector $\mathbf{r}_t \in \mathbb{R}^n$, the encoder outputs the mean and log-variance of a Gaussian latent distribution,
    \[
    (\boldsymbol{\mu}_t, \log \boldsymbol{\sigma}^2_t) = f_{\phi}(\mathbf{r}_t),
    \]
    and a latent sample is obtained via the reparameterization trick,
    \[
    \mathbf{z}_t = \boldsymbol{\mu}_t + \exp\!\big(\tfrac{1}{2}\log \boldsymbol{\sigma}^2_t\big)\odot \boldsymbol{\epsilon}_t,\qquad \boldsymbol{\epsilon}_t \sim \mathcal{N}(\mathbf{0}, I).
    \]
   Here, $I$ is an identity matrix. The decoder $g_{\theta}$ maps $\mathbf{z}_t$ back to the input space to reconstruct $\hat{\mathbf{r}}_t=g_{\theta}(\mathbf{z}_t)$. 
   
    Training minimizes a reconstruction term plus a Kullback–Leibler divergence that regularizes the latent distribution towards the standard normal:
    \[
    \mathcal{L}_{\text{VAE}}(\theta,\phi)
    = \sum_{t=1}^T \underbrace{\big\|\mathbf{r}_t-\hat{\mathbf{r}}_t\big\|_2^2}_{\text{reconstruction}}
    \;+\; \lambda_{\mathrm{KL}}\,\underbrace{\sum_{j=1}^{d}\tfrac{1}{2}\!\left(\mu_{t,j}^2+\sigma_{t,j}^2-\log\sigma_{t,j}^2-1\right)}_{\mathrm{KL}\big(\mathcal{N}(\boldsymbol{\mu}_t,\mathrm{diag}(\boldsymbol{\sigma}^2_t))\,\|\,\mathcal{N}(\mathbf{0},I)\big)},
    \]
    where $d$ is the latent dimension, and $\lambda_{\mathrm{KL}}>0$ controls the strength of the KL regularization (in code: \texttt{lambda\_kl}). 
    This probabilistic bottleneck encourages smooth, disentangled latent factors that capture common market structure while preventing overfitting.

    In implementation, we use a two-layer encoder to produce $(\boldsymbol{\mu}_t,\log\boldsymbol{\sigma}^2_t)$, apply the reparameterization $\mathbf{z}_t$, and a two-layer decoder to reconstruct $\hat{\mathbf{r}}_t$. 
    

\end{enumerate}

All autoencoder variants follow the same training and selection pipeline. The models are trained using standardized asset return data, minimizing the mean squared reconstruction error between original and reconstructed returns. For each asset, the average reconstruction loss is computed and used to rank assets according to their similarity to the overall market structure. To balance representativeness and diversity, a mixed selection rule is applied—selecting 70\% of assets with the smallest reconstruction errors (communal) and 30\% with the largest errors (idiosyncratic). The final selected assets are then passed to the portfolio optimization stage for weight estimation.



Once a subset of assets is selected (denote by the set as \(\mathcal{S}\)), the tracking portfolio is determined by an optimization model that minimizes the MSE between the benchmark index returns and the returns generated by the selected assets (see MSE minimization in Section \ref{sec: optimization framework}). 


\bigskip

\item \textbf{Deep Learning} Recently, deep learning techniques, particularly deep neural networks (DNNs), have emerged as powerful tools for asset selection and portfolio construction in index tracking. In this context, \cite{Ouyang_deepNN_2019} proposed a hybrid two-stage approach integrating deep autoencoders and deep neural networks. This method leverages the representation power of autoencoders for asset selection and the predictive capability of neural networks for portfolio optimization. The framework involves the following sequential steps:

\subsubsection*{Deep neural network framework}
\cite{Ouyang_deepNN_2019} propose a deep learning-based two-stage method for index tracking, combining asset selection via a deep autoencoder and portfolio optimization through a deep neural network (DNN). The methodology consists of the following steps:

\begin{enumerate}
    \item \textbf{Asset selection with a deep autoencoder} 

\noindent A deep autoencoder is employed to identify representative assets by reconstructing historical asset returns and selecting those with minimal reconstruction errors. Specifically, given historical asset returns \(R \in \mathbb{R}^{T\times n}\), the autoencoder learns parameters \(\theta\) that minimize the reconstruction loss:
\[
\min_{\theta} \|R - \hat{R}\|^2_{F},
\]
where \(\hat{R} = g_{\theta}(R)\) denotes the reconstructed returns, parameterized by weights and biases \(\theta\). After training, the reconstruction error for each asset \(i\) is computed as:
\[
d_i = \frac{1}{T}\sum_{t=1}^{T}(r_{it}-\hat{r}_{it})^2.
\]
The top \(h\) assets with the lowest reconstruction errors \(d_i\) are selected to form the reduced candidate set \(\mathcal{S}\).

\item \textbf{Portfolio weight optimization using DNN}

\noindent Using the selected assets \(\mathcal{S}\), a feed-forward (DNN) is trained to predict benchmark index returns from the corresponding asset returns. The network parameters \(\phi\) are optimized by minimizing the mean squared error loss:
\[
\min_{\phi} \frac{1}{T}\sum_{t=1}^{T}(I_t-f_{\phi}(R_{t,\mathcal{S}}))^2,
\]
where \(R_{t,\mathcal{S}}\) denotes the returns of selected assets at time \(t\), and \(f_{\phi} (\cdot)\) the DNN prediction function parameterized by \(\phi\). This step enables the model to capture complex nonlinear dependencies between the selected assets and the benchmark index.

\item \textbf{Determining portfolio weights through sensitivity analysis}

\noindent The trained network is then used to obtain portfolio weights by performing a gradient-based sensitivity of the predicted index return with respect to each input feature. The average absolute gradient for asset \(i\) is computed as
\[
s_i = \frac{1}{T}\sum_{t=1}^{T}\left|\frac{\partial f_{\phi}(R_{t,\mathcal{S}})}{\partial r_{it}}\right|,
\quad i \in \mathcal{S}.
\]
These sensitivity scores are normalized to obtain portfolio weights satisfying
\[
w_i = \frac{s_i}{\sum_{j \in \mathcal{S}} s_j}, \qquad 
\sum_{i \in \mathcal{S}} w_i = 1, \quad w_i \geq 0.
\]
\end{enumerate}

This approach effectively leverages deep learning to dynamically select and weight assets, demonstrating strong empirical performance with significantly reduced tracking error and portfolio complexity compared to traditional methods.

\subsubsection*{Deep neural network with fixed noise (Deep NNF)}
The DNN with fixed noise (Deep-NNF) approach, introduced by \cite{KwakSongLee_NN_2021}, provides an innovative method for portfolio selection and index tracking.  Unlike conventional feed-forward neural networks that take asset returns as direct inputs, the Deep-NNF model generates portfolio weights through a latent noise vector that is independent of the observed market data. This framework uniquely removes dependency on explicit asset returns as inputs, instead using a fixed, random noise vector as the sole input to a DNN. This noise-driven method inherently emphasizes portfolio diversification and prevents overfitting to historical data.

The Deep-NNF methodology involves two principal stages, detailed as follows:

\noindent \textbf{Stage 1: Asset selection via fixed-noise neural network.} \\
Initially, a deep feed-forward neural network \(f(\cdot)\) is constructed with a fixed, randomly generated noise vector \( \mathbf{\xi} \in \mathbb{R}^{n}\) serving as its only input. The network architecture consists of multiple fully connected layers with dropout and ReLU activations, formulated as:
\[
\mathbf{w} = f(\mathbf{\xi}; \theta),
\]
where \(\theta\) denotes the network parameters (weights and biases). The output layer employs a softmax activation function to generate valid portfolio weights:
\[
w_i = \frac{e^{u_i}}{\sum_{j=1}^{N} e^{u_j}}, \quad i=1,\dots,N,
\]
where \( u_i \) represents the pre-activation output of the final layer for asset \( i \). 
The network parameters $\theta$ are learned by minimizing the mean squared error between the realized portfolio return and the benchmark index return as
\[
\mathcal{L}(\theta) = \frac{1}{T}\sum_{t=1}^{T}\big(I_t - R_{wt}\big)^2.
\]
After training, the top \( h \) assets are selected based on the magnitude of their corresponding portfolio weights.

\noindent \textbf{Stage 2: Partial replication and portfolio refinement.} \\
In the second stage, the neural network is re-trained exclusively on the returns of the selected \( h \) assets. Specifically, the asset return matrix \( \mathbf{r}_t^{(h)} \in \mathbb{R}^{h} \) corresponding to the chosen assets is used, along with a new fixed noise vector \( \mathbf{\xi}^{(h)} \in \mathbb{R}^{h} \). The model again minimizes the mean squared tracking error:
\[
\min_{\theta_h}\frac{1}{T}\sum_{t=1}^{T}\left(I_t - \mathbf{w}_h^\top \mathbf{r}_t^{(h)}\right)^2,
\]
to obtain the refined optimal portfolio weights \(\mathbf{w}_h\).
Finally, these optimized weights are expanded back into the full-dimensional asset space by assigning zeros to the unselected assets, thus completing the partial replication strategy.
This two-stage Deep-NNF framework leverages the randomness inherent in the fixed-noise input and the flexibility of deep neural networks to achieve robust and diversified portfolios, demonstrating superior tracking accuracy and generalization compared to traditional index-tracking methods \citep{KwakSongLee_NN_2021}.

\end{enumerate}

\section{Numerical study}
\label{Sec4}

To illustrate the practical implications of the reviewed index tracking (IT) models, we conduct a comparative numerical study on the S\&P~500 index. The sample comprises daily closing prices of the index and its constituents over a ten-year horizon, from 10 December 2012 to 11 August 2022, covering $T=2524$ trading days. For asset $i \in \{1,\ldots,n\}$ and trading day $t \in \{1,\ldots,T\}$, let $p_{i,t}$ denote the closing price, with simple daily return
\begin{equation}
r_{i,t}=\frac{p_{i,t}-p_{i,t-1}}{p_{i,t-1}} \, .
\end{equation}
Constituents with complete price histories over this horizon are retained, yielding $n=462$ assets. The period was chosen to maximize the number of constituents with data availability.


\subsection{Methodology}
We evaluate $29$ representative IT models spanning three modeling paradigms: (i) optimization-based approaches ($8$ models), (ii) statistical-based methods ($7$ models), and (iii) data-driven methods ($14$ models). Each model is evaluated using a rolling-window design: a two-year in-sample window ($504$ trading days) followed by a three-month out-of-sample test window ($63$ trading days). The window advances by $63$ days, producing a total of $32$ evaluation windows. To ensure comparability, we impose a uniform cardinality constraint of $K=45$ (approximately $10\%$ of the total number of assets). 

Out-of-sample performance is assessed using multiple measures detailed below. In addition, we perform pairwise statistical tests on model tracking errors to evaluate whether observed performance differences are significant. Finally, we conduct a two-stage comparison: (i) within-paradigm assessment to identify the leading model in each class, followed by (ii) a cross-paradigm evaluation of the best representatives. This approach highlights not only absolute performance but also the relative strengths and limitations of optimization-, statistical-, and data-driven frameworks in index tracking.




\subsection{Performance Metrics}
\label{sec: performance metrics}

The performance of the tracking portfolios is evaluated using five complementary categories of measures: (i) tracking performance, (ii) return performance, (iii) risk performance, (iv) risk-adjusted performance, and (v) asset-weight characteristics. Let $r_{p,t}$ and $r_{b,t}$ denote the portfolio and benchmark returns at time $t$, respectively, over an out-of-sample evaluation horizon of length $H$. The average portfolio return is denoted by $\bar r_p = \tfrac{1}{H}\sum_{t=1}^H r_{p,t}$.


\begin{enumerate}
    \item \textbf{Tracking Performance:} To evaluate how closely the portfolio replicates the S\&P~500, we use two complementary measures: (i) tracking error, which captures the magnitude of deviations, and (ii) correlation, which reflects directional co-movement.

    \begin{enumerate}
        \item \textbf{Tracking error (TE):} 
    TE measures the variability of the active return series $(r_{p,t} - r_{b,t})$, i.e., how much the portfolio deviates from the benchmark on average. 
    It is given by
    \[
    \text{TE} = \sqrt{\frac{1}{H-1}\sum_{t=1}^{H} \Big( (r_{p,t} - r_{b,t}) - \overline{(r_p-r_b)} \Big)^2},
    \]
    where $\overline{(r_p-r_b)} = \tfrac{1}{H}\sum_{t=1}^H (r_{p,t}-r_{b,t})$. 
    Smaller TE values indicate closer replication of the benchmark.

    \item \textbf{Correlation with the index:} 
    Correlation assesses the strength of linear co-movement between the portfolio and the benchmark. 
    It is computed as
    \[
    \rho(r_p,r_b) = \frac{\operatorname{Cov}(r_p,r_b)}{\sigma_p\sigma_b},
    \]
    where $\sigma_p$ and $\sigma_b$ are the standard deviations of $r_{p,t}$ and $r_{b,t}$. 
\end{enumerate}
    \item \textbf{Return Performance:} To assess the growth potential of the tracking portfolio, we report three simple measures: the average return, the worst observed return, and the best observed return over the evaluation horizon.

\begin{enumerate}
    \item \textbf{Average return:} 
    The mean return $\bar r_p$ (already defined above) provides a measure of the typical performance delivered by the tracking portfolio.

    \item \textbf{Minimum return:} 
    The lowest single-period return observed,
    \[
    r_{\min} = \min_{1 \leq t \leq H} r_{p,t},
    \]
    capturing the downside extremum.

    \item \textbf{Maximum return:} 
    The highest single-period return observed,
    \[
    r_{\max} = \max_{1 \leq t \leq H} r_{p,t},
    \]
    indicating the best realized outcome.
\end{enumerate}



    
    
    \item  \textbf{Risk Performance:} To evaluate the stability of the tracking portfolio, we consider two measures: volatility (overall variability of returns) and average drawdown (typical losses from peak values).

\begin{enumerate}
   \item \textbf{Standard Deviation:} A measure of the portfolio's return volatility, denoted by $\sigma_p$, is calculated as
    \[
    \sigma_p = \sqrt{\frac{1}{H-1}\sum_{t=1}^H (r_{p,t} - \bar r_p)^2}.
    \]
    Lower values indicate greater stability in portfolio performance.

    \item \textbf{Average drawdown:} 
    Let the cumulative wealth process be $W_t = \prod_{s=1}^t (1+r_{p,s})$. 
    The drawdown at time $t$ is defined as
    \[
    DD_t = \frac{W_t - \max_{1\leq s \leq t} W_s}{\max_{1\leq s \leq t} W_s} \leq 0.
    \]
    The average drawdown is then
    \[
    \overline{DD} = \frac{1}{H}\sum_{t=1}^H |DD_t|,
    \]
    which reflects the mean proportional loss from peak levels over the evaluation horizon.
\end{enumerate}
    

        \item \textbf{Risk-adjusted performance:} 
Absolute returns alone can be misleading without accounting for risk exposure. 
To obtain a fairer comparison, we evaluate performance ratios that relate excess return to different notions of risk. All ratios are interpreted meaningfully only when the portfolio achieves a positive excess return.

\begin{enumerate}
   \item \textbf{Sharpe ratio:} 
    Relates average excess return to standard deviation,
    \[
    \text{Sharpe Ratio} = \frac{\bar r_p - r_f}{\sigma_p}, \quad \bar r_p - r_f > 0,
    \]
    where $\sigma_p$ is the portfolio standard deviation. Higher values indicate better risk-adjusted returns overall.

    \item \textbf{Sortino ratio:} 
    Focuses on downside risk by replacing total volatility with downside deviation,
    \[
    \text{Sortino Ratio} = \frac{\bar r_p - r_f}{\sigma_d}, \quad \bar r_p - r_f > 0,
    \]
    where $\sigma_d$ is the standard deviation of negative returns. Larger values emphasize protection against losses.

    \item \textbf{Treynor ratio:} 
    Measures excess return per unit of systematic risk,
    \[
    \text{Treynor Ratio} = \frac{\bar r_p - r_f}{\beta_p}, \quad \bar r_p - r_f > 0,
    \]
    where $\beta_p$ is the portfolio beta with respect to the benchmark. It highlights performance relative to market exposure.

    \item \textbf{Information ratio:} 
    Evaluates active return relative to tracking error,
    \[
    \text{Information Ratio} = \frac{\bar r_p - \bar r_b}{\text{TE}}, \quad \bar r_p - \bar r_b > 0,
    \]
    where $\bar r_b$ is the average benchmark return. Higher values imply more efficient benchmark outperformance.
\end{enumerate}

 \item \textbf{Turnover and time complexity:} To evaluate cost efficiency and sparsity of the constructed portfolios, we consider three measures, two based on the portfolio weights and one based on the time taken. 

\begin{enumerate}
    \item \textbf{Turnover ratio:} 
    Captures the average trading activity across rebalancing windows. 
    Higher turnover implies greater transaction costs. It is defined as
    \[
    \text{TR} = \frac{1}{N-1} \sum_{k=1}^{N-1}\sum_{i=1}^{n}\big|w_{i,k+1}-w_{i,k}\big|,
    \]
    where $w_{i,k}$ is the weight of asset $i$ in the $k$-th window and $N$ is the total number of rolling windows.

    \item \textbf{No. of Assets:} 
    The average number of assets with non-zero weights in the tracking portfolio. This reflects portfolio sparsity, with smaller values indicating more concentrated portfolios.

    \item \textbf{Solver time:} The total CPU time taken by the solver to obtain optimal or near-optimal solutions across all rolling windows. This metric reflects the computational tractability and scalability of each modeling framework, thus providing a practical perspective on the trade-off between model complexity and real-world applicability.
    
\end{enumerate}
\end{enumerate}

\subsection{Statistical test}
\label{sec: stat_test}

To assess whether differences in tracking error across models are statistically significant, 
we perform \textit{pairwise one-sided paired $t$-tests}. 
For each model pair $(i,j)$, let $\text{TE}_{i,k}$ and $\text{TE}_{j,k}$ denote the out-of-sample tracking errors in window $k$, for $k=1,\ldots,N$. 
The null and alternative hypotheses are specified as
\[
H_0 : \bar{\text{TE}}_i = \bar{\text{TE}}_j 
\quad \text{vs.} \quad 
H_A : \bar{\text{TE}}_i < \bar{\text{TE}}_j,
\]
where $\bar{\text{TE}}_i = \tfrac{1}{N}\sum_{k=1}^N \text{TE}_{i,k}$ is the mean tracking error of model $i$. 

The paired setting accounts for temporal dependence, since both models are evaluated on the same rolling windows. 
The resulting $p$-values are summarized in a matrix, which highlights statistically significant dominance relationships among competing models.

\subsection{S\&P 500 Index Features}
\label{sec: index features}

We characterize the S\&P~500 index and its constituents over the study period 
(12 October 2012 to 8 November 2022) using three descriptive statistics: mean returns, standard deviation (volatities), and correlations with the index.

\begin{enumerate}
    \item \textbf{Distribution of daily mean returns:} Figure~\ref{fig:MR_SP500} reports the cross-sectional distribution of mean daily returns across all constituents. Most assets cluster around modest positive levels, with the distribution right-skewed due to a small number of high-return outliers. The vertical dashed lines indicate the cross-sectional mean (red) and cross-sectional median (blue) of constituent mean returns. The fact that the median lies below the mean reflects the influence of a few extreme outliers in pulling the average upward, suggesting that relatively few stocks drive the right tail of the return distribution while the majority deliver more modest performance. 
    

\begin{figure}[htp!]
\centering
\includegraphics[height=7cm, width=14cm]{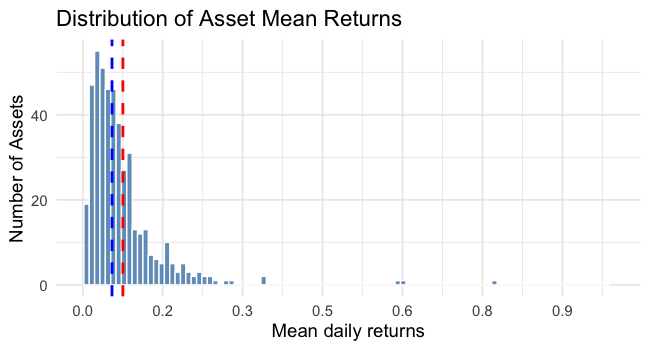}
\caption{Distribution of mean daily returns for the S\&P 500 constituents over the sample period 2012-12-10 to 2022-08-11. The red dashed line denotes the cross-sectional mean of constituent mean returns, while the blue dashed line denotes the cross-sectional median.}
\label{fig:MR_SP500}
\end{figure}

\item \textbf{Distribution of daily volatility:} Figure \ref{fig:SD_SP500} shows the distribution of standard deviation of daily returns, serving as a measure of asset volatility. Most assets exhibit volatility between 0.2 and 0.5, while the distribution is right-skewed with a small number of highly volatile stocks exhibiting standard deviations greater than 1. The vertical dashed lines denote the cross-sectional mean (red) and median (blue) volatilities across constituents. The fact that the mean lies to the right of the median highlights the impact of a few extreme outliers in inflating the average, while the median better represents the typical stock’s volatility level.


\begin{figure}[htp!]
\centering
\includegraphics[height=7cm, width=14cm]{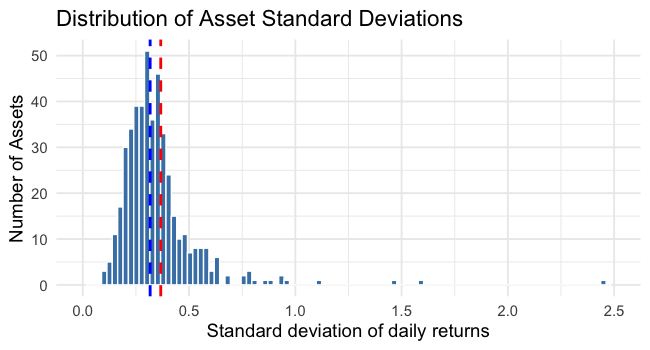}
\caption{Distribution of daily return volatilities (standard deviations) for the S\&P 500 constituents over the sample period 2012-12-10 to 2022-08-11. The red dashed line denotes the cross-sectional mean volatility across constituents, while the blue dashed line denotes the cross-sectional median volatility}
\label{fig:SD_SP500}
\end{figure}

\item \textbf{Distribution of asset correlations:} Figure \ref{fig:CORR_SP500} displays the distribution of correlations between each asset’s returns and the S\&P~500 index return. Most constituents exhibit relatively high correlations in the range of 0.6 to 0.9, indicating strong co-movement with the market benchmark. The distribution is moderately left-skewed, with a non-trivial number of assets displaying weak or even negative correlations. The vertical dashed lines denote the cross-sectional mean (red) and median (blue) correlations across assets. The median lying slightly below the mean highlights the influence of a subset of highly correlated assets in pulling the average upward.


\begin{figure}[htp!]
\centering
\includegraphics[height=7cm, width=14cm]{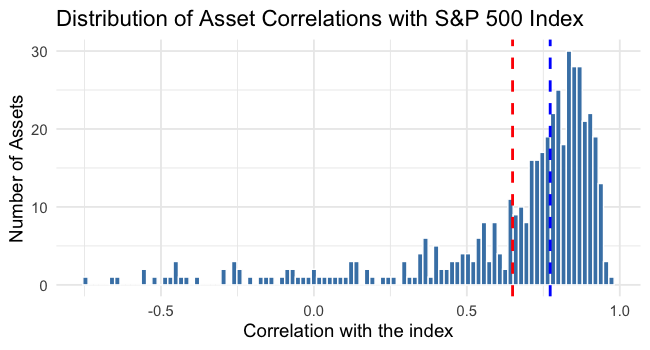}
\caption{Distribution of daily asset correlations with the S\&P 500 index over the 10-year period from 10-12-2012 to 11-08-2022. The red dashed line denotes the mean correlation, and the blue dashed line denotes the median correlation across assets.}
\label{fig:CORR_SP500}
\end{figure}
\end{enumerate}




\subsection{Out-of-sample analysis}
We now turn to the comparative evaluation of index-tracking models. 
The analysis focuses on out-of-sample performance, assessed over the rolling windows described earlier. 
For clarity, Table~\ref{tab:models} provides a complete listing of the models considered, grouped into three broad paradigms: optimization-based, statistical, and data-driven approaches and Table \ref{tab:model specifications} provides the model specifications, listing the parameter values and other specifications considered for the empirical study.

\begin{table}[h!]
\centering
\caption{Index-tracking models evaluated in the study, grouped into three categories: optimization-based, statistical, and data-driven. 
This taxonomy provides the framework for the comparative out-of-sample analysis.}
\footnotesize

\begin{tabular}{p{0.28\linewidth} p{0.30\linewidth} p{0.38\linewidth}}

\toprule
\textbf{Optimization models} & \textbf{Statistical models} & \textbf{Data-driven models} \\
\midrule
1. Mean Squared Error (MSE) & 1. Least Squares Regression (LSR) & 1. Clustering model 1 (Clust~1) \\ 
2. Sum of Errors Squared (SES) & 2. Quantile Regression (QR) & 2. Clustering model 2 (Clust~2) \\
3. Mean Absolute Deviation (MAD) & 3. Regression + Non-Negative LASSO (NNL) & 3. Sparse $\epsilon$-SVR model ($\epsilon$-SVR) \\ 
4. Mean Absolute Downside Deviation (MADD) & 4. Regression + Elastic Net (NNEN) & 4. Sparse $\nu$-SVR model ($\nu$-SVR) \\
5. MinMax & 5. Co-integration + Simulation (Coint-Sim) & 5. RF clustering + ridge (RF-Clust) \\
6. Downside MinMax (DMinMax) & 6. Convex Co-integration (Cvx-CoInt) & 6. RF regression + ridge (RF-Reg) \\ 
7. Tracking Error Variance (TEV) & 7. Factor-Based Model (FBM) & 7. Deep Autoencoders (six variants: SH-AE, Con-AE, DEN-AE, SP-AE, STCK-AE, VAR-AE) \\ 
8. Tailed Mixed CVaR (TMCVaR) & & 8. Deep Neural Network (DNN) \\
 & & 9. DNN with Fixed Noise (DNNF) \\ 
\bottomrule
\end{tabular}
\label{tab:models}
\end{table}

\begin{table}[h!]
\centering
\caption{Index-tracking models evaluated in the study with their specifications}
\footnotesize
\begin{tabular}{p{0.15\linewidth} p{0.80\linewidth} }
\toprule
\textbf{Model Name} & \textbf{Model Specifications} \\
\hline 
MSE & Cardinality constraint (no. of assets), $K=45$ \\
SES & Cardinality constraint (no. of assets), $K=45$ \\
MAD & Cardinality constraint (no. of assets), $K=45$ \\
MADD & Cardinality constraint (no. of assets), $K=45$ \\
MinMax & Cardinality constraint (no. of assets), $K=45$ \\
DMinMax & Cardinality constraint (no. of assets), $K=45$ \\
TEV & $K=45$, benchmark index weights are chosen using least squares regression and then normalized \\
TMCVaR & $K=45$, $\alpha=(0.9, 0.75, 0.5, 0.1, 0.01)$, $\delta=0.5$ \\
\hline
LSR & $K=45$\\
QR & $K=45$, $\tau=0.5$  \\ 
NNL & Cardinality $K=45$, regularization parameter $\lambda$ obtained using bisection method, starting with $0$ and $100$ \\
NNEN & Cardinality $K=45$, \\
FBM & Number of factors, $k=5$, Cardinality, $K=45$ assets \\
Coint-Sim & $K=45$, number of iterations$=10000$, significance level for ADF test $=0.05$, maximum lag in ADF test$=1$, autolag criterion in ADF test='AIC'\\
Cvx-CoInt & Cardinality, $K=45$;
convex MIQP solved using Gurobi \\
\hline
Clust~1 & Number of clusters, $K=45$ \\ 
Clust~2 & Number of clusters, $K=45$\\
$\epsilon$-SVR & $K=45, C_1=\{ 0.1, 1, 10, 50 \}, \epsilon=\{0.001, 0.005, 0.01, 0.05\}$ \\ 
$\nu$-SVR & $K=45, C_1=\{ 0.1, 1, 10, 50 \}, \epsilon=\{0.001, 0.005, 0.01, 0.05\}, C_2=1$ \\ 
RF-Clust& number of trees=100, cardinality $K=45$, $\alpha=[10^{-4}, 10^{-2}]$, cross-validation folds=5 \\
RF-Reg & number of trees=100, cardinality $K=45$, $\alpha=[10^{-4}, 10^{-2}]$, cross-validation folds=5 \\ 
SH-AE & Hidden layer: 16 neurons (ReLU), optimizer: Adam, loss: MSE, epochs: 50, batch size: 32, $K=45$ \\
SP-AE & Hidden layer: 16 neurons (ReLU), sparsity penalty $\lambda=10^{-4}$ (L1), epochs: 50, batch size: 32, $K=45$ \\
CON-AE & Hidden layer: 16 neurons (ReLU), contractive penalty $\lambda=10^{-4}$, epochs: 50, batch size: 32, $K=45$ \\
STCK-AE & Encoder: [64, 32, 16], decoder: [32, 64], activation: ReLU, epochs: 50, batch size: 32, $K=45$ \\
DEN-AE & Hidden layer: 16 neurons (ReLU), Gaussian noise (0.1), epochs: 50, batch size: 32, $K=45$ \\
VAR-AE & Latent dimension: 16, KL regularization $\lambda_{KL}=10^{-4}$, epochs: 50, batch size: 32, $K=45$ \\
DNN & Asset selection via stacked autoencoder (encoder: 64,32,16;          decoder: 32,64; activation: ReLU); 
      optimizer: Adam; loss: MSE; epochs: 100; batch size: 32; \\
      & Deep NN (layers: 64,32,1; activation: ReLU); optimizer: Adam (lr = 0.01); loss: MSE; epochs: 100; batch size: 16\\
DNNF & $K=45$, hidden layers=6, neurons per layer=64, activation=ReLU, dropout=0.5, 
optimizer=Adam (lr=0.01), epochs=100, batch size=16, output=Softmax normalization \\ 
\bottomrule
\end{tabular}
\label{tab:model specifications}
\end{table}

Note that for fair comparison, each machine learning model was trained under its own empirically optimal configuration, ensuring convergence of training loss. The number of epochs and batch sizes were not forced to be identical across models, as the learning objectives and convergence characteristics differ by architecture. However, all models were evaluated on identical rolling windows, same cardinality constraint and benchmarked using consistent evaluation metrics, as described in Section \ref{sec: performance metrics}. Across all optimization models, the values of $\epsilon_{i}0$ and $\delta_i$ is set to 0 and 1, respectively for all assets $i$.

\subsubsection{Performance evaluation of optimization-based models} 
Table~\ref{tab:OSP_optimization} reports the out-of-sample performance of the eight optimization-based index-tracking portfolios. 
For clarity, the discussion below highlights key results grouped by performance criteria mentioned previously.
\begin{table*}[t!]
\centering
\caption{Out-of-sample performance of optimization-based index-tracking portfolios. 
Panel~I reports performance measures for the models; Panel~II reports pairwise $p$-values from one-sided paired $t$-tests on tracking error. 
Best values are shown in \textbf{bold} and worst in \textit{italics}.}
\label{tab:OSP_optimization}
\scalebox{0.75}{%
\begin{tabular}{l|rrrrrrrrr}
\toprule
\multicolumn{9}{c}{\textbf{Panel I: Out-of-sample performance}} \\
\midrule
\textbf{Models} & MSE & SES & MAD & MADD & MinMax & DMinMax & TEV & TMCVaR & S\&P 500\\
\midrule
\multicolumn{10}{l}{\textbf{Tracking performance}} \\
\quad Tracking error & 0.00146 & 0.01490 & 0.00146 & 0.00150 & 0.00145 & 0.00148 & \textbf{0.00142} & \textit{0.00159} & 0.00000 \\
\quad Correlation & 99.19\% & \textit{69.21\%} & 99.19\% & 99.18\% & 99.20\% & 99.18\% & \textbf{99.25\%} & 99.09\% & 100\% \\
\midrule
\multicolumn{10}{l}{\textbf{Return performance}} \\
\quad Average return & 0.00050 & \textbf{0.00124} & 0.00045 & 0.00047 & 0.00048 & 0.00046 & \textit{0.00044} & \textit{0.00044} & 0.00042 \\
\quad Minimum return & -0.11905 & \textit{-0.17794} & -0.12037 & -0.12355 & -0.12187 & \textbf{-0.11693} & -0.12619 & -0.12235 & -0.11984 \\ 
\quad Maximum return & 0.10101 & \textbf{0.18139} & 0.09855 & 0.10533 & 0.09841 & \textit{0.09811} & 0.10090 & 0.10920 & 0.09383 \\
\midrule
\multicolumn{10}{l}{\textbf{Risk performance}} \\
\quad Standard deviation & 0.01147 & \textit{0.02032} & \textbf{0.01143} & 0.01168 & 0.01154 & 0.01154 & 0.01162 & 0.01172 & 0.01142 \\
\quad Average drawdown & \textbf{0.03496} & \textit{0.08524} & 0.03618 & 0.03586 & 0.03584 & 0.03562 & 0.03675 & 0.03735 & 0.03835 \\
\midrule
\multicolumn{10}{l}{\textbf{Risk-adjusted performance}} \\
\quad Sharpe ratio & 0.04316 & \textbf{0.06084} & 0.03944 & 0.04054 & 0.04175 & 0.04004 & 0.03816 & \textit{0.03791} & 0.03650 \\
\quad Sortino ratio & 0.05872 & \textbf{0.08670} & 0.05349 & 0.05525 & 0.05689 & 0.05435 & \textit{0.05179} & 0.05191 & 0.04931 \\
\quad Treynor ratio & 0.00050 & \textbf{0.00100} & 0.00045 & 0.00047 & 0.00048 & 0.00046 & \textit{0.00044} & \textit{0.00044} & 0.00042 \\
\quad Information ratio & 0.05365 & \textbf{0.05499} & 0.02348 & 0.03779 & 0.04484 & 0.03064 & 0.01875 & \textit{0.01742} & -- \\
\midrule
\multicolumn{10}{l}{\textbf{Turnover and Time complexity}} \\
\quad Turnover (TR) & 1.15647 & \textit{1.91312} & 1.18348 & \textbf{1.13976} & 1.21212 & 1.23969 & 1.18920 & 1.31755 & -- \\
\quad No. of assets & 45 & 2 & 45 & 45 & 45 & 45 & 45 & 45 & -- \\
\quad Solver time & 17.52h & 1.81s & 17.58h & 17.52h & 17.76h & 18.04h & 18.01h & 18.22h & -- \\
\midrule
\multicolumn{9}{c}{\textbf{Panel II: Pairwise $p$-values of tracking error $t$-tests}} \\
\midrule
$p$-values & MSE & SES & MAD & MADD & MinMax & DMinMax & TEV & TMCVaR \\ 
\midrule
MSE & ****** & 1.00000 & 0.55610 & 0.95268 & 0.68441 & 0.67954 & 0.06967 & 0.89609 \\
SES & 1.4E-16 & ****** & 2.3E-16 & 1.8E-16 & 1.6E-16 & 1.4E-16 & 1.3E-16 & 1.4E-16 \\
MAD & 0.44390 & 1.00000 & ****** & 0.92645 & 0.62114 & 0.64187 & 0.06082 & 0.89603 \\ 
MADD & 0.04732 & 1.00000 & 0.07355 & ****** & 0.09033 & 0.12427 & 0.00278 & 0.61012 \\
MinMax & 0.31559 & 1.00000 & 0.37886 & 0.90967 & ****** & 0.55783 & 0.06581 & 0.85050 \\
DMinMax & 0.32046 & 1.00000 & 0.35813 & 0.87573 & 0.44217 & ****** & 0.06129 & 0.83036 \\
TEV & 0.93033 & 1.00000 & 0.93918 & 0.99722 & 0.93419 & 0.93871 & ****** & 0.98849 \\
TMCVaR & 0.10391 & 1.00000 & 0.10397 & 0.38988 & 0.14950 & 0.16964 & 0.01151 & ****** \\
\bottomrule
\end{tabular}
}
\end{table*}

\begin{enumerate}
    \item \textbf{Tracking performance:}
    \begin{itemize}
        \item With the exception of SES, all models achieve tight tracking errors in the range 0.0014–0.0016. The TEV model records the lowest TE (0.00142), followed by MinMax (0.00145). In contrast, SES performs poorly, with a TE of 0.01490.  

        \item Consistent with the TE results, most models exhibit very high correlations (above 99\%) with the S\&P~500 index. 
        TEV again performs best (99.25\%), closely followed by MinMax. 
        SES is the clear outlier, with correlation dropping to 69.21\%.
    \end{itemize}

\item \textbf{Return performance:}
\begin{itemize}
    \item Owing to its very small portfolio size, the SES model achieves the highest average return (0.00124) and maximum return (0.18139) among all optimization models. However, it also records the lowest minimum return ($-0.17794$), corresponding to the largest maximum loss. This combination reflects a highly volatile and unstable return profile compared to its peers. 

    \item For the remaining models, average returns are tightly clustered around the index mean return of 0.00042, with values between 0.00044 and 0.00050. The MSE portfolio records the highest mean return (0.00050), followed by MinMax (0.00048). Similarly, maximum returns range narrowly from 0.098 to 0.109, and minimum returns from $-0.116$ to $-0.126$, closely matching the benchmark values (0.0938 and $-0.1198$). This demonstrates that, aside from SES, the optimization-based portfolios reproduce the benchmark’s return profile with only small deviations. 

    \item Notably, the TEV portfolio, which achieved the lowest tracking error and highest correlation, also produces an average return (0.00044) that is very close to the index mean, reinforcing its consistency across both tracking and return dimensions. 
\end{itemize}

\item \textbf{Risk performance:}
\begin{itemize}
    \item The SES model again stands out as an outlier, producing the largest volatility (0.02032), almost double that of the index and the other optimization portfolios (0.0114–0.0117). It also suffers the largest drawdown (0.08524). These results are linked to its extremely sparse asset selection (two assets despite the cardinality constraint), which limits diversification and increases instability.

    \item In contrast, all other portfolios exhibit volatilities comparable to the index value of 0.01142. The MSE portfolio has the lowest volatility (0.01147), while MADD records the highest (0.01168). Average drawdowns are also consistently smaller than the index benchmark of 0.03835, ranging from 0.03496 (MSE) to 0.03735 (TMCVaR). This illustrates the benefit of optimization-based selection, which produces diversified portfolios with lower downside risk than holding the full index. 
\end{itemize}

\item \textbf{Risk-adjusted performance:} 
\begin{itemize}
    \item The SES portfolio attains the highest values across all risk-adjusted ratios (Sharpe $0.0608$, Sortino $0.0867$, Treynor $0.0010$, Information $0.0550$), largely because of its unusually high mean return. Nevertheless, SES is not a desirable tracker: its extreme concentration (two holdings) results in materially higher risk, both volatility ($0.0203$) and drawdown ($0.0852$), reflecting poor diversification.

    \item Among the remaining, more stable trackers, MSE delivers the strongest set of ratios overall (Sharpe $0.0432$, Sortino $0.0587$, Treynor $0.00050$, Information $0.0537$), followed by MinMax; TMCVaR tends to be the weakest (e.g., Information $0.0174$). Excluding SES, all optimization portfolios exceed the index on Sharpe, Sortino, and Treynor (ranges $0.0379$–$0.0432$, $0.0518$–$0.0587$, and $0.00044$–$0.00050$ versus $0.0365$, $0.0493$, and $0.00042$, respectively), supporting the effectiveness of optimization-based selection out of sample. Within this group, TEV and TMCVaR track the index’s ratios most closely. 
    \end{itemize}

\item \textbf{Turnover and asset count:}  
All optimization-based portfolios satisfy the cardinality constraint of 45 assets, except SES, which selects only two assets. 
This extreme concentration explains SES’s unstable performance despite favorable risk-adjusted ratios. SES also exhibits the highest turnover (1.913), indicating frequent rebalancing, whereas the other models maintain turnover in a narrower range of approximately 1.14–1.32.

\item \textbf{Computational efficiency:} We evaluate the computational efficiency of the optimization-based models in terms of their total runtime over 32 rolling windows. All optimization models were coded in R and solved using the Gurobi Optimizer through its R library interface, with a uniform time limit of 1800 seconds per rolling window. With the exception of the SES model, all optimization models consistently reached this time limit in each window, yielding an average runtime of approximately 1800 seconds per window. This amounts to a cumulative computational effort of about 16 hours across 32 windows. In practice, the observed wall-clock time was slightly higher (17–18 hours) due to solver overhead and system resource allocation. As a result, these models are computationally expensive\footnote{For the cardinality-constrained TMCVaR model, each rolling window was solved with the 1800-second time limit. While most windows produced feasible portfolios with the desired cardinality of 45 assets, three windows (8, 10, and 28) yielded degenerate solutions with significantly fewer assets. When these windows were re-run independently, full-cardinality solutions were obtained, suggesting that the batch run terminated prematurely due to time-limited convergence. This highlights both the sensitivity of mixed-integer portfolio models to solver resource constraints and the numerical conditioning of certain windows.} under the current resource setting . It is worth noting that, since all underlying formulations are convex programs, the solutions reported are globally optimal when convergence occurs. With access to more sophisticated computing infrastructure, the runtime burden could be substantially reduced.

\item \textbf{Statistical significance (Panel II):}  
Panel~II reports the $p$-values from the pairwise one-sided $t$-tests described in Section~\ref{sec: stat_test}, evaluating whether one portfolio achieves a lower tracking error than another. A model with many significant entries in its column is interpreted as dominating other portfolios in terms of tracking accuracy. The TEV column contains multiple significant entries: its tracking error is lower than SES, MADD, and TMCVaR at the 5\% level, and lower than MSE, MAD, MinMax, and DMinMax at the 10\% level. 
At the same time, the TEV row shows no significant rejections, confirming that no competing model outperforms it. Together, these results indicate that TEV achieves the most robust tracking accuracy among the optimization-based formulations at conventional significance levels.

\end{enumerate}

As a synthesis of the findings, clear distinctions emerge among the eight optimization-based portfolios.  
The SES model is the weakest tracker: with only two holdings, it achieves limited diversification and records a correlation of just 69.21\% with the index. Its exceptionally high mean return is offset by elevated volatility and drawdown, making it unstable and unreliable.  

Among the remaining portfolios, MSE delivers the strongest overall performance on mean return and risk-adjusted ratios, followed closely by MinMax. However, their statistical advantage in tracking error is limited, being significantly better only relative to SES and MADD.  

By contrast, TEV consistently dominates. It achieves the lowest tracking error, the highest correlation (99.25\%), and statistically outperforms all peers in terms of tracking accuracy. At the same time, its return and risk profile closely match that of the index, making TEV the most effective and reliable optimization-based index-tracking model in this comparison.

\begin{figure}[ht!]
\centering
\includegraphics[height=8cm, width=\textwidth]{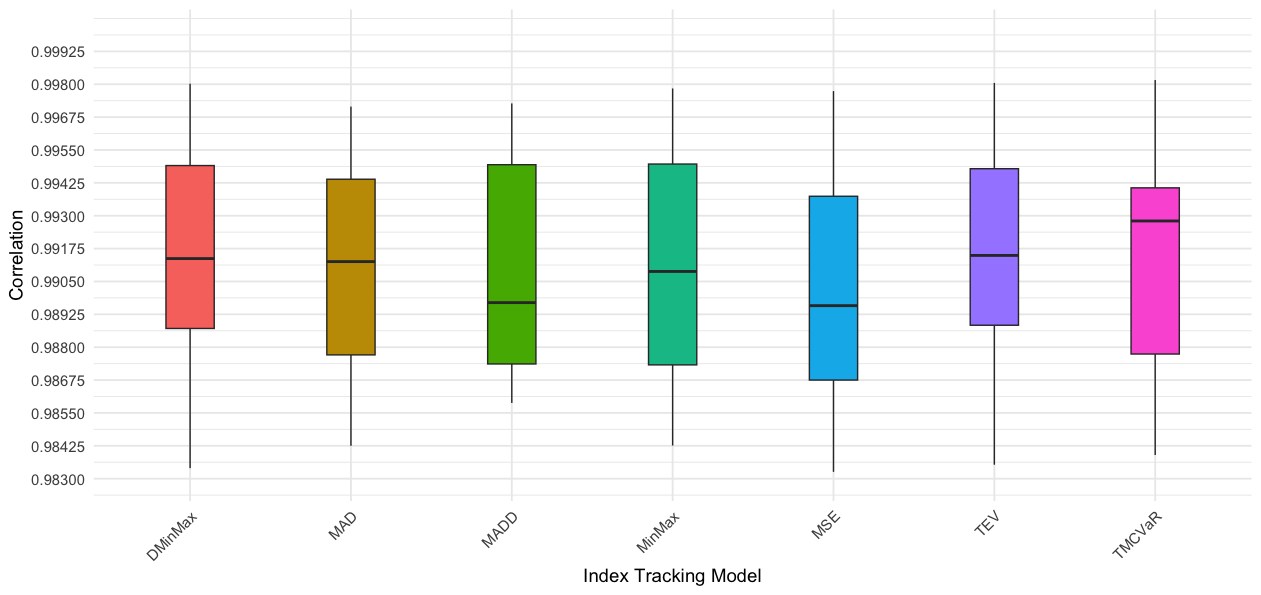}
\caption{Boxplots of correlations between the S\&P~500 index and optimization-based tracking portfolios across 32 rolling out-of-sample windows.}
\label{fig:corr1_opt}
\end{figure}

\begin{figure}[ht!]
\centering
\includegraphics[height=8cm, width=\textwidth]{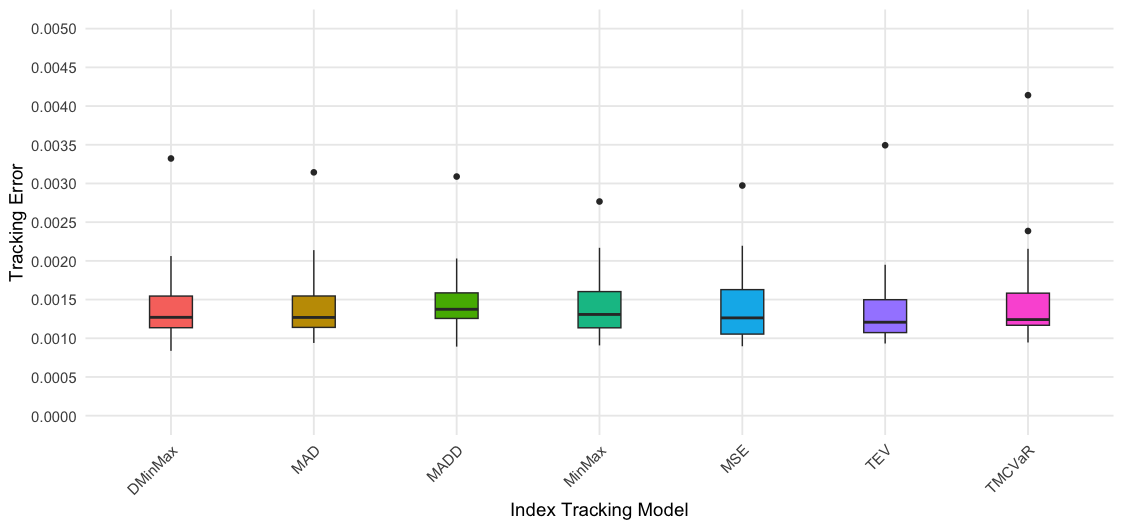}
\caption{Boxplots of tracking errors for optimization-based index-tracking portfolios across 32 rolling out-of-sample windows.}
\label{fig:te_opt}
\end{figure}

Figure~\ref{fig:corr1_opt} presents the distribution of correlations with the index for the optimization-based models across rolling windows. 
SES is excluded because its correlations are dispersed over a much wider range, which would distort the visualization. For the remaining models, correlations are tightly clustered around 0.99, indicating consistently strong tracking accuracy. Among them, TMCVaR attains the highest median correlation, while TEV exhibits the narrowest interquartile range, reflecting greater stability across windows. The correlation ranges of TEV and TMCVaR overlap; TEV is more stable, whereas TMCVaR has a slightly higher median value. By comparison, models such as DMinMax, MAD, and MSE show slightly wider spreads and lower minima, pointing to weaker performance in certain windows. Overall, while all models (excluding SES) deliver robust results, TEV stands out as the most reliable tracker, with TMCVaR in second place.

Figure~\ref{fig:te_opt} presents the distribution of tracking errors for the optimization-based models across rolling windows, with SES excluded due to its extremely large error range. All remaining portfolios achieve very low errors, generally clustered between 0.001 and 0.0015, underscoring their effectiveness in replicating the benchmark index. Among them, TEV and MinMax record the lowest median errors with relatively narrow spreads, indicating strong and consistent performance. 
TEV stands out further, as its full range lies below that of MinMax, making it the most accurate tracker in terms of error minimization. By contrast, MSE and TMCVaR exhibit greater variability and occasional high outliers, suggesting less stable performance across windows. MAD, MADD, and DMinMax occupy the middle ground, with tracking errors that are stable but not among the best. Overall, excluding SES, all optimization models demonstrate credible tracking accuracy, with TEV emerging as the most reliable, followed by MinMax, while MSE and TMCVaR are comparatively more dispersed.

Having examined the optimization-based formulations, we next evaluate the statistical-based models.

\subsubsection{Performance evaluation of Statistical based index tracking optimization models}
\begin{table*}[t!]
\centering
\caption{Out-of-sample performance of statistical-based index-tracking portfolios. 
Panel~I reports performance measures, while Panel~II shows pairwise $p$-values from one-sided paired $t$-tests on tracking error. 
Within each block, the best values among Lasso- and Elastic-Net–type penalties are in \textbf{bold}, and the overall best values are in \textit{italics}.}
\label{tab:OSP_stat}
\scalebox{0.75}{%
\begin{tabular}{l|cccccccc}
\toprule
\multicolumn{9}{c}{\textbf{Panel I: Out-of-sample performance}} \\
\midrule
\textbf{Models} & LSR & QR & NNL & NNEN & Factor & CoInt-Sim & Cvx-CoInt & Index \\
\midrule
\multicolumn{9}{l}{\textbf{Tracking performance}} \\
\quad Tracking error & \textit{0.01322} & 0.01307 & 0.00303 & 0.00206 & 0.00312 & 0.00246 & \textbf{0.00148} & 0.00000 \\
\quad Correlation & \textit{63.00\%} & 64.70\% & 97.77\% & 98.73\% & 96.75\% & 98.01\% & \textbf{99.17\%} & 100\% \\
\midrule
\multicolumn{9}{l}{\textbf{Return performance}} \\
\quad Average return & \textbf{0.00069} & 0.00105 & 0.00055 & 0.00050 & 0.00053 & \textit{0.00047} & 0.00049 & 0.00042 \\
\quad Minimum return & \textit{-0.13449} & \textbf{-0.09626} & -0.12917 & -0.12493 & -0.11511 & -0.12531 & -0.11753 & -0.11984 \\
\quad Maximum return & \textbf{0.18836} & 0.16664 & 0.12265 & 0.11057 & 0.10422 & 0.11434 & \textit{0.09713} & 0.09383 \\
\midrule
\multicolumn{9}{l}{\textbf{Risk performance}} \\
\quad Volatility & 0.01700 & \textit{0.01713} & 0.01301 & 0.01225 & 0.01223 & 0.01214 & \textbf{0.01148} & 0.01142 \\
\quad Avg drawdown & \textit{0.21365} & 0.08094 & 0.04244 & 0.03771 & 0.04687 & 0.04270 & \textbf{0.03626} & 0.03835 \\
\midrule
\multicolumn{9}{l}{\textbf{Risk-adjusted performance}} \\
\quad Sharpe ratio & 0.04077 & \textbf{0.06146} & 0.04217 & 0.04056 & 0.04297 & \textit{0.03836} & 0.04226 & 0.03650 \\
\quad Sortino ratio & 0.05964 & \textbf{0.08938} & 0.05775 & 0.05556 & 0.05808 & \textit{0.05205} & 0.05740 & 0.04931 \\
\quad Treynor ratio & 0.00074 & \textbf{0.00108} & 0.00049 & 0.00047 & 0.00051 & \textit{0.00045} & 0.00049 & 0.00042 \\
\quad Information ratio & 0.02091 & \textbf{0.04869} & 0.04357 & 0.03889 & 0.03493 & \textit{0.01998} & 0.04629 & -- \\
\midrule
\multicolumn{9}{l}{\textbf{Turnover and complexity}} \\
\quad Turnover (TR) & 1.49575 & 1.17703 & \textbf{0.56941} & 0.58561 & 1.04997 & \textit{2.02480} & 1.12265 & -- \\
\quad No. of assets & 3 & 3 & 32 & 38 & 38 & 43 & 45 & -- \\
\quad Solver time & 3.98s & 6.02s & 1.90s & 7.46s & 7.62s & 19.49m & 18.10h & -- \\
\midrule
\multicolumn{8}{c}{\textbf{Panel II: Pairwise $p$-values of tracking error $t$-tests}} \\
\midrule
$p$-values & LSR & QR & NNL & NNEN & Factor & CoInt-Sim & Cvx-CoInt \\
\midrule
LSR & ****** & 0.64100 & 1.4E-12 & 6.3E-13 & 2.4E-12 & 3.5E-13 & 1.2E-13 \\
QR & 0.35900 & ****** & 4.8E-16 & 2.0E-16 & 4.4E-16 & 8.5E-17 & 6.0E-17 \\
NNL & 1.00000 & 1.00000 & ****** & 4.5E-07 & 0.81668 & 5.6E-09 & 1.6E-10 \\
NNEN & 1.00000 & 1.00000 & 1.00000 & ****** & 1.00000 & 0.99830 & 2.4E-10 \\
Factor & 1.00000 & 1.00000 & 0.18332 & 3.4E-10 & ****** & 1.3E-05 & 5.7E-14 \\
CoInt-Sim & 1.00000 & 1.00000 & 0.99994 & 0.00169 & 0.99998 & ****** & 2.8E-10 \\
Cvx-CoInt & 1.00000 & 1.00000 & 1.00000 & 1.00000 & 1.00000 & 1.00000 & ****** \\
\bottomrule
\end{tabular}
}
\end{table*}

Table~\ref{tab:OSP_stat} reports the out-of-sample performance of the seven statistical-based index-tracking models introduced in Table \ref{tab:models}. Several clear patterns emerge from the results.

\begin{enumerate}
    \item \textbf{Tracking performance:}  
    \begin{itemize}
        \item \textit{Tracking error:} The convex co-integration model (Cvx\_CoInt) achieves the lowest tracking error (0.00148), followed by the NNEN model (0.00206). The cointegration with simulations also delivers a comparably low error (0.00246). In contrast, regression-based baselines such as LSR and QR record substantially higher errors, reflecting their limited diversification and weaker ability to replicate index movements.  

        \item \textit{Correlation with the index:} Unlike the optimization-based models, where correlations with the benchmark consistently exceed 99\%, only Cvx\_CoInt approaches this level in the statistical group (99.17\%). NNEN and CoInt\_Sim follow with correlations of 98.73\% and 98.01\%, respectively. By contrast, LSR and QR again perform poorly, with correlations of only 63–65\%, underscoring their inability to capture index dynamics reliably. Intermediate performance is observed for NNL and the Factor model, which achieve correlations in the 96–98\% range.  
\end{itemize}
\item \textbf{Return performance:}  
\begin{itemize}
    \item The regression-based models (LSR and QR) stand out with unusually high average returns, with QR recording the highest among all models (0.00105). However, this comes at the cost of poor diversification, as both models select only three assets across rolling windows. Consequently, they exhibit extreme outcomes, with LSR attaining the largest maximum return (0.18836) but also the deepest minimum return ($-0.13449$).  

    \item In contrast, the remaining five models yield average returns within a narrow range ($\approx 0.00047$–0.00055), moderately above the index average (0.00042). Among these, NNL produces the highest mean return, closely followed by the Factor model. NNL also achieves the strongest maximum return (0.12265) within this diversified group, though it simultaneously suffers the most severe loss, indicating greater volatility in its outcomes.  

    \item Notably, Cvx\_CoInt, which already demonstrates the best tracking accuracy, also provides the closest alignment with the index in terms of return distribution. Its average return (0.00049) lies just above the benchmark’s (0.00042), and both its maximum and minimum returns remain tightly clustered around those of the index. This indicates that the Cvx\_CoInt portfolio not only minimizes tracking error but also replicates the benchmark’s return profile more faithfully than its peers.  
\end{itemize}

\item \textbf{Risk performance:}  
\begin{itemize}
    \item The regression-based models (LSR and QR), constrained by their limited diversification, exhibit the highest volatility and drawdowns. Their concentrated asset allocations expose them to large swings in portfolio value, underscoring their fragility as index trackers.  

    \item Among the diversified portfolios, NNL emerges as the riskiest, reporting the highest standard deviation (0.01301) and the second-largest average drawdown (0.04244).  

    \item By contrast, the co-integration–based approaches (Cvx\_CoInt and CoInt\_Sim) and the NNEN record the lowest and most stable risk values. In particular, Cvx\_CoInt not only outperforms its peers but also aligns most closely with the benchmark index, reinforcing its reliability in terms of stability.  
\end{itemize}

\item \textbf{Risk-adjusted performance:}  
\begin{itemize}
    \item Driven by its exceptionally high average return, QR attains the highest Sharpe, Sortino, and Treynor ratios. However, these elevated ratios are misleading, as they arise from extreme return realizations combined with poor diversification and high volatility. LSR, on the other hand, fails to deliver competitive ratios: its low mean return relative to elevated risk prevents it from achieving meaningful risk-adjusted gains.  

    \item Within the well-diversified group, the Factor model and Cvx\_CoInt delivers the strongest performance across all risk-adjusted measures, closely followed by the NNL. Notably, all five diversified trackers surpass the benchmark in terms of Sharpe, Sortino, and Treynor ratios. This demonstrates the ability of statistical frameworks, when properly specified, to provide superior risk-adjusted performance compared to direct index replication.  
\end{itemize}

\item \textbf{Turnover and asset count:}  
Although the desired cardinality was set at 45 assets, the regression-based portfolios (LSR and QR) consistently select only 2–3 assets. This limited diversification inflates their turnover ratios, reflecting frequent and concentrated reallocations. By contrast, the remaining five models allocate between 32 and 45 assets, resulting in substantially lower transaction costs. Among these, NNL and NNEN deliver the most favorable turnover values. Notably, Cvx\_CoInt stands out within the co-integration family, as it both attains relatively low turnover and reliably selects the full 45 assets, thereby adhering most closely to the cardinality constraint.  

 \item \textbf{Computational efficiency:}  
The regression-based models (LSR and QR) and the penalized regressions (NNL and NNEN) are extremely fast, producing solutions within only a few seconds across all rolling windows. By contrast, the co-integration-based models are considerably more computationally demanding. CoInt\_Sim requires on average about 20 minutes to solve across 32 windows, while Cvx\_CoInt frequently hits the maximum solver time limit of 30 minutes per window. This disparity highlights an important trade-off: while co-integration models provide superior tracking accuracy and stability, they do so at the expense of substantially higher computational costs.  

\item \textbf{Statistical significance (Panel II):}  
The $p$-value matrix reveals that Cvx\_CoInt delivers statistically lower tracking errors than nearly all competing models, with highly significant rejections ($p < 0.05$ and often $p < 0.001$) against regression-based baselines (LSR, QR) and other statistical trackers (NNL, NNEN, Factor). Crucially, the row corresponding to Cvx\_CoInt contains no significant rejections, confirming that no alternative model consistently surpasses it. NNEN emerges as the second-strongest performer, showing significant improvements in TE relative to most trackers but not relative to Cvx\_CoInt. Finally, CoInt\_Sim demonstrates competitive performance, producing significantly lower errors than several peers but falling short of the NNEN and its convex counterpart.  
\end{enumerate}


\begin{figure}[htp!]
\centering
\includegraphics[height=7cm, width=\textwidth]{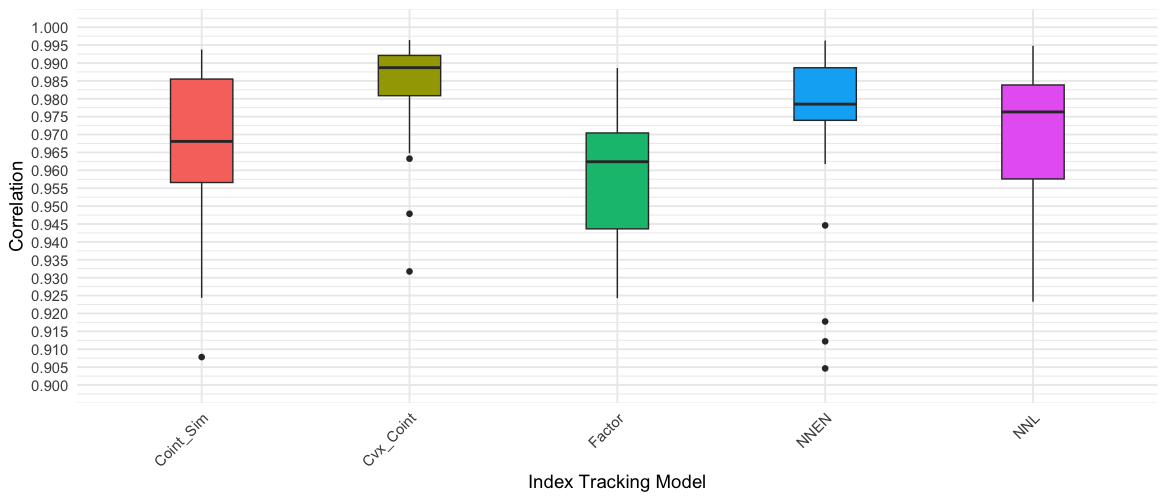}
\caption{Box plot distribution of correlation with index for the statistical-based index tracking models}
\label{fig:corr1_stat}
\end{figure}

Overall, the analysis of the statistical-based index tracking models highlights clear contrasts in performance. The regression-based approaches (LSR and QR) fail to construct diversified portfolios, allocating to only a handful of assets and thereby achieving correlations of merely 63–65\% with the benchmark. Among the five well-diversified models, the co-integration-based portfolios emerge as the most effective trackers, with Cvx\_CoInt consistently outperforming its peers. While the NNL model delivers the highest average return, this comes at the cost of elevated volatility and drawdowns, reducing its attractiveness as a practical strategy. In contrast, both Cvx\_CoInt and CoInt\_Sim generate lower risk values, with Cvx\_CoInt offering the closest alignment with the index across return and risk measures. The statistical significance tests further reinforce this dominance: Cvx\_CoInt produces significantly lower tracking errors than all competitors, with no model found to outperform it. Taken together, these findings establish Cvx\_CoInt as the strongest candidate for index-tracking funds within the statistical modeling paradigm. From an investment perspective, these results suggest that co-integration-based strategies, particularly Cvx\_CoInt, provide the most reliable replication of the index. While they demand greater computational resources, they deliver statistically validated tracking advantages that regression-based approaches cannot match.

Figure \ref{fig:corr1_stat} shows the distribution of correlations with the index for the statistical-based models, excluding LSR and QR to avoid distortion from their substantially lower correlations. Among the remaining trackers, the co-integration models (CoInt\_Sim and Cvx\_CoInt) exhibit high correlations but with different stability. Cvx\_CoInt demonstrates the most robust behavior, with tightly clustered correlations around 0.99, and a narrow interquartile range compared to CoInt\_Sim. NNEN and NNL also achieve strong median correlations near 0.98, though they exhibit slightly wider variability and a few downward outliers. By contrast, the Factor model shows a broader distribution and a lower median correlation (around 0.96), underscoring its weaker and less stable tracking ability. Overall, co-integration-based models deliver the most reliable correlation performance in this category, with Cvx\_CoInt clearly setting the benchmark.


\begin{figure}[htp!]
\centering
\includegraphics[height=8cm, width=\textwidth]{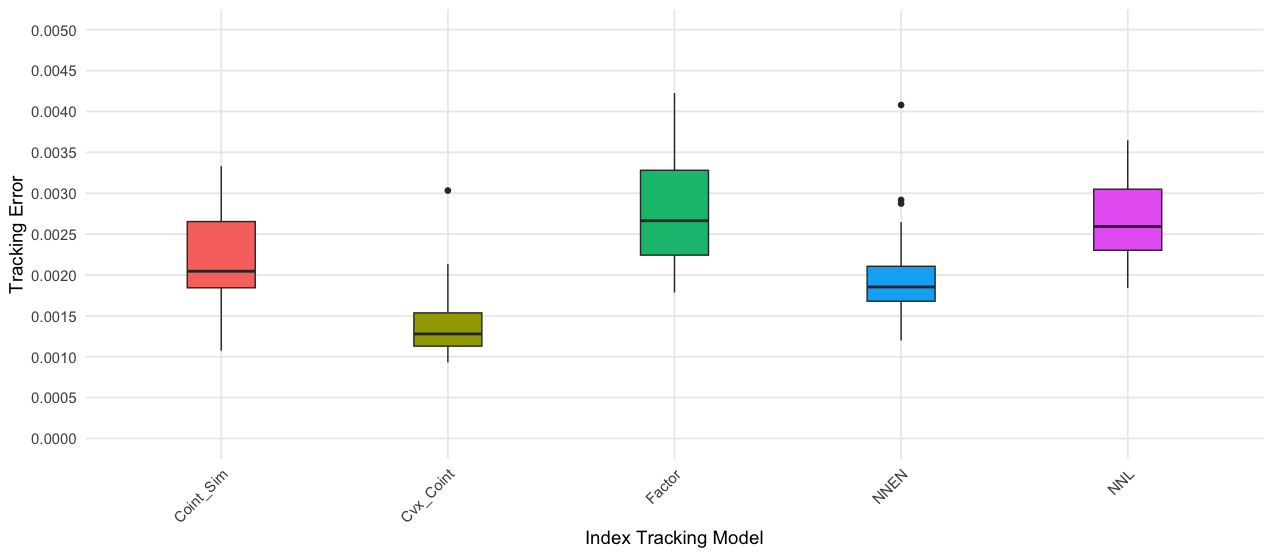}
\caption{Box plot distribution of tracking error for the statistical-based index tracking models}
\label{fig:TE_stat}
\end{figure}

Next, Figure \ref{fig:TE_stat} presents the distribution of tracking errors for the statistical-based models, again excluding LSR and QR to avoid distortion from their much larger error ranges. Among the remaining approaches, Cvx\_CoInt achieves the lowest median tracking error and the narrowest interquartile spread, underscoring its reliability as the most effective statistical formulation. CoInt\_Sim also performs well, though with slightly higher variability, it still maintains consistently low error levels. In contrast, the Factor, NNEN, and NNL exhibit higher medians, wider dispersions, with more frequent outliers, indicating weaker accuracy and less stability relative to the co-integration models.  

Taken together, the graphical evidence confirms that while CoInt\_Sim is a competitive tracker, Cvx\_CoInt clearly delivers the most accurate and consistent tracking performance among all statistical-based methods. The next subsection turns to the data-driven category, where machine learning algorithms provide alternative approaches to index tracking.

\subsubsection{Performance evaluation of Data-driven index tracking models}

\begin{sidewaystable}[htp!]
\centering
\caption{Out-of-sample performance of data-driven index-tracking portfolios. 
Panel~I reports performance measures; Panel~II shows pairwise $p$-values from one-sided paired $t$-tests on tracking error. 
Best values are shown in \textbf{bold} and worst in \textit{italics}.}
\label{tab:OSP_ML}
\tiny
\begin{tabular}{l|ccccccccccccccc}
\toprule
\multicolumn{16}{c}{\textbf{Panel I: Out-of-sample performance}} \\
\midrule
\textbf{Models} & SH-AE & Con-AE & DEN-AE & SP-AE & STCK-AE & VAR-AE & $\epsilon$-SVR & $\nu$-SVR & RF-Clust & RF-Reg & DeepNN & DeepNNF & Clust1 & Clust2 & Index \\
\midrule
\multicolumn{16}{l}{\textbf{Tracking performance}} \\
\quad Tracking error & 0.00402 & 0.00403 & 0.00406 & 0.00415 & 0.00388 & 0.00380 & 0.00475 & 0.00405 & 0.00288 & 0.00271 & \textit{0.00498} & \textbf{0.00153} & 0.00271 & 0.00262 & 0.00000 \\
\quad Correlation (\%) & 94.10 & 93.99 & 94.12 & 93.71 & 94.48 & 94.70 & 91.60 & 93.51 & 97.22 & 98.10 & \textit{90.14} & \textbf{99.11} & 97.23 & 97.39 & 100.00 \\
\midrule
\multicolumn{16}{l}{\textbf{Return performance}} \\
\quad Average return & 0.00053 & 0.00053 & 0.00049 & 0.00062 & 0.00057 & 0.00046 & 0.00046 & 0.00047 & 0.00056 & 0.00049 & 0.00040 & 0.00047 & 0.00042 & 0.00045 & 0.00042 \\
\quad Minimum return & -0.13007 & -0.12409 & -0.13379 & -0.12429 & -0.12524 & -0.12589 & -0.11366 & \textbf{-0.09710} & \textit{-0.13389} & -0.12335 & -0.12952 & -0.11643 & -0.11955 & -0.11149 & -0.11984 \\
\quad Maximum return & 0.12554 & 0.12444 & \textbf{0.12906} & 0.12884 & 0.12376 & 0.11690 & 0.12005 & \textit{0.09370} & 0.10342 & 0.12204 & 0.10793 & 0.10149 & 0.11475 & 0.09608 & 0.09383 \\
\midrule
\multicolumn{16}{l}{\textbf{Risk performance}} \\
\quad Volatility & 0.01186 & 0.01175 & 0.01202 & 0.01187 & 0.01182 & 0.01182 & 0.01171 & \textbf{0.01057} & 0.01216 & \textit{0.01277} & 0.01086 & 0.01144 & 0.01156 & 0.01152 & 0.01142 \\
\quad Avg drawdown & 0.04079 & 0.03891 & \textit{0.04438} & \textbf{0.03478} & 0.03780 & 0.04365 & 0.04230 & 0.03528 & 0.03885 & 0.04385 & 0.03782 & 0.03650 & 0.03859 & 0.03870 & 0.03835 \\
\midrule
\multicolumn{16}{l}{\textbf{Risk-adjusted performance}} \\
\quad Sharpe ratio & 0.04443 & 0.04522 & 0.04113 & \textbf{0.05234} & 0.04846 & 0.03891 & 0.03960 & 0.04405 & 0.04586 & 0.03838 & 0.03725 & 0.04065 & \textit{0.03635} & 0.03892 & 0.03650 \\
\quad Sortino ratio & 0.06085 & 0.06192 & 0.05635 & \textbf{0.07198} & 0.06668 & 0.05325 & 0.05429 & 0.05919 & 0.06151 & 0.05257 & 0.05099 & 0.05537 & \textit{0.04938} & 0.05308 & 0.04931 \\
\quad Treynor ratio & 0.00054 & 0.00055 & 0.00050 & \textbf{0.00064} & 0.00059 & 0.00047 & 0.00049 & 0.00054 & 0.00054 & 0.00045 & 0.00047 & 0.00047 & \textit{0.00043} & 0.00046 & 0.00042 \\
\quad Information ratio & 0.02742 & 0.02843 & 0.01910 & \textbf{0.04925} & 0.04022 & 0.01141 & 0.00989 & 0.01215 & 0.04907 & 0.02702 & \textit{-0.00247} & 0.03167 & 0.00133 & 0.01214 & -- \\
\midrule
\multicolumn{16}{l}{\textbf{Turnover and complexity}} \\
\quad Turnover (TR) & 0.82699 & 0.85497 & 0.82795 & 0.81242 & 0.85659 & 0.87958 & 1.72386 & 1.68549 & 1.26397 & 1.11341 & \textbf{0.59538} & 0.59882 & 1.77349 & \textit{1.85803} & -- \\
\quad No. of assets & 45 & 45 & 45 & 45 & 45 & 45 & 45 & 45 & 45 & 45 & 45 & 45 & 42 & 40 & -- \\
\quad Solver time & 32.05s & 31.79s & 32.01s & 33.08s & 46.67s & 43.14s & 27.84s & 48.91s & 34.47s & 37.46s & 2.20m & 51.63m & 4.99s & 5.02s & -- \\
\midrule
\multicolumn{16}{c}{\textbf{Panel II: Pairwise $p$-values of tracking error $t$-tests}} \\
\midrule
$p$-values & SH-AE & Con-AE & DEN-AE & SP-AE & STCK-AE & VAR-AE & $\epsilon$-SVR & $\nu$-SVR & RF-Clust & RF-Reg & DeepNN & DeepNNF & Clust1 & Clust2 \\
\midrule
SH-AE     & ******  & 0.45387 & 0.34268 & 0.84042 & 0.06520 & 0.19645 & 0.94554 & 0.71912 & 2.52E-05 & 1.95E-07 & 0.99999 & 4.27E-11 & 3.90E-07 & 1.80E-06 \\
Con-AE    & 0.54613 & ******  & 0.37443 & 0.88885 & 0.10054 & 0.21707 & 0.92788 & 0.74956 & 1.29E-04 & 8.34E-07 & 0.99999 & 8.68E-11 & 8.73E-07 & 3.03E-06 \\
DEN-AE    & 0.65732 & 0.62557 & ******  & 0.93842 & 0.19752 & 0.30417 & 0.96628 & 0.77674 & 2.14E-04 & 3.05E-06 & 0.99999 & 8.03E-10 & 3.82E-06 & 2.37E-05 \\
SP-AE     & 0.15958 & 0.11115 & 0.06158 & ******  & 0.02025 & 0.06822 & 0.90975 & 0.55734 & 5.20E-05 & 1.09E-06 & 0.99997 & 1.64E-10 & 7.19E-07 & 3.33E-06 \\
STCK-AE   & 0.93480 & 0.89946 & 0.80248 & 0.97975 & ******  & 0.60735 & 0.98350 & 0.92454 & 3.38E-04 & 6.74E-07 & 1.00000 & 1.59E-10 & 2.81E-06 & 1.35E-05 \\
VAR-AE    & 0.80355 & 0.78293 & 0.69583 & 0.93178 & 0.39265 & ******  & 0.95216 & 0.89787 & 2.01E-05 & 6.30E-07 & 0.99999 & 1.70E-13 & 2.92E-08 & 1.29E-08 \\
Eps-SVR   & 0.05446 & 0.07212 & 0.03372 & 0.09025 & 0.01650 & 0.04784 & ******  & 0.15963 & 1.16E-04 & 1.22E-05 & 0.98727 & 6.56E-08 & 2.32E-05 & 1.06E-04 \\
Nu-SVR    & 0.28088 & 0.25044 & 0.22326 & 0.44266 & 0.07546 & 0.10213 & 0.84037 & ******  & 1.40E-06 & 2.58E-07 & 0.99991 & 7.58E-13 & 1.15E-09 & 4.42E-08 \\
RF-Clust  & 0.99997 & 0.99987 & 0.99979 & 0.99995 & 0.99966 & 0.99998 & 0.99988 & 0.99999 & ******  & 0.04684 & 1.00000 & 1.06E-12 & 0.05163 & 0.05130 \\
RF-Reg    & 1.00000 & 1.00000 & 0.99999 & 0.99999 & 0.99999 & 0.99999 & 0.99999 & 1.00000 & 0.95316 & ******  & 1.00000 & 1.78E-09 & 0.63888 & 0.53653 \\
DeepNN    & 7.73E-06 & 5.61E-06 & 4.58E-06 & 3.19E-05 & 5.75E-07 & 3.38E-06 & 0.01273 & 8.53E-05 & 1.21E-09 & 5.94E-12 & ******  & 7.30E-14 & 4.14E-12 & 4.21E-10 \\
DeepNNF   & 1.00000 & 1.00000 & 1.00000 & 1.00000 & 1.00000 & 1.00000 & 1.00000 & 1.00000 & 1.00000 & 1.00000 & 1.00000 & ******  & 1.00000 & 1.00000 \\
Clust1    & 1.00000 & 1.00000 & 0.99999 & 0.99999 & 0.99999 & 1.00000 & 0.99998 & 1.00000 & 0.94837 & 0.36112 & 1.00000 & 6.77E-12 & ******  & 0.38878 \\
Clust2    & 1.00000 & 1.00000 & 0.99998 & 0.99999 & 0.99999 & 1.00000 & 0.99989 & 1.00000 & 0.94870 & 0.46347 & 1.00000 & 9.91E-15 & 0.61122 & ****** \\
\bottomrule
\end{tabular}
\end{sidewaystable}

Table \ref{tab:OSP_ML} reports the out-of-sample performance of 14 data-driven index-tracking portfolios, spanning five methodological families: clustering (Clust 1, Clust 2), support-vector regression ($\epsilon$-SVR, $\nu$-SVR), random forests (RF-Clust, RF-Reg), deep autoencoders (SH-AE, Con-AE, DEN-AE, SP-AE, STCK-AE, VAR-AE), and deep learning (DNN, DNNF). In contrast to regression-based models in the previous section, all data-driven approaches produce well-diversified portfolios, typically allocating to 40–45 assets per window. The detailed tracking performance is summarized below.  
\begin{enumerate}

\item \textbf{Tracking performance:}  
Among these models, DNNF delivers the strongest results, achieving the lowest tracking error (0.00153) and the highest correlation with the index (99.11\%), thereby standing out as the best data-driven tracker. RF-Reg ranks second, with a TE of 0.00271 and a correlation of 98.10\%, showcasing the effectiveness of tree-based ensemble methods in capturing nonlinear relationships.  

By contrast, the baseline deep neural network (DNN) performs poorly relative to its enhanced counterpart, with the highest TE (0.00498) and the lowest correlation (90.14\%). The SVR-based models also underperform, recording relatively high tracking errors ($\epsilon$-SVR: 0.00475; $\nu$-SVR: 0.00405) and weaker correlations (91.60\% and 93.51\%, respectively).  

Within the autoencoder family, VAR-AE emerges as the best variant, with the lowest TE (0.00380) and the highest correlation (94.70\%), followed closely by STCK-AE (0.00388). The remaining autoencoder variants exhibit comparable performance, with correlations clustered in the 94–95\% range, offering moderate improvements over SVR but falling short of RF-Reg and DNNF.

 \item \textbf{Return performance:}  
Similar to the optimization and statistical based portfolios, nearly all data-driven models (with the exception of DNN) generate higher average returns than the index. The strongest performer is SP-AE, which achieves the highest mean return of 0.00062, followed by STCK-AE at 0.00057.  

In terms of extreme outcomes, the SVR-based models deliver relatively mild downside risk, with $\epsilon$-SVR and $\nu$-SVR recording the least severe minimum returns ($-0.1137$ and $-0.0971$, respectively). By contrast, RF-Clust exhibits the deepest drawdown, with a minimum return of $-0.1339$. On the upside, DEN-AE produces the highest maximum return (0.1291), closely followed by SP-AE (0.1288), reflecting the potential of autoencoder architectures to capture strong return episodes.  

Notably, DNNF—the best tracker in terms of TE and correlation closely mirrors the benchmark’s return profile, with an average return of 0.00047, a maximum of 0.1015, and a minimum of $-0.1164$. Clust 2, one of the clustering-based methods, also aligns closely with the index across all return measures, making it a stable, though less distinctive performer in this category.

\item \textbf{Risk performance:}  
Among the data-driven models, $\nu$-SVR achieves the lowest volatility, with a standard deviation of 0.01057, while SP-AE records the smallest average drawdown at 0.03478. The deep learning portfolios also perform well on risk measures: DNN and DNNF register standard deviations of 0.01086 and 0.01144, respectively, placing them second and third overall. Both models also rank among the portfolios with lower drawdowns, with DNNF in particular aligning closely with the benchmark’s volatility (index: 0.01142).  

By contrast, the random forest models are more volatile, with RF-Reg showing the highest standard deviation (0.01277), making it the riskiest portfolio in this group. The autoencoder-based trackers fall within a moderate range (0.01175–0.0120), with Con-AE at the lower end and DEN-AE at the upper end. Notably, DEN-AE not only exhibits the highest standard deviation among the autoencoders (0.01202) but also suffers the largest drawdowns, reflecting greater exposure to downside risk.

\item \textbf{Risk-adjusted performance:}  
Across the board, all data-driven models outperform the index on risk-adjusted metrics (Sharpe, Sortino, Treynor, and Information ratios), underscoring their ability to generate superior returns relative to risk. SP-AE stands out as the top performer, achieving the highest values across all ratios (Sharpe $=0.0523$, Sortino $=0.0720$, Treynor $=0.00064$, Information $=0.0493$). Its strong performance reflects the combination of elevated mean returns with relatively moderate risk levels, making it the most attractive model in this category from an investment perspective.  

STCK-AE emerges as the closest competitor, consistently ranking second across the ratios. Indeed, most autoencoder-based models (except VAR-AE) maintain above-average ratio values, largely attributable to their higher mean returns.  

At the other end of the spectrum, Clust 1 records the weakest performance, with the lowest Sharpe, Sortino, and Treynor ratios. The baseline DNN also underperforms, as its average return falls below that of the index, rendering its Information ratio undefined. In contrast, DNNF provides more balanced outcomes: although its Sharpe (0.0407), Sortino (0.0554), and Treynor (0.00047) ratios are mid-ranked, it achieves the fourth-best Information ratio (0.0317), confirming its stable and well-diversified profile.

\item \textbf{Turnover and asset count:}  
The autoencoder-based portfolios maintain relatively moderate turnover ratios ($\approx 0.81–0.88$), reflecting stable allocation dynamics across windows. By contrast, the clustering-based models incur the highest turnover costs (Clust 1 $=$1.77, Clust 2 $=$ 1.86), and the SVR-based models also lie on the higher side ($\approx 1.68–1.72$).  

The deep learning models (DNN and DNNF) stand out with the lowest turnover values ($\approx$ 0.60), making them the most cost-efficient strategies among all 14 portfolios. Importantly, all models satisfy the cardinality constraint of 45 assets, except the clustering approaches, which allocate to slightly fewer stocks (42 for Clust 1, 40 for Clust 2). This limited diversification, combined with their elevated turnover, highlights the relative inefficiency of clustering-based trackers in a practical investment setting.

 \item \textbf{Computational efficiency:}  
Solver times vary substantially across the 14 data-driven approaches. Clustering models are the fastest ($\approx 5$s per window), while most autoencoder, SVR, and random forest models solve within a practical range of 27–49s. By contrast, the deep learning models are considerably more demanding: DNN requires 2.2 minutes, and DNNF takes as long as 51.6 minutes across all 32 windows. These results highlight the computational trade-off in training deep architectures, which offer improved tracking precision but at significantly higher time costs relative to shallower machine learning or statistical learners.  

\item \textbf{Statistical significance (Panel II):}  
The $p$-value matrix demonstrates that DNNF clearly dominates this category: its column contains overwhelmingly small values ($p < 10^{-10}$) against all competitors, confirming its significant superiority in terms of tracking error. Within the autoencoder family, SP-AE achieves statistically significant improvements over several peers ($p < 0.05$ in multiple comparisons), reinforcing its strong relative performance. In contrast, the baseline DNN fails to statistically outperform any model, consistent with its weak empirical results. RF-Reg delivers mixed outcomes, performing competitively against some autoencoders but falling short of DNNF, illustrating that while ensemble methods are robust, they cannot match the precision of deep learning with noise regularization.

\end{enumerate}

\begin{figure}[htp!]
\centering
\includegraphics[height=7cm, width=\textwidth]{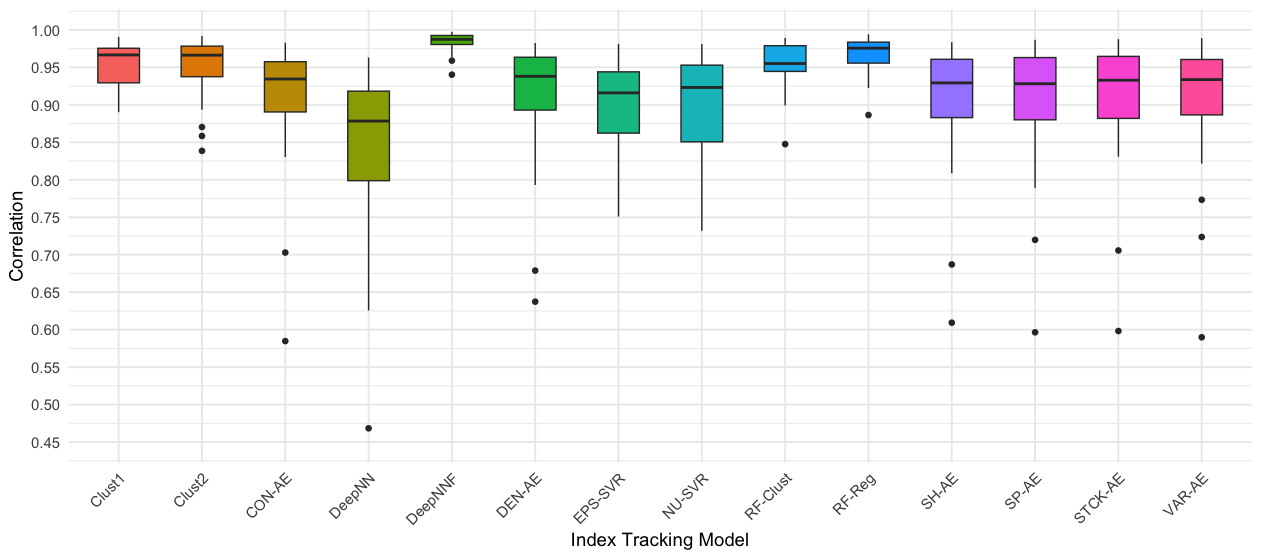}
\caption{Box plot distribution of correlation with index for the data-driven index tracking models}
\label{fig:corr_ml}
\end{figure}
Overall, DNNF emerges as the most effective data-driven index-tracking model, combining the lowest tracking error, the highest correlation, strong statistical significance, and the lowest turnover cost. SP-AE stands out within the autoencoder family, balancing close tracking accuracy with superior Sharpe and Sortino ratios, supported by statistical validation. $\nu$-SVR and RF-Reg appear as competitive alternatives, though both incur higher turnover costs. By contrast, the baseline DNN consistently underperforms across metrics, and is the only model to generate an average return lower than the index.  

Figure \ref{fig:corr_ml} presents the distribution of correlations with the index across the 14 data-driven models. DNNF clearly dominates, with correlations clustering tightly above 0.98, reflecting exceptional reliability across all evaluation windows. The clustering-based approaches (Clust 1 and Clust 2) also perform strongly, maintaining stable correlations in the range of 0.93–0.97. Similarly, RF-Reg and RF-Clust achieve median correlations around 0.95, comparable to clustering methods.  

The autoencoder variants (Con-AE, SH-AE, SP-AE, STCK-AE, VAR-AE, DEN-AE) and the SVR-based models ($\epsilon$-SVR and $\nu$-SVR) deliver moderately high correlations (0.90-0.95), but with wider dispersion and more variability than DNNF. In contrast, the baseline DNN exhibits the weakest performance, with highly unstable behavior and occasional outliers below 0.5. Taken together, these results highlight DNNF as the most robust and reliable data-driven tracker, followed by clustering- and random forest-based approaches, while simpler autoencoders and SVRs remain middle-tier options.

\begin{figure}[htp!]
\centering
\includegraphics[height=7cm, width=\textwidth]{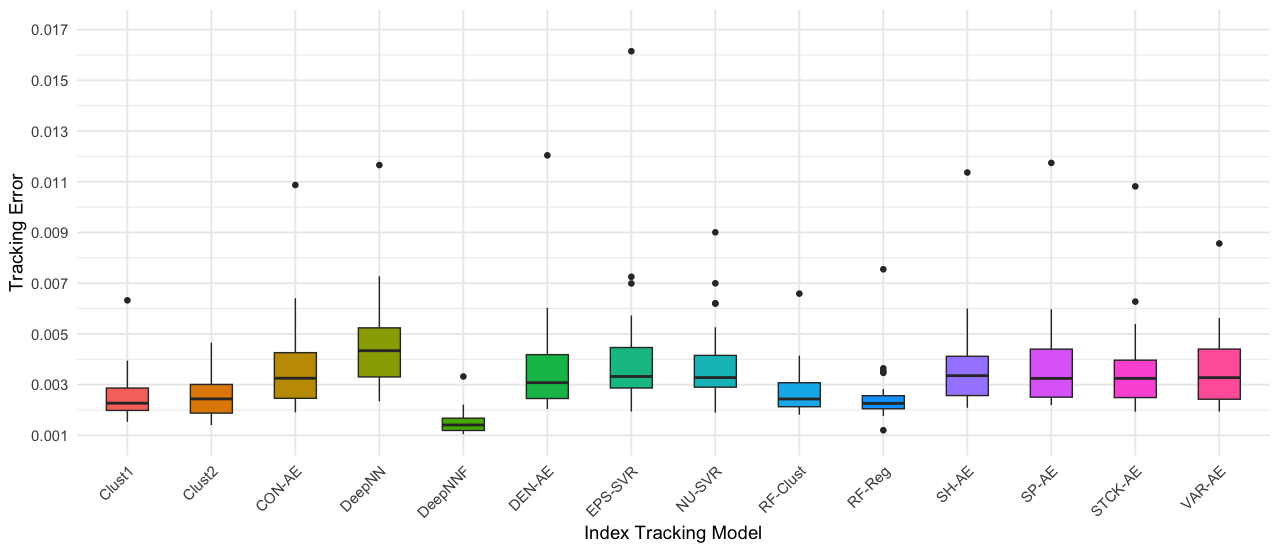}
\caption{Box plot distribution of out-of-sample tracking error for the data-driven index tracking models}
\label{fig:TE_ML}
\end{figure}

Figure \ref{fig:TE_ML} presents the distribution of tracking errors for the data-driven models. DNNF clearly outperforms all others, achieving the lowest and most stable tracking error, with values tightly concentrated around very small magnitudes—underscoring its robustness across evaluation windows. The clustering-based models (Clust 1 and Clust 2) also perform competitively, maintaining low medians with modest spreads. In contrast, DNN and the SVR-based approaches ($\epsilon$-SVR and $\nu$-SVR) record higher median errors alongside wide variability and numerous outliers, indicating weaker and less reliable tracking accuracy. The autoencoder-based variants (SH-AE, SP-AE, STCK-AE, VAR-AE, Con-AE, DEN-AE) fall in the middle, producing acceptable but clearly inferior error distributions compared to DNNF and clustering methods.  

Taken together, these results reinforce DNNF as the most accurate and stable data-driven tracker, closely followed by clustering-based approaches, while SVR and baseline DNN remain the least effective within this category.

\noindent
\textbf{Which modeling framework is generally better suited for tracking the S\&P 500?}  
We now provide a comparative perspective across the three modeling paradigms—optimization-based models (8 variants), statistical-based models (7 variants), and data-driven models (14 variants)—to assess their relative effectiveness in replicating the S\&P 500 index.  

\begin{itemize}
    \item \textit{Tracking error and correlation with the index:}  
    Excluding SES, which is insufficiently diversified (allocating to only 2–3 assets), the remaining optimization-based portfolios achieve exceptionally tight tracking errors in the range $0.00142$–$0.00159$ (median $=0.00146$) and very high correlations of 99.09\%–99.25\% (median $=99.09\%$).  
    Statistical-based models (excluding the poorly diversified LSR and QR) deliver somewhat weaker performance, with TE values between $0.00148$–$0.00303$ (median $=0.00206$) and correlations of 96.75\%–99.17\% (median $=98.64\%$).  
    Data-driven models are the most heterogeneous: while the best (DNNF) achieves TE $=0.00153$ and correlation $=99.11\%$, the category overall spans a wider TE range of $0.00153$–$0.00498$ (median $=0.00388$) and correlation range of 90.14\%–99.11\% (median $=94.3\%$).  
    These results indicate that optimization formulations, as a group, provide the most accurate and consistent index replication, with statistical approaches ranking second and data-driven approaches showing greater variability despite some standout performers.  

    \item \textit{Information ratio:}  
    In terms of risk-adjusted excess return, optimization models generate information ratios between $0.0174$–$0.0550$ (median $=0.0342$), while statistical models achieve $0.0209$–$0.0545$ (median $=0.0436$). Data-driven models, by contrast, span $0.0099$–$0.0493$ (median $=0.0270$).  
    Although the single highest information ratio is produced by an optimization-based model, the statistical frameworks provide the strongest performance overall, delivering the highest median and a tighter concentration of values. This suggests that statistical formulations are particularly effective when the objective is not just tight index replication, but also improving risk-adjusted returns.  

    \item \textit{Turnover ratio:}  
    Turnover, a proxy for transaction costs, further differentiates the paradigms. Optimization-based portfolios fall in the range $1.14$–$1.91$ (median $=1.20$), implying frequent rebalancing and higher trading costs. Statistical-based models report a similar but slightly lower median of $1.18$. In contrast, data-driven models are substantially more cost-efficient, spanning $0.60$–$1.86$ with a median of $0.87$.  
    These results highlight a fundamental trade-off: optimization models achieve the tightest tracking but at the expense of higher turnover, whereas data-driven models strike a better balance between tracking accuracy and cost efficiency.  

    \item \textit{Computational efficiency:}  
    A final distinction emerges in solver times. All optimization models (except SES) are computationally intensive, requiring 17–18 hours on average per evaluation window due to the complexity of cardinality-constrained formulations. Among statistical models, only Cvx\_CoInt incurs comparable costs (18 hours), while the others are solved in seconds. Data-driven models are generally the most efficient, with clustering, random forest, SVR, and autoencoder methods completing in seconds, though deep learning models (DNN and DNNF) remain computationally demanding, requiring minutes rather than seconds.  
    From an implementation standpoint, statistical and most machine learning methods are far more practical for large-scale or high-frequency applications, whereas optimization-based formulations may be prohibitive without access to substantial computing resources.  
\end{itemize}

Taken together, the comparative evidence reveals clear trade-offs across the three paradigms: optimization models deliver the most precise index replication but at high turnover and computational cost; statistical models stand out in risk-adjusted performance, particularly information ratios; and data-driven methods offer scalable, low-cost solutions with moderate tracking accuracy. The ``best” framework thus depends on the investor’s priority, tight replication, enhanced return–risk trade-offs, or efficiency and scalability. The next section sharpens these contrasts through a head-to-head comparison of the top-performing models from each category.

\subsubsection{Comparative analysis of the best models from the three paradigms}  

This section compares the strongest performers from each modeling paradigm—optimization, statistical, and data-driven—based on their out-of-sample tracking ability, return–risk characteristics, and statistical robustness. The three models identified are: (1) \textbf{TEV} from optimization-based formulations, (2) \textbf{Cvx\_CoInt} from statistical approaches, and (3) \textbf{DNNF} from data-driven methods.

\begin{table*}[htp!]
\centering
\caption{Comparison of the best index-tracking models across modeling paradigms (optimization: TEV; statistical: Cvx-CoInt; data-driven: DNNF) and the S\&P~500 index. 
Panel~I reports performance measures; Panel~II shows pairwise $p$-values from one-sided paired $t$-tests on tracking error. 
Best values among the models are in \textbf{bold}; worst are in \textit{italics}.}
\label{tab:OSP_comp}
\scalebox{0.80}{%
\begin{tabular}{l|cccc}
\toprule
\multicolumn{5}{c}{\textbf{Panel I: Out-of-sample performance}} \\
\midrule
\textbf{Models} & TEV & Cvx\_CoInt & DNNF & S\&P~500 \\
\midrule
\multicolumn{5}{l}{\textbf{Tracking performance}} \\
\quad Tracking error & \textbf{0.00142} & 0.00148 & \textit{0.00153} & -- \\
\quad Correlation (\%) & \textbf{99.25} & 99.17 & \textit{99.11} & -- \\
\midrule
\multicolumn{5}{l}{\textbf{Return performance}} \\
\quad Average return & 0.00044 & \textbf{0.00049} & 0.00047 & \textit{0.00042} \\
\quad Minimum return & \textit{-0.12619} & -0.11753 & \textbf{-0.11643} & -0.11984 \\
\quad Maximum return & 0.10090 & 0.09713 & \textbf{0.10149} & \textit{0.09383} \\
\midrule
\multicolumn{5}{l}{\textbf{Risk performance}} \\
\quad Volatility & \textit{0.01162} & 0.01148 & 0.01144 & \textbf{0.01142} \\
\quad Avg drawdown & 0.03675 & \textbf{0.03626} & 0.03650 & \textit{0.03835} \\
\midrule
\multicolumn{5}{l}{\textbf{Risk-adjusted performance}} \\
\quad Sharpe ratio & 0.03816 & \textbf{0.04226} & 0.04065 & \textit{0.03650} \\
\quad Sortino ratio & 0.05179 & \textbf{0.05740} & 0.05537 & \textit{0.04931} \\
\quad Treynor ratio & 0.00044 & \textbf{0.00049} & 0.00047 & \textit{0.00042} \\
\quad Information ratio & 0.01875 & \textbf{0.04629} & 0.03167 & -- \\
\midrule
\multicolumn{5}{l}{\textbf{Turnover and complexity}} \\
\quad Turnover (TR) & \textit{1.18920} & 1.20571 & \textbf{0.59882} & -- \\
\quad No. of assets & 45 & 45 & 45 & -- \\
\quad Solver time & 18.01h & \textit{18.10h} & \textbf{51.63m} & -- \\
\midrule
\multicolumn{5}{c}{\textbf{Panel II: Pairwise $p$-values of tracking error $t$-tests}} \\
\midrule
$p$-values & TEV & Cvx\_CoInt & DNNF & \\
\midrule
TEV        & ****** & 0.95202 & 0.99344 & \\
Cvx\_CoInt & 0.04798 & ****** & 0.81170 & \\
DNNF       & 0.00656 & 0.18829 & ******  & \\
\bottomrule
\end{tabular}
}
\end{table*}

Table \ref{tab:OSP_comp} reports their comparative performance. All three deliver very tight replication of the S\&P 500, with correlations exceeding 99\% and extremely low tracking errors. Among them, the optimization model TEV provides the most accurate benchmark replication, achieving both the lowest tracking error (0.00142) and the highest correlation (99.25\%). The statistical model Cvx\_CoInt follows closely, but distinguishes itself with the best balance of return and risk, recording the highest Sharpe, Sortino, and Information ratios. The data-driven DNNF model, while slightly weaker on pure tracking precision, is the most cost-effective and computationally efficient, combining low turnover with solver times in minutes rather than hours. The lower panel of Table \ref{tab:OSP_comp} presents the pairwise $p$-values from the tracking error tests. At the 5\% significance level, TEV statistically outperforms both Cvx\_CoInt ($p=0.04798$) and DNNF ($p=0.00656$). The difference between Cvx\_CoInt and DNNF is not statistically significant ($p=0.18829$), indicating that their tracking errors are comparable in practice. Taken together, TEV is the most precise replicator, Cvx\_CoInt the strongest risk-adjusted performer, and DNNF the most efficient in cost and computation.

\begin{figure}[htp!]
\centering
\includegraphics[height=8cm, width=\textwidth]{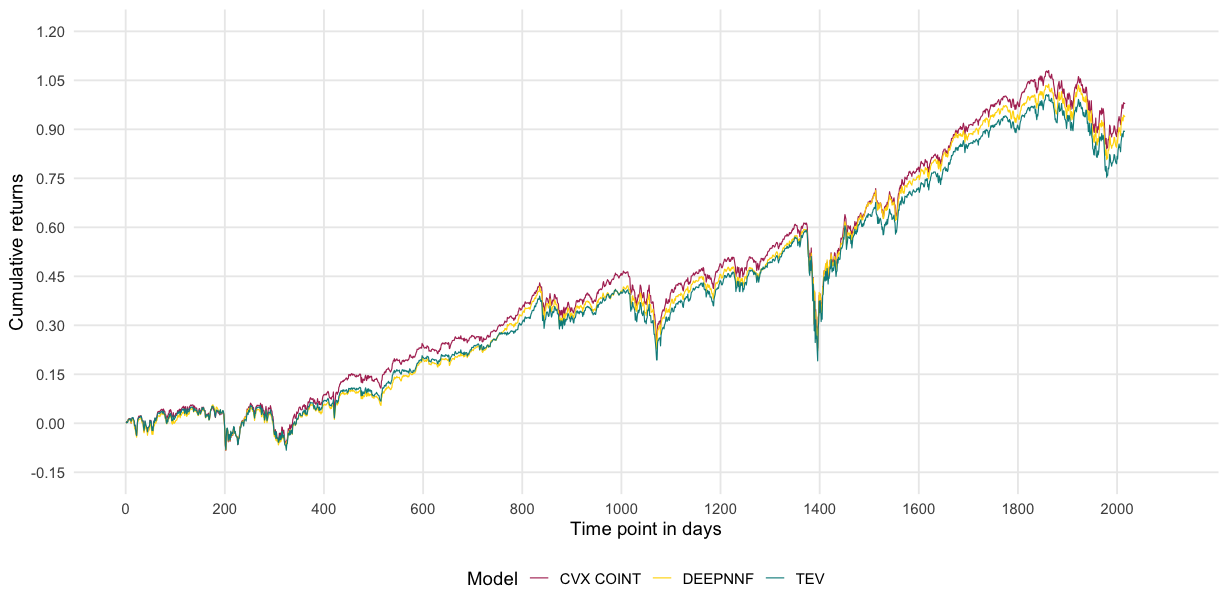}
\caption{Cumulative returns of the best tracking portfolios across the three modeling paradigms for index tracking along with the S\&P 500 index}
\label{fig:CR_best}
\end{figure}

Figure \ref{fig:CR_best} illustrates the cumulative returns of the S\&P 500 index alongside the three benchmark models over 2019 trading days (December 2014–August 2022). All three models track the index closely, moving almost in lockstep. TEV remains the closest to the benchmark throughout, underscoring its superior replication precision. Cvx\_CoInt consistently compounds slightly higher returns during prolonged market upswings, reflecting its stronger risk–return balance. DNNF traces a trajectory nearly identical to Cvx\_CoInt, albeit with slightly lower peaks, while offering the lowest turnover and fastest computation. Overall, this head-to-head comparison confirms that each paradigm brings distinct strengths: TEV sets the standard in pure tracking accuracy, Cvx\_CoInt achieves the most favorable return–risk trade-off, and DNNF demonstrates how modern machine learning can achieve competitive accuracy with unparalleled cost and computational efficiency.

\section{Conclusion}
\label{Sec5}

This review has examined the literature on index tracking by systematically classifying existing approaches into three paradigms: optimization-based models, statistical frameworks, and data-driven machine learning methods. Each paradigm contributes distinct strengths: optimization-based models provide rigorous formulations for precise benchmark replication; statistical methods exploit structural relationships in asset returns; and data-driven approaches introduce flexibility and scalability to capture nonlinear dependencies and complex market dynamics.

Alongside the conceptual survey, a large-scale empirical analysis was conducted on 29 representative models applied to the S\&P~500 index. The findings reveal clear contrasts across paradigms. The tracking-error variance (TEV) model delivers the most accurate replication, achieving the lowest tracking error and the highest correlation with the index. The convex cointegration (Cvx\_CoInt) model attains the strongest risk-adjusted performance, as reflected in superior Sharpe, Sortino, and information ratios. In contrast, the deep neural network with fixed noise (DNNF) achieves competitive replication accuracy while offering substantial advantages in terms of turnover and computational efficiency. These results highlight that the optimal choice of methodology depends on the investor’s objective, whether minimizing replication error, maximizing return–risk trade-offs, or improving scalability.

The review also identifies several promising directions for future research. Important extensions include dynamic model selection across market regimes, explicit incorporation of transaction costs and liquidity constraints, and the development of multi-period rebalancing policies. Moreover, explainable and robust machine learning approaches remain largely unexplored in this context and represent a natural avenue for further study.

To encourage reproducibility and comparability, we provide an open-source repository containing implementations of the models reviewed here together with the experimental framework. This resource allows researchers and practitioners to benchmark new methods directly against established ones, supporting continued progress in the design and evaluation of index-tracking strategies.

\section*{Data Declaration}
The data that support the findings of this study were sourced from Bloomberg data stream and are available from the corresponding author on reasonable request. The codes (R/python) used for the analysis are publicly accessible through the project’s GitHub repository \url{https://github.com/vrindadhingra/index-tracking-review}.

\section*{Declaration of Generative AI and AI-assisted technologies in the writing process}

While preparing this work, the authors used GPT-5 to refine the language and enhance the clarity of the manuscript. After using this tool, the authors reviewed and edited the content as needed and take full responsibility for the content of the publication.

\section*{Disclosure of Interest}

The authors declare no conflict of interest.

\bibliographystyle{sn-basic}
\bibliography{main_AIR}

\end{document}